\documentclass[a4paper,11pt]{article}

\usepackage{jcappub}
\allowdisplaybreaks

\newcommand{\be}{\begin{equation}}
\newcommand{\ee}{\end{equation}}
\newcommand{\bea}{\begin{eqnarray}}
\newcommand{\eea}{\end{eqnarray}}
\newcommand{\nn}{\nonumber}
\def\nl{\\ & \quad}
\def\nlq{\\ & \quad \qquad}

\def\nnl{\nn\\ & \quad}

\def\vct#1{\vec{#1}}

\def\gravthree{G}
\def\rel{r}
\def\dd{\text{d}}
\DeclareMathOperator{\Order}{O}

\title{Equivalence of ADM Hamiltonian and\\ 
Effective Field Theory approaches at\\ 
next-to-next-to-leading order\\ 
spin1-spin2 coupling of binary inspirals}

\author[a,b]{Michele Levi}
\author[c]{and Jan Steinhoff}

\affiliation[a]{Universit\'e Pierre et Marie Curie, CNRS-UMR 7095, 
Institut d'Astrophysique de Paris,\\ 98 bis Boulevard Arago, 75014 Paris, France} 
\affiliation[b]{Sorbonne Universit\'es, Institut Lagrange de Paris,\\ 98 bis Boulevard Arago, 75014 Paris, France} 
\affiliation[c]{Centro Multidisciplinar de Astrofisica, Instituto Superior Tecnico, Universidade de Lisboa,\\ Avenida Rovisco Pais 1, 1049-001 Lisboa, Portugal}

\emailAdd{michele.levi@upmc.fr}
\emailAdd{jan.steinhoff@ist.utl.pt}

\abstract{The next-to-next-to-leading order spin1-spin2 potential for an inspiralling binary, that is essential for accuracy to fourth post-Newtonian order, if both components in the binary are spinning rapidly, has been recently derived independently via the ADM Hamiltonian and the Effective Field Theory approaches, using different gauges and variables. Here we show the complete physical equivalence of the two results, thereby we first prove the equivalence of the ADM Hamiltonian and the Effective Field Theory approaches at next-to-next-to-leading order with the inclusion of spins. The main difficulty in the spinning sectors, which also prescribes the manner in which the comparison of the two results is tackled here, is the existence of redundant unphysical spin degrees of freedom, associated with the spin gauge choice of a point within the extended spinning object for its representative worldline. After gauge fixing and eliminating the unphysical degrees of freedom of the spin and its conjugate at the level of the action, we arrive at curved spacetime generalizations of the Newton-Wigner variables in closed form, which can also be used to obtain further Hamiltonians, based on an Effective Field Theory formulation and computation. Finally, we make use of our validated result to provide gauge invariant relations among the binding energy, angular momentum, and orbital frequency of an inspiralling binary with generic compact spinning components to fourth post-Newtonian order, including all known sectors up to date.} 

\begin{document}

\maketitle

\flushbottom

\section{Introduction}

Gravitational Waves (GWs) constitute one of the most fundamental predictions of the theory of General Relativity (GR). 
Within the coming few years second-generation ground-based interferometers, such as Advanced LIGO in the US \cite{LIGO}, 
Advanced Virgo in Europe \cite{Virgo}, and KAGRA in Japan \cite{Kagra}, will start to operate with unprecedented sensitivity, 
making the anticipated direct detection of GWs a realistic prospect.   
Inspiralling binaries of compact objects are among the most promising astrophysical sources in the accessible frequency band 
of such experiments, and have an analytical description in terms of the post-Newtonian (PN) approximation of GR \cite{Blanchet:2013haa}.
Since the search for GWs from these sources employs the matched-filtering technique, a bank of accurate theoretical template waveforms 
is crucial to obtain a successful detection. It turns out that even at relative high order beyond Newtonian Gravity, 
such as the fourth PN (4PN) order, there is a phenomenological impact on the theoretical waveform templates for the binary inspiral.
Moreover, accuracy to 4PN order is crucial in order to gain information about the inner structure of the constituents of the binary through future 
gravitational wave observations \cite{Yagi:2013baa}. Furthermore, astrophysical objects are expected to have large spins \cite{McClintock:2011zq}, 
which may enhance GW signals and increase the expected event rates in the GW detectors \cite{Reisswig:2009vc}. Therefore, PN corrections involving 
spins for rapidly rotating compact objects, which enter already at 1.5 PN order, should be completed to similar high orders as in the non spinning case, 
which was recently completed to 4PN order within the ADM Hamiltonian formalism \cite{Damour:2014jta}.

Indeed, various methods have made much progress in recent years in the obtainment of higher order PN spin effects for the binary inspiral. 
The next-to-leading order (NLO) spin-orbit (SO) interaction at 2.5 PN was first approached at the level of the equations of motion (EOM) 
in \cite{Tagoshi:2000zg}, and later completed in \cite{Faye:2006gx,Blanchet:2006gy}, with the corresponding Hamiltonian derived in 
\cite{Damour:2007nc}. Following the introduction of a novel Effective Field Theory (EFT) approach for the binary inspiral in \cite{Goldberger:2004jt,Goldberger:2007hy}, and its extension to spinning objects in \cite{Porto:2005ac}, the NLO spin1-spin2 (S$_1$S$_2$) 
interaction at 3PN was treated in \cite{Porto:2006bt,Levi:2008nh,Porto:2008tb}, while its corresponding Hamiltonian was found in \cite{Steinhoff:2007mb}.
The latter was the first application of the ADM Hamiltonian formalism with spin, which was successively developed in \cite{Steinhoff:2008zr,Steinhoff:2009ei,Steinhoff:2010zz}.
The NLO spin-squared potential was also computed via the EFT approach in \cite{Porto:2008jj}, and its 
corresponding Hamiltonian was found in \cite{Steinhoff:2008ji,Hergt:2008jn,Hergt:2010pa}.
Following the EFT treatment of the radiative sector in the non spinning case in \cite{Goldberger:2009qd}, 
the S$_1$S$_2$ and spin-squared components of multipole moments were computed to NLO in \cite{Porto:2010zg}, 
where NLO SO radiative effects were obtained via traditional methods in \cite{Blanchet:2006gy}, 
and have been pushed to include tail effects in \cite{Blanchet:2011zv}. The next-to-next-to-leading order (NNLO) SO Hamiltonian 
at 3.5PN was found in \cite{Hartung:2011te}, where its corresponding EOM were computed independently in \cite{Marsat:2012fn,Bohe:2012mr}. 
NNLO SO radiative effects, and related NLO tail effects were treated in \cite{Bohe:2013cla} and \cite{Marsat:2013caa}, respectively.
It should be noted that all the aforementioned radiative effects concern only the energy flux and phasing of the GWs.
The full waveform, including the GW amplitude, is known to 2PN order \cite{Buonanno:2012rv}, 
with preliminary results to 2.5PN order \cite{Porto:2012as}.
Finally, the conservative NNLO S$_1$S$_2$ sector at 4PN, which is in the focus of this work, was computed simultaneously 
in terms of a Hamiltonian in \cite{Hartung:2011ea,Hartung:2013dza}, and an EFT potential in \cite{Levi:2011eq}.

Thus two independent approaches, that stand out as efficient methods to obtain higher order PN spin effects for the binary inspiral, 
are the ADM Hamiltonian approach and the EFT approach. Each of these approaches has its strengths and weaknesses, and in many respects 
they complement each other. Starting from the conceptual level, the EFT approach provides a systematic methodology to construct the 
action to arbitrarily high accuracy, in terms of operators ordered by their relevance and their Wilson coefficients, which is invaluable 
beyond the point particle approximation. Moreover, the EFT approach also provides a natural framework to handle the regularization required 
for higher order corrections in the PN approximation with the standard renormalization scheme, and applies the efficient standard tools
of Quantum Field Theory, such as Feynman diagrams, and thus benefits from existent developed Feynman integral calculus at its disposal. 
In particular, a notable difference between the two approaches is that in the EFT approach all integrations are performed in momentum space, 
whereas in the ADM Hamiltonian approach they are traditionally performed in position space. Another advantageous practice in the EFT approach 
is the use of the non relativistic gravitational (NRG) fields, which were introduced in \cite{Kol:2007bc} for this purpose. The NRG decomposition 
of spacetime is essentially a reduction over the time dimension, and therefore it is the sensible decomposition for the PN limit. The ADM decomposition 
of spacetime on the other hand, is a reduction over the spatial dimensions, and it has been explicitly demonstrated to function less efficiently in 
the context of the PN EFT approach, compared to the NRG decomposition \cite{Kol:2010ze}.
However, while the EFT approach is based on a functional integration in configuration space, the ADM canonical formalism uses phase space variables, 
i.e.~loosely speaking it is integrating out the spatial metric and its canonical conjugate. Higher-order time derivatives then start to appear only 
at 3PN order, compared to 2PN order in the EFT action approach.
Finally, considering the end result, that each of these two methods produce, the advantageous usefulness of the Hamiltonian 
for the straightforward obtainment of gauge-invariant quantities, and for implementations within the effective one-body 
formulation \cite{Buonanno:1998gg}, is indisputable.

In this paper we set out to examine the equivalence of the ADM Hamiltonian and the EFT approaches in the spin-dependent case at NNLO. In particular, 
we compare the NNLO $S_1S_2$ Hamiltonian computed in \cite{Hartung:2011ea,Hartung:2013dza} with the potential computed via the EFT approach in \cite{Levi:2011eq}.  
The task we have here is to some extent similar to that, which was carried out in the non spinning 3PN sector \cite{damour:2000ni, deAndrade:2000gf}. 
However, the inclusion of spins makes this comparison much more intricate. First, it is clear that a mapping between canonical variables, which are the ones used in the Hamiltonian approach, and the EFT variables would be required in order to compare the two results. The other central issue here is that the spin gauge is not fixed prior to the EFT computation, but instead, the unphysical redundant degrees of freedom, associated with the temporal components of the spin tensor $S^{i0}$, are left as independent degrees of freedom. Yet to our advantage, such a mapping to canonical variables was already considered at the level of the effective EFT action in \cite{Hergt:2011ik}, based on the insertion of gauge constraints, that eliminate the redundant degrees of freedom of both the spin and its conjugate, the angular velocity. This fact together with the higher order time derivatives found at NNLO, actually prescribed the manner in which this comparison was tackled. The elimination of higher order time derivatives entails a redefinition of the variables, see e.g.~\cite{Barker:1980, Schafer:1984mr, Damour:1985mt, Damour:1990jh}, 
that would have modified the mapping to canonical variables, and possibly even the spin constraint itself. This indicated that in order to conveniently apply the mapping to canonical variables, we should start by eliminating the temporal spin components $S^{i0}$, which would be necessary, even if we derived the EOM first. 
We should then transform to the canonical variables and potential, and only then eliminate the higher order time derivatives. After performing all of these 
necessary steps the most direct and complete way to compare the two results is just to obtain the corresponding EFT Hamiltonian via a straightforward Legendre transform. We should note that a comparison between ADM and harmonic coordinate approaches at the level of the EOM was made for the NNLO SO sector in \cite{Marsat:2012fn,Bohe:2012mr}, and that comparisons more similar to the one we make here, i.e.~at the level of the effective potentials were made at NLO in \cite{Hergt:2010pa, Levi:2008nh, Levi:2010zu, Hergt:2011ik}, but due to the lower level of complexity at NLO, the general steps noted here, could have been, 
and were indeed, followed in a different order. Hence, an important outcome of the present paper is that in the course of this comparison, we present a general method to transform EFT potentials with spin into Hamiltonians.

The outline of the paper is therefore as follows. In section \ref{EFTpot} we consider the EFT potentials required for this comparison, 
starting with the NNLO S$_1$S$_2$ result from \cite{Levi:2011eq}, in the form which still contains higher order time derivatives of velocities
and spins, and the temporal spin components $S^{0i}$, associated with the redundant degrees of freedom. In section \ref{spinGF} we proceed 
to consider these unphysical spin degrees of freedom, clarify the obtainment of EOM in the EFT for spin, and find the required $S^{0i}$ components 
via an additional EFT computation of the metric. In section \ref{VSXcan} we proceed to redefine the spin and position variables, and the potential 
at the level of the action, such that they correspond to canonical variables, where we arrive at curved spacetime generalizations of the Newton-Wigner 
variables in closed form. In section \ref{htd} we make use of variable redefinitions to eliminate the higher order time derivatives of velocities and 
spins, via the substitution of lower order EOM, where we explicitly extend this treatment for the case of a spin variable. Then in section \ref{EFTH} 
we make a Legendre transform, and obtain the corresponding EFT Hamiltonians, where we also collect the ADM Hamiltonians required for the comparison. 
In section \ref{CT} we then resolve the difference between the EFT and ADM Hamiltonians, using higher order PN canonical transformations, 
and taking into account lower order ones. We arrive at a complete agreement between the ADM and EFT results. Thus in section \ref{GI}
we make use of our validated result, and provide gauge invariant relations among the binding energy, angular momentum, and orbital frequency
of an inspiralling binary with generic compact spinning components to 4PN order, including all known results up to date. 
In section \ref{theendmyfriend} we summarize our main conclusions.

Throughout this paper we use $c\equiv1$, and $\eta_{\mu\nu}\equiv \text{Diag}[1,-1,-1,-1]$. 
Greek letters denote indices in the global coordinate frame, lowercase Latin letters 
from the beginning of the alphabet denote indices in the local Lorentz frame, 
and upper case Latin letters from the beginning of the alphabet denote the body fixed, corotating Lorentz frame.
The spin variables are always considered in the local frame.  
All indices run from 0 to 3, while spatial tensor indices from 1 to 3, are denoted with lowercase Latin letters 
from the middle of the alphabet. Uppercase Latin letters from the middle of the alphabet denote particle labels.
The notation $\int_{\vec{k}} \equiv \int \frac{d^d\vec{k}}{(2\pi)^d}$ is used for abbreviation,
where $d$ is the number of spatial dimensions. 
The scalar triple product appears here with no brackets, i.e.~$\vec{a}\times\vec{b}\cdot\vec{c}\equiv(\vec{a}\times\vec{b})\cdot\vec{c}$. 
The notation $[\,\,\,\,]$ over indices extracts the totally antisymmetric part of the tensor, e.g.~$T_{[ab]}=\frac{1}{2}(T_{ab}-T_{ba})$.

\section{EFT potentials}\label{EFTpot}

In this section and the next we build on the work in \cite{Levi:2011eq} and on sections II and III of \cite{Levi:2010zu}, following similar definitions and conventions as those that were used there.
In order to prove the equivalence of the NNLO S$_1$S$_2$ interaction potential for a binary system of compact spinning objects 
at 4PN order, computed in \cite{Levi:2011eq}, with the Hamiltonian computed in \cite{Hartung:2011ea}, we would also need 
to consider the non spinning Lagrangian up to 2PN order, and the linear in spin potentials up to NLO, 
that are consistent with the field gauge and spin gauge, which were employed in the computation in \cite{Levi:2011eq}.  
Thus, we would need the sum given by  
\be \label{lagdef}
L = L_{\text{N}} + L_{\text{1PN}} + L_{\text{2PN}} -V_{\text{SO}}^{\text{LO}} -V_{\text{S$_1$S$_2$}}^{\text{LO}} -V_{\text{SO}}^{\text{NLO}} - V_{\text{S$_1$S$_2$}}^{\text{NLO}} - V_{\text{S$_1$S$_2$}}^{\text{NNLO}},
\ee 
where we proceed now to detail each term of the sum. 

We begin with the recent NNLO S$_1$S$_2$ interaction potential at 4PN order, computed in \cite{Levi:2011eq}, which reads  
\begin{align}
&-V_{\text{S$_1$S$_2$}}^{\text{NNLO}}=-\frac{G}{8r^3}\left[3\vec{S}_1\cdot\vec{S}_2v_1^2v_2^2
+2\vec{S}_1\cdot\vec{S}_2(\vec{v}_1\cdot\vec{v}_2)^2
-2\vec{S}_1\cdot\vec{v}_1\vec{S}_2\cdot\vec{v}_1v_2^2
\right.
\nn\\
&+4\vec{S}_1\cdot\vec{v}_1\vec{S}_2\cdot\vec{v}_2\vec{v}_1\cdot\vec{v}_2
-4\vec{S}_1\cdot\vec{v}_2\vec{S}_2\cdot\vec{v}_1\vec{v}_1\cdot\vec{v}_2
-2\vec{S}_1\cdot\vec{v}_2\vec{S}_2\cdot\vec{v}_2v_1^2
-9\vec{S}_1\cdot\vec{S}_2(\vec{v}_1\cdot\vec{n})^2v_2^2
\nn\\
&-9\vec{S}_1\cdot\vec{S}_2(\vec{v}_2\cdot\vec{n})^2v_1^2
-24\vec{S}_1\cdot\vec{S}_2\vec{v}_1\cdot\vec{v}_2\vec{v}_1\cdot\vec{n}\vec{v}_2\cdot\vec{n}
+6\vec{S}_1\cdot\vec{v}_1\vec{S}_2\cdot\vec{v}_1(\vec{v}_2\cdot\vec{n})^2
\nn\\
&+24\vec{S}_1\cdot\vec{v}_2\vec{S}_2\cdot\vec{v}_1\vec{v}_1\cdot\vec{n}\vec{v}_2\cdot\vec{n}
+6\vec{S}_1\cdot\vec{v}_2\vec{S}_2\cdot\vec{v}_2(\vec{v}_1\cdot\vec{n})^2
+6\vec{S}_1\cdot\vec{v}_1\vec{S}_2\cdot\vec{n}\vec{v}_1\cdot\vec{n}v_2^2
\nn\\
&+6\vec{S}_1\cdot\vec{v}_2\vec{S}_2\cdot\vec{n}\vec{v}_2\cdot\vec{n}v_1^2
+6\vec{S}_1\cdot\vec{n}\vec{S}_2\cdot\vec{v}_1\vec{v}_1\cdot\vec{n}v_2^2
+6\vec{S}_1\cdot\vec{n}\vec{S}_2\cdot\vec{v}_2\vec{v}_2\cdot\vec{n}v_1^2
+3\vec{S}_1\cdot\vec{n}\vec{S}_2\cdot\vec{n}v_1^2v_2^2
\nn\\
&-6\vec{S}_1\cdot\vec{n}\vec{S}_2\cdot\vec{n}(\vec{v}_1\cdot\vec{v}_2)^2
-12\vec{S}_1\times\vec{v}_1\cdot\vec{v}_2\vec{S}_2\times\vec{v}_1\cdot\vec{n}\vec{v}_2\cdot\vec{n}
\nn\\
&-12\vec{S}_1\times\vec{v}_1\cdot\vec{v}_2\vec{S}_2\times\vec{v}_2\cdot\vec{n}\vec{v}_1\cdot\vec{n}
-12\vec{S}_1\times\vec{v}_1\cdot\vec{n}\vec{S}_2\times\vec{v}_2\cdot\vec{v}_1\vec{v}_2\cdot\vec{n}
\nn\\
&-12\vec{S}_1\times\vec{v}_2\cdot\vec{n}\vec{S}_2\times\vec{v}_2\cdot\vec{v}_1\vec{v}_1\cdot\vec{n}
-12\vec{S}_1\times\vec{v}_1\cdot\vec{n}\vec{S}_2\times\vec{v}_2\cdot\vec{n}\vec{v}_1\cdot\vec{v}_2
\nn\\
&+12\vec{S}_1\times\vec{v}_2\cdot\vec{n}\vec{S}_2\times\vec{v}_1\cdot\vec{n}\vec{v}_1\cdot\vec{v}_2
+45\vec{S}_1\cdot\vec{S}_2(\vec{v}_1\cdot\vec{n})^2(\vec{v}_2\cdot\vec{n})^2
\nn\\
&-30\vec{S}_1\cdot\vec{v}_1\vec{S}_2\cdot\vec{n}\vec{v}_1\cdot\vec{n}(\vec{v}_2\cdot\vec{n})^2
-30\vec{S}_1\cdot\vec{v}_2\vec{S}_2\cdot\vec{n}\vec{v}_2\cdot\vec{n}(\vec{v}_1\cdot\vec{n})^2
\nn\\
&-30\vec{S}_1\cdot\vec{n}\vec{S}_2\cdot\vec{v}_1\vec{v}_1\cdot\vec{n}(\vec{v}_2\cdot\vec{n})^2
-30\vec{S}_1\cdot\vec{n}\vec{S}_2\cdot\vec{v}_2\vec{v}_2\cdot\vec{n}(\vec{v}_1\cdot\vec{n})^2
\nn\\
&-15\vec{S}_1\cdot\vec{n}\vec{S}_2\cdot\vec{n}(\vec{v}_1\cdot\vec{n})^2v_2^2
-15\vec{S}_1\cdot\vec{n}\vec{S}_2\cdot\vec{n}(\vec{v}_2\cdot\vec{n})^2v_1^2
\nn\\
&+60\vec{S}_1\times\vec{v}_1\cdot\vec{n}\vec{S}_2\times\vec{v}_2\cdot\vec{n}\vec{v}_1\cdot\vec{n}\vec{v}_2\cdot\vec{n}
-60\vec{S}_1\times\vec{v}_2\cdot\vec{n}\vec{S}_2\times\vec{v}_1\cdot\vec{n}\vec{v}_1\cdot\vec{n}\vec{v}_2\cdot\vec{n}
\nn\\
&+105\vec{S}_1\cdot\vec{n}\vec{S}_2\cdot\vec{n}(\vec{v}_1\cdot\vec{n})^2(\vec{v}_2\cdot\vec{n})^2
-4\left[3S_1^{0i}S_2^{0i}\left(\vec{v}_1\cdot\vec{v}_2+\vec{v}_1\cdot\vec{n}\vec{v}_2\cdot\vec{n}\right)
+S_1^{0i}S_2^{0j}\left(5v_1^iv_2^j-3v_2^iv_1^j\right.\right.
\nn\\
&\left. -9v_1^in^j\vec{v}_2\cdot\vec{n}+3v_2^in^j\vec{v}_1\cdot\vec{n}+3n^iv_1^j\vec{v}_2\cdot\vec{n}
-9n^iv_2^j\vec{v}_1\cdot\vec{n}+3n^in^j\left(\vec{v}_1\cdot\vec{v}_2+5\vec{v}_1\cdot\vec{n}\vec{v}_2\cdot\vec{n}\right)\right)
\nn\\
&-\left(S_1^{0i}\left((\vec{S}_2\times\vec{v}_1)^i\vec{v}_1\cdot\vec{v}_2-(\vec{S}_2\times\vec{v}_1)^iv_2^2
-(\vec{S}_2\times\vec{v}_2)^i\vec{v}_1\cdot\vec{v}_2\right.-v_1^i(\vec{S}_2\times\vec{v}_2)\cdot\vec{v}_1\right.
\nn\\
&+v_2^i(\vec{S}_2\times\vec{v}_2)\cdot\vec{v}_1-3(\vec{S}_2\times\vec{v}_1)^i\vec{v}_1\cdot\vec{n}\vec{v}_2\cdot\vec{n}
+3(\vec{S}_2\times\vec{v}_1)^i(\vec{v}_2\cdot\vec{n})^2
\nn\\
&+9(\vec{S}_2\times\vec{v}_2)^i\vec{v}_1\cdot\vec{n}\vec{v}_2\cdot\vec{n}
+3(\vec{S}_2\times\vec{n})^i\vec{v}_1\cdot\vec{n}v_2^2-3v_1^i\vec{S}_2\times\vec{v}_1\cdot\vec{n}\vec{v}_2\cdot\vec{n}
\nn\\
&-3v_1^i\vec{S}_2\times\vec{v}_2\cdot\vec{n}\vec{v}_2\cdot\vec{n}
-3v_2^i\vec{S}_2\times\vec{v}_1\cdot\vec{n}\vec{v}_1\cdot\vec{n}
+6v_2^i\vec{S}_2\times\vec{v}_1\cdot\vec{n}\vec{v}_2\cdot\vec{n}
+3v_2^i\vec{S}_2\times\vec{v}_2\cdot\vec{n}\vec{v}_1\cdot\vec{n}
\nn\\
&+3n^i\vec{S}_2\times\vec{v}_2\cdot\vec{v}_1\vec{v}_1\cdot\vec{n}
+3n^i\vec{S}_2\times\vec{v}_2\cdot\vec{v}_1\vec{v}_2\cdot\vec{n}
-3n^i\vec{S}_2\times\vec{v}_1\cdot\vec{n}\vec{v}_1\cdot\vec{v}_2
+3n^i\vec{S}_2\times\vec{v}_2\cdot\vec{n}\vec{v}_1\cdot\vec{v}_2
\nn\\
&\left. -15(\vec{S}_2\times\vec{n})^i\vec{v}_1\cdot\vec{n}(\vec{v}_2\cdot\vec{n})^2
+15n^i\vec{S}_2\times\vec{v}_1\cdot\vec{n}\vec{v}_1\cdot\vec{n}\vec{v}_2\cdot\vec{n}
-15n^i\vec{S}_2\times\vec{v}_2\cdot\vec{n}\vec{v}_1\cdot\vec{n}\vec{v}_2\cdot\vec{n}\right)
\nn\\
&\bigl.\bigl.\biggl.
+(1\longleftrightarrow2)\biggr)\biggr]\biggr]
\nn\\
&-\frac{G}{8r^2}\left[3\vec{S}_1\cdot\vec{S}_2\vec{a}_1\cdot\vec{n}v_2^2
+2\vec{S}_1\cdot\vec{S}_2\vec{a}_1\cdot\vec{v}_2\vec{v}_2\cdot\vec{n}-2\vec{S}_1\cdot\vec{v}_2\vec{S}_2\cdot\vec{v}_2\vec{a}_1\cdot\vec{n}\right.
\nn\\
&+2\vec{S}_1\cdot\vec{a}_1\vec{S}_2\cdot\vec{v}_2\vec{v}_2\cdot\vec{n}
+2\vec{S}_1\cdot\vec{n}\vec{S}_2\cdot\vec{v}_2\vec{a}_1\cdot\vec{v}_2
-6\vec{S}_1\cdot\vec{v}_2\vec{S}_2\cdot\vec{a}_1\vec{v}_2\cdot\vec{n}
+2\vec{S}_1\cdot\vec{v}_2\vec{S}_2\cdot\vec{n}\vec{a}_1\cdot\vec{v}_2
\nn\\
&-\vec{S}_1\cdot\vec{a}_1\vec{S}_2\cdot\vec{n}v_2^2
-\vec{S}_1\cdot\vec{n}\vec{S}_2\cdot\vec{a}_1v_2^2
-4\vec{S}_1\times\vec{v}_2\cdot\vec{a}_1\vec{S}_2\times\vec{v}_2\cdot\vec{n}
+4\vec{S}_1\times\vec{v}_2\cdot\vec{n}\vec{S}_2\times\vec{v}_2\cdot\vec{a}_1
\nn\\
&-9\vec{S}_1\cdot\vec{S}_2(\vec{v}_2\cdot\vec{n})^2\vec{a}_1\cdot\vec{n}
+6\vec{S}_1\cdot\vec{v}_2\vec{S}_2\cdot\vec{n}\vec{v}_2\cdot\vec{n}\vec{a}_1\cdot\vec{n}
+6\vec{S}_1\cdot\vec{n}\vec{S}_2\cdot\vec{v}_2\vec{v}_2\cdot\vec{n}\vec{a}_1\cdot\vec{n}
\nn\\
&+3\vec{S}_1\cdot\vec{n}\vec{S}_2\cdot\vec{a}_1(\vec{v}_2\cdot\vec{n})^2
+3\vec{S}_1\cdot\vec{a}_1\vec{S}_2\cdot\vec{n}(\vec{v}_2\cdot\vec{n})^2
+3\vec{S}_1\cdot\vec{n}\vec{S}_2\cdot\vec{n}\vec{a}_1\cdot\vec{n}v_2^2
\nn\\
&-6\vec{S}_1\cdot\vec{n}\vec{S}_2\cdot\vec{n}\vec{a}_1\cdot\vec{v}_2\vec{v}_2\cdot\vec{n}
-15\vec{S}_1\cdot\vec{n}\vec{S}_2\cdot\vec{n}(\vec{v}_2\cdot\vec{n})^2\vec{a}_1\cdot\vec{n}
\nn\\
&+12\vec{S}_1\times\vec{n}\cdot\vec{a}_1\vec{S}_2\times\vec{v}_2\cdot\vec{n}\vec{v}_2\cdot\vec{n}
-12\vec{S}_1\times\vec{v}_2\cdot\vec{n}\vec{S}_2\times\vec{n}\cdot\vec{a}_1\vec{v}_2\cdot\vec{n}
+4\left[S_1^{0i}\left((\vec{S}_2\times\vec{a}_1)^i\vec{v}_2\cdot\vec{n}\right.\right.
\nn\\
&\left. -v_2^i\vec{S}_2\times\vec{n}\cdot\vec{a}_1-n^i\vec{S}_2\times\vec{v}_2\cdot\vec{a}_1
+3n^i\vec{S}_2\times\vec{n}\cdot\vec{a}_1\vec{v}_2\cdot\vec{n}\right)
-S_2^{0i}\left((\vec{S}_1\times\vec{v}_2)^i\vec{a}_1\cdot\vec{n}\right.
\nn\\
&-(\vec{S}_1\times\vec{n})^i\vec{a}_1\cdot\vec{v}_2+2(\vec{S}_1\times\vec{a}_1)^i\vec{v}_2\cdot\vec{n}
+v_2^i\vec{S}_1\times\vec{n}\cdot\vec{a}_1-n^i\vec{S}_1\times\vec{v}_2\cdot\vec{a}_1
+2a_1^i\vec{S}_1\times\vec{v}_2\cdot\vec{n}
\nn\\
&\left. -3(\vec{S}_1\times\vec{n})^i\vec{v}_2\cdot\vec{n}\vec{a}_1\cdot\vec{n}
+3n^i\vec{S}_1\times\vec{n}\cdot\vec{a}_1\vec{v}_2\cdot\vec{n}\right)
+\dot{S}_1^{0i}\left(S_2^{0i}\vec{v}_2\cdot\vec{n}+v_2^iS_2^{0j}n^j-3n^iS_2^{0j}v_2^j\right.
\nn\\
&+3n^iS_2^{0j}n^j\vec{v}_2\cdot\vec{n}+(\vec{S}_2\times\vec{v}_1)^i\vec{v}_2\cdot\vec{n}
-3(\vec{S}_2\times\vec{v}_2)^i\vec{v}_2\cdot\vec{n}-(\vec{S}_2\times\vec{n})^iv_2^2
+v_2^i\vec{S}_2\times\vec{v}_1\cdot\vec{n}
\nn\\
&-v_2^i\vec{S}_2\times\vec{v}_2\cdot\vec{n}-n^i\vec{S}_2\times\vec{v}_2\cdot\vec{v}_1
+3(\vec{S}_2\times\vec{n})^i(\vec{v}_2\cdot\vec{n})^2
-3n^i\vec{S}_2\times\vec{v}_1\cdot\vec{n}\vec{v}_2\cdot\vec{n}
\nn\\
&\left.\left.\left.+3n^i\vec{S}_2\times\vec{v}_2\cdot\vec{n}\vec{v}_2\cdot\vec{n}\right)\right]\right]
+[1\longleftrightarrow2]
\nn\\
&+\frac{G}{8r}
\left[\vec{S}_1\cdot\vec{S}_2\vec{a}_1\cdot\vec{a}_2-3\vec{S}_1\cdot\vec{a}_1\vec{S}_2\cdot\vec{a}_2
+5\vec{S}_1\cdot\vec{a}_2\vec{S}_2\cdot\vec{a}_1+3\vec{S}_1\cdot\vec{S}_2\vec{a}_1\cdot\vec{n}\vec{a}_2\cdot\vec{n}\right.
\nn\\
&-\vec{S}_1\cdot\vec{a}_1\vec{S}_2\cdot\vec{n}\vec{a}_2\cdot\vec{n}
-\vec{S}_1\cdot\vec{a}_2\vec{S}_2\cdot\vec{n}\vec{a}_1\cdot\vec{n}
-\vec{S}_1\cdot\vec{n}\vec{S}_2\cdot\vec{a}_1\vec{a}_2\cdot\vec{n}
-\vec{S}_1\cdot\vec{n}\vec{S}_2\cdot\vec{a}_2\vec{a}_1\cdot\vec{n}
\nn\\
&+3\vec{S}_1\cdot\vec{n}\vec{S}_2\cdot\vec{n}\vec{a}_1\cdot\vec{a}_2
+4\vec{S}_1\times\vec{n}\cdot\vec{a}_1\vec{S}_2\times\vec{n}\cdot\vec{a}_2
-4\vec{S}_1\times\vec{n}\cdot\vec{a}_2\vec{S}_2\times\vec{n}\cdot\vec{a}_1
\nn\\
&+3\vec{S}_1\cdot\vec{n}\vec{S}_2\cdot\vec{n}\vec{a}_1\cdot\vec{n}\vec{a}_2\cdot\vec{n}-4\left[\dot{S}_1^{0i}\dot{S}_2^{0i}
+\dot{S}_1^{0i}n^i\dot{S}_2^{0j}n^j-\dot{S}_1^{0i}\left(2(\vec{S}_2\times\vec{a}_2)^i
-(\vec{S}_2\times\vec{n})^i\vec{a}_2\cdot\vec{n}\right.\right.
\nn\\
&\left.\left.\left. +n^i\vec{S}_2\times\vec{n}\cdot\vec{a}_2\right)
-\dot{S}_2^{0i}\left(2(\vec{S}_1\times\vec{a}_1)^i-(\vec{S}_1\times\vec{n})^i\vec{a}_1\cdot\vec{n}
+n^i\vec{S}_1\times\vec{n}\cdot\vec{a}_1\right)\right]\right]
\nn\\
&+\frac{G^2m_2}{2r^4}\left[11\vec{S}_1\cdot\vec{S}_2v_2^2+3\vec{S}_1\cdot\vec{v}_1\vec{S}_2\cdot\vec{v}_2
+5\vec{S}_1\cdot\vec{v}_2\vec{S}_2\cdot\vec{v}_1-16\vec{S}_1\cdot\vec{v}_2\vec{S}_2\cdot\vec{v}_2\right.
\nn\\
&-22\vec{S}_1\cdot\vec{S}_2(\vec{v}_2\cdot\vec{n})^2+16\vec{S}_1\cdot\vec{S}_2\vec{v}_1\cdot\vec{n}\vec{v}_2\cdot\vec{n}
+2\vec{S}_1\cdot\vec{v}_1\vec{S}_2\cdot\vec{n}\vec{v}_2\cdot\vec{n}
\nn\\
&-24\vec{S}_1\cdot\vec{v}_2\vec{S}_2\cdot\vec{n}\vec{v}_1\cdot\vec{n}
+40\vec{S}_1\cdot\vec{v}_2\vec{S}_2\cdot\vec{n}\vec{v}_2\cdot\vec{n}
+2\vec{S}_1\cdot\vec{n}\vec{S}_2\cdot\vec{v}_1\vec{v}_2\cdot\vec{n}
\nn\\
&
+16\vec{S}_1\cdot\vec{n}\vec{S}_2\cdot\vec{v}_2\vec{v}_2\cdot\vec{n}+7\vec{S}_1\cdot\vec{n}\vec{S}_2\cdot\vec{n}v_2^2
-16\vec{S}_1\cdot\vec{n}\vec{S}_2\cdot\vec{n}\vec{v}_1\cdot\vec{v}_2
\nn\\
&-4\vec{S}_1\times\vec{v}_1\cdot\vec{n}\,\vec{S}_2\times\vec{v}_2\cdot\vec{n}
+10\vec{S}_1\times\vec{v}_2\cdot\vec{n}\,\vec{S}_2\times\vec{v}_1\cdot\vec{n}
-8\vec{S}_1\times\vec{v}_2\cdot\vec{n}\,\vec{S}_2\times\vec{v}_2\cdot\vec{n}
\nn\\
&\left.-42\vec{S}_1\cdot\vec{n}\vec{S}_2\cdot\vec{n}(\vec{v}_2\cdot\vec{n})^2
-12\vec{S}_1\cdot\vec{n}\vec{S}_2\cdot\vec{n}\vec{v}_1\cdot\vec{n}\vec{v}_2\cdot\vec{n}
-4S_1^{0i}S_2^{0i}+16S_1^{0i}n^iS_2^{0j}n^j\right.
\nn\\
&+2S_1^{0i}\left(2(\vec{S}_2\times\vec{v}_1)^i-6(\vec{S}_2\times\vec{v}_2)^i
-2(\vec{S}_2\times\vec{n})^i\vec{v}_1\cdot\vec{n}+11(\vec{S}_2\times\vec{n})^i\vec{v}_2\cdot\vec{n}\right.
\nn\\
&\left.-5n^i\vec{S}_2\times\vec{v}_1\cdot\vec{n}+6n^i\vec{S}_2\times\vec{v}_2\cdot\vec{n}\right)
-2S_2^{0i}\left(2(\vec{S}_1\times\vec{v}_1)^i+2(\vec{S}_1\times\vec{v}_2)^i\right.
\nn\\
&\left.\left.-6(\vec{S}_1\times\vec{n})^i\vec{v}_1\cdot\vec{n}-3(\vec{S}_1\times\vec{n})^i\vec{v}_2\cdot\vec{n}
-2n^i\vec{S}_1\times\vec{v}_1\cdot\vec{n}+n^i\vec{S}_1\times\vec{v}_2\cdot\vec{n}\right)\right]
+[1\longleftrightarrow2]
\nn\\ 
&+\frac{G^3(m_1^2+m_2^2)}{2r^5}\left(11\vec{S}_1\cdot\vec{S}_2-35\vec{S}_1\cdot\vec{n}\vec{S}_2\cdot\vec{n}\right)
+3\frac{G^3m_1m_2}{r^5}\left(7\vec{S}_1\cdot\vec{S}_2-27\vec{S}_1\cdot\vec{n}\vec{S}_2\cdot\vec{n}\right),           \label{vnnloss}
\end{align}
where $[1 \leftrightarrow 2]$ refers to all terms given previously with the particle labels 1 and 2 interchanged.
As we noted in \cite{Levi:2011eq} this result contains higher order time derivatives, e.g.~acceleration and precession dependent terms, 
and $S^{0i}$ dependent terms, such that the $S^{0i}$ components are left as independent degrees of freedom, 
and the additional contributions of the field corrections in the $S^{0i}$ components are not taken into account explicitly in this form. 
We will elaborate on these two key issues in section \ref{htd} and sections \ref{spinGF}, \ref{VSXcan}, respectively, and as they start 
to appear in the lower order potentials from eq.~(\ref{lagdef}), that we will now present.

First then we have $L_{\text{N}}$ which is just the Newtonian Lagrangian given by 
\be
L_{\text{N}} = \frac{1}{2}\sum_{I=1}^{2}m_Iv_I^2 + \frac{\gravthree m_1m_2}{\rel},
\ee
and $L_{\text{1PN}}$ is the Einstein-Infeld-Hoffmann 1PN correction Lagrangian, that is the leading order (LO) correction to Newtonian gravity, 
given by 
\be
L_{\text{1PN}}= \frac{1}{8}\sum_{I=1}^{2}m_Iv_I^4 + \frac{\gravthree m_1m_2}{2\rel}  \left[3v_1^2+3v_2^2-7\vec{v}_1\cdot\vec{v}_2-(\vec{v}_1\cdot\vec{n})(\vec{v}_2\cdot\vec{n})\right]
-\frac{\gravthree^2 m_1m_2(m_1+m_2)}{2\rel^2}.
\ee
$L_{\text{2PN}}$ is the 2PN Lagrangian, which is gauge dependent already, hence we take here the result computed in the harmonic gauge in 
\cite{Gilmore:2008gq} given by
\begin{align}\label{eq:2pn}
 L_{\text{2PN}} = & \frac{m_1 v_1^6}{16}  \nn \\
           + & \frac{G m_1 m_2}{r} \Bigg(\frac{7}{8} v_1^4
             - \frac{5}{4} v_1^2 \vec{v}_1 \cdot \vec{v}_2 
             - \frac{3}{4} v_1^2 \vec{v}_1 \cdot \vec{n} \vec{v}_2 \cdot \vec{n} 
             + \frac{3}{16} v_1^2 v_2^2 
             + \frac{1}{8} (\vec{v}_1 \cdot \vec{v}_2)^2 \nn \\
           & {} \hspace*{40pt} - \frac{1}{8} v_1^2 (\vec{v}_2 \cdot \vec{n})^2 
           + \frac{3}{4} \vec{v}_1 \cdot \vec{v}_2 \vec{v}_1 \cdot \vec{n} \vec{v}_2 \cdot \vec{n} 
           + \frac{3}{16} (\vec{v}_1 \cdot \vec{n})^2 (\vec{v}_2 \cdot \vec{n} )^2 \Bigg)\nn \\
         + & \underbrace{G m_1 m_2 \Bigg(\frac{1}{8} v_2^2 \vec{a}_1 \cdot \vec{n}  
           + \frac{3}{2} \vec{v}_1 \cdot \vec{a}_1 \vec{v}_2 \cdot \vec{n}  
           - \frac{7}{4} \vec{v}_2 \cdot \vec{a}_1 \vec{v}_2 \cdot \vec{n}
           - \frac{1}{8} \vec{a}_1 \cdot \vec{n} (\vec{v}_2 \cdot \vec{n})^2 \Bigg)}_{\text{HTD dependent}} \nn \\
         + & \underbrace{G m_1 m_2 r \Bigg(\frac{15}{16} \vec{a}_1 \cdot \vec{a}_2 
           - \frac{1}{16} \vec{a}_1 \cdot \vec{n} \vec{a}_2 \cdot \vec{n} \Bigg)}_{\text{HTD dependent}} \nn \\
         + & \frac{G^2 m_1 m_2^2}{r^2} \left(\frac{7}{4} v_1^2 + 2 v_2^2 
           - \frac{7}{2} \vec{v}_1 \cdot \vec{v}_2 + \frac{1}{2}(\vec{v}_1 \cdot \vec{n})^2\right) \nn \\
         + & \frac{G^3 m_1 m_2^3}{2 r^3} + \frac{3 G^3 m_1^2 m_2^2}{2 r^3} + [1 \longleftrightarrow 2].
\end{align}
Note that here, on the NLO correction to Newtonian gravity, there appears a part, which depends on accelerations, 
i.e.~higher order time derivatives of the velocities. Hence, we denote such part as ``HTD dependent'' here and in the following. 
These higher order time derivatives of velocities will eventually be eliminated via variable redefinitions,
a common procedure in the PN approximation, see e.g.~\cite{Barker:1980,Schafer:1984mr,Damour:1985mt,Damour:1990jh}, of which we make repeated use throughout this work. In section \ref{htd} we will elaborate on this procedure, in which higher order time derivatives are substituted with equations of motion (EOM) from lower PN orders, and we also extend it explicitly for higher order time derivatives of spin, which also appear in spin dependent potentials as of NLO as we will see here below. 
We should stress that although this is a non-spinning sector, the HTD dependent terms here contribute 
eventually to the NNLO S$_1$S$_2$ sector upon the substitution of lower order EOM from the LO spin sectors. 
In contrast to eq.~(\ref{vnnloss}) here, these terms were included in \cite{Levi:2011eq}.

We proceed to the linear in spin potentials, where $V_{\text{SO}}^{\text{LO}}$ is the LO SO interaction potential at 1.5PN order, 
as given in eq.~(70) of \cite{Levi:2010zu}
\be \label{vloso}
V_{\text{SO}}^{\text{LO}} = -\frac{\gravthree m_2}{\rel^2} \vec{S}_1\cdot\left(\vec{v}_1\times\vec{n} - 2\vec{v}_2\times\vec{n}\right) 
              -\underbrace{\frac{\gravthree m_2}{\rel^2}S_1^{0i}n^i}_{\text{spin gauge dependent}} + [1 \leftrightarrow 2].
\ee
Note that here, at the lowest order spin-dependent correction, there is a spin gauge dependence, namely a dependence 
in the choice of the point \emph{within} the extended spinning object, whose evolution we follow along the worldline \cite{Tulczyjew:1959,Tulczyjew:1959c,Hanson:1974qy}.      
$V_{\text{S$_1$S$_2$}}^{\text{LO}}$ is the LO S$_1$S$_2$ interaction potential at 2PN order, given by
\be
V_{\text{S$_1$S$_2$}}^{\text{LO}} = -\frac{\gravthree}{\rel^3}(\vec{S}_1\cdot\vec{S}_2-3\vec{S}_1\cdot\vec{n}\vec{S}_2\cdot\vec{n}).
\ee
$V_{\text{SO}}^{\text{NLO}}$ is the NLO SO interaction potential at 2.5PN, as computed in \cite{Levi:2010zu}
\begin{align}\label{vnloso}
V_{\text{SO}}^{\text{NLO}} = & -\frac{\gravthree m_2}{\rel^2}\vec{S}_1\cdot 
\left[\vec{v}_1\times\vec{n}\left(\frac{1}{2}\vec{v}_1\cdot\vec{v}_2-\frac{1}{2}v_2^2-\frac{3}{2}\vec{v}_1\cdot\vec{n}\vec{v}_2\cdot\vec{n}\right) 
\right. 
\nn\\
&\left.
+\vec{v}_2\times\vec{n}\biggl(\vec{v}_1\cdot\vec{v}_2-v_2^2+3\vec{v}_1\cdot\vec{n}\vec{v}_2\cdot\vec{n}\biggr)+\vec{v}_1\times\vec{v}_2\left(\frac{1}{2}\vec{v}_1\cdot\vec{n}+\vec{v}_2\cdot\vec{n}\right)\right]
\nn\\ 
&+\frac{\gravthree^2 m_2}{\rel^3}\vec{S}_1\cdot\left[\vec{v}_1\times\vec{n}\left(m_1-\frac{1}{2}m_2\right)+\vec{v}_2\times\vec{n}\left(\frac{5}{2}m_2\right)\right]
\nn\\
&-\underbrace{\frac{\gravthree m_2}{\rel}\left[\vec{S}_1\cdot\left(\vec{v}_1\times\vec{a}_2+\frac{1}{2}\vec{v}_2\times\vec{a}_1+\frac{1}{2}\vec{a}_1\times\vec{n}\vec{v}_2\cdot\vec{n}+\vec{a}_2\times\vec{n}\vec{v}_1\cdot\vec{n}\right)\right.}_{\text{HTD dependent}} 
\nn\\
&\underbrace{\left.+\dot{\vec{S}}_1\cdot \left(-\frac{1}{2}\vec{v}_1\times\vec{v}_2+\frac{1}{2}\vec{v}_1\times\vec{n}\vec{v}_2\cdot\vec{n}-\vec{v}_2\times\vec{n}\vec{v}_2\cdot\vec{n}\right)\right]+\gravthree m_2 \dot{\vec{S}}_1\cdot \vec{a}_2\times\vec{n}}_{\text{HTD dependent}} 
\nn\\
& -\underbrace{\frac{3}{2}\frac{\gravthree m_2}{\rel^2}S_1^{0i}\left[v_1^i\vec{v}_2\cdot\vec{n}-v_2^i\vec{v}_1\cdot\vec{n} -n^i\left(\vec{v}_1\cdot\vec{v}_2-v_2^2+\vec{v}_1\cdot\vec{n}\vec{v}_2\cdot\vec{n}\right)\right]}_{\text{spin gauge dependent}}
\nn\\
&
\underbrace{+\frac{\gravthree^2 m_2}{\rel^3}S_1^{0i}n^i\left(m_1+2m_2\right)}_{\text{spin gauge dependent}}-\underbrace{\frac{\gravthree m_2}{2\rel}\dot{S}_1^{0i}\left(3v_2^i+n^i\vec{v}_2\cdot\vec{n}\right)}_{\text{spin gauge \& HTD dependent}}+\,\,\,\,\,\,\, [1 \longleftrightarrow 2].
\end{align}
Note that here in addition to the spin gauge dependence, there appear higher order time derivative terms as in the NLO non spinning case at 2PN, but here there are also higher order time derivatives of spin. Also note, that there appears a part here, which is both HTD and spin gauge dependent. It contains a time derivative on the $S^{0i}$ components, which as we explain in section \ref{htd} is an undesirable complication. However, the $\dot{S}^{0i}$ factor can be avoided here and at NLO in general, by flipping the time derivative from it. 
We include in this expression all terms, which arise from the computation in \cite{Levi:2010zu}, even those which do not contribute at NLO, 
upon the substitution of higher order time derivatives, and temporal spin components $S^{0i}$. 
Here also, there are contributions to the NNLO S$_1$S$_2$ sector from the HTD dependent terms upon the substitution of lower order EOM from 
the LO spin sectors, which were included in \cite{Levi:2011eq}.

$V_{\text{S$_1$S$_2$}}^{\text{NLO}}$ is the NLO S$_1$S$_2$ interaction potential at 3PN, as computed in \cite{Levi:2008nh}
\begin{align}\label{vnloss}
V_{S_1S_2}^{NLO} = &\frac{\gravthree}{2\rel^3}\left[ \vec{S}_1\cdot
\vec{S}_2\left(3\vec{v}_1\cdot\vec{v}_2-3\vec{v}_1\cdot{\vec{n}}\vec{v}_2\cdot{\vec{n}}\right)
+\vec{S}_1\cdot\vec{v}_1\vec{S}_2\cdot\vec{v}_2-3\vec{S}_1\cdot\vec{v}_2\vec{S}_2\cdot\vec{v}_1
\right.
\nn\\
&
+3\vec{S}_1\cdot\vec{v}_1\vec{S}_2\cdot\vec{n}\vec{v}_2\cdot\vec{n}
+3\vec{S}_2\cdot\vec{v}_2\vec{S}_1\cdot\vec{n}\vec{v}_1\cdot\vec{n}
+3\vec{S}_1\cdot\vec{v}_2\vec{S}_2\cdot\vec{n}\vec{v}_1\cdot\vec{n}
+3\vec{S}_2\cdot\vec{v}_1\vec{S}_1\cdot\vec{n}\vec{v}_2\cdot\vec{n}
\nn\\
& 
-3\vec{S}_1\cdot\vec{n}\vec{S}_2\cdot\vec{n}\left(\vec{v}_1\cdot\vec{v}_2+5\vec{v}_1\cdot\vec{n}\vec{v}_2\cdot\vec{n}\right)
-6(\vec{S}_1\times\vec{v}_1)\cdot\vec{n}(\vec{S}_2\times\vec{v}_2)\cdot\vec{n}
\nn\\
& 
\left.+6(\vec{S}_1\times\vec{v}_2)\cdot\vec{n}(\vec{S}_2\times\vec{v}_1)\cdot\vec{n}\right]
+3\frac{\gravthree^2(m_1+m_2)}{\rel^4}\left[\vec{S}_1\cdot\vec{S}_2-3\vec{S}_1\cdot\vec{n}\vec{S}_2\cdot\vec{n}\right]
\nn\\
&
\underbrace{+\frac{G}{2r^2}\left[\dot{\vec{S}}_1\cdot\vec{S}_2\vec{v}_2\cdot\vec{n}-\dot{\vec{S}}_1\cdot\vec{v}_2\vec{S}_2\cdot\vec{n}-\dot{\vec{S}}_1\cdot\vec{n}\vec{S}_2\cdot\vec{v}_2+3\dot{\vec{S}}_1\cdot\vec{n}\vec{S}_2\cdot\vec{n}\vec{v}_2\cdot\vec{n}\right.}_{\text{HTD dependent}}
\nn\\
&
\underbrace{\left.-\dot{\vec{S}}_2\cdot\vec{S}_1\vec{v}_1\cdot\vec{n}+\dot{\vec{S}}_2\cdot\vec{v}_1\vec{S}_1\cdot\vec{n}+\dot{\vec{S}}_2\cdot\vec{n}\vec{S}_1\cdot\vec{v}_1-3\dot{\vec{S}}_2\cdot\vec{n}\vec{S}_1\cdot\vec{n}\vec{v}_1\cdot\vec{n}\right]}_{\text{HTD dependent}}
\nn\\
&
-\underbrace{\frac{\gravthree }{\rel^3}\left[S_1^{0i}S_2^{0i} - 3S_1^{0i}n^iS_2^{0j}n^j + S_1^{0i}\left((\vec{S}_2\times\vec{v}_2)^i - 3n^i\vec{S}_2\times\vec{v}_2\cdot\vec{n}\right)\right.}_{\text{spin gauge dependent}} 
\nn\\
&
\underbrace{+ S_2^{0i}\left((\vec{S}_1\times\vec{v}_1)^i - 3n^i\vec{S}_1\times\vec{v}_1\cdot\vec{n}\right)+3S_1^{0i}\left(n^i\vec{S}_2\times\vec{v}_1\cdot\vec{n}-(\vec{S}_2\times\vec{n})^i\vec{v}_1\cdot\vec{n}\right)}_{\text{spin gauge dependent}}
\nn\\
&
\underbrace{\left.+3S_2^{0i}\left(n^i\vec{S}_1\times\vec{v}_2\cdot\vec{n}-(\vec{S}_1\times\vec{n})^i\vec{v}_2\cdot\vec{n}\right)\right]}_{\text{spin gauge dependent}} -\underbrace{\frac{G}{r^2}\left[\dot{S}_1^{0i}(\vec{S}_2\times\vec{n})^i-\dot{S}_2^{0i}(\vec{S}_1\times\vec{n})^i\right]}_{\text{spin gauge \& HTD dependent}}.
\end{align}

Note that here also there are both higher order time derivative terms, and spin gauge dependent part, as well as a part which depends on both with $\dot{S}^{0i}$. 
Also here, HTD dependent terms, which arise from the computation in \cite{Levi:2008nh}, also contribute 
to the NNLO S$_1$S$_2$ sector, and were included in \cite{Levi:2011eq}.

Finally, all the spin gauge dependent parts from lower order linear in spin sectors, i.e.~those in eqs.~(\ref{vloso}), (\ref{vnloso}), and (\ref{vnloss}), 
should also be taken into account as possible contributions to the NNLO S$_1$S$_2$ sector, similar to the situation in the NLO spin-dependent sectors.  
As in \cite{Levi:2011eq} these terms are not repeated in eq.~(\ref{vnnloss}).

\section{Elimination of unphysical spin degrees of freedom}\label{spinGF}

As we noted in the previous section, the EFT potentials with spin contain spin gauge dependent parts, since the redundant unphysical degrees of freedom, associated with the temporal spin components $S^{0i}$, were not eliminated. This is a gauge freedom in the choice of the internal point in the extended spinning object for the representative worldline  \cite{Tulczyjew:1959,Tulczyjew:1959c,Hanson:1974qy}. In order to eliminate the redundant degrees of freedom associated with the spin, we need to apply a spin supplementary condition (SSC), where it should be stressed that such SSC should fix the representative worldline of the body uniquely, which is not guaranteed for every SSC in curved spacetime. Such an appropriate SSC is the covariant one \cite{Schattner:1979vp,Schattner:1979vn}, formulated by Tulczyjew \cite{Tulczyjew:1959c}, 
which for the linear in spin case, and in the local frame can be written as
\begin{align}
S_{1ab} u^{b}_1 = 0 \quad \Leftrightarrow \quad 
S^{(0)(i)}_1 = - S^{(i)(j)}_1 \frac{u^{(j)}_1}{u^{(0)}_1}, \label{covssc}
\end{align}
where we have denoted the local basis explicitly by round brackets, that is we are considering the spin, when projected on the local reference tetrad, 
see eq.~(54) in \cite{Levi:2010zu}. We apply the covariant SSC in the local frame since this is the frame, where it physically makes sense to trace the spin. In fact, the spin always appears in the local frame in this paper (for the sake of brevity, we thus omit the round brackets on the spatial indices of the spin in some sections, including the previous section, where the spin dependent potentials were presented). 
Notice that we have here exactly 3 independent conditions to eliminate the 3 redundant degrees of freedom of spin.

Actually, general covariance provides a straightforward way to formulate a general SSC for curved spacetime --- 
the only such, known up to date.
In addition, it should be emphasized, that since the EFT approach relies on consistency with the symmetry of general covariance 
for the construction of its initial effective action, the covariant SSC is in fact also the only sensible choice for gauge fixing 
the spin in EFT after the covariant effective action has been constructed, that makes manifest which operators in the action are physically 
redundant, considering the rotational degrees of freedom, and would be eliminated if gauge fixing is done at the level of the action, 
prior to integrating out the fields.  

\subsection{Spin in the EFT approach} \label{EFTofspin}

One possible way, which is seemingly straightforward, to compare the EFT potential in eq.~(\ref{vnnloss}) with the Hamiltonian in 
\cite{Hartung:2011ea}, is to derive the EOM for the positions and spins of the objects, and try to relate them. If we follow the 
Routhian approach in \cite{Porto:2008tb} for spin in the EFT approach,  then the $S^{0i}$ components are left as independent degrees of freedom till after the obtainment of EOM, and we should derive the EOM for the spin in terms of the Poisson brackets of the so$(1,3)$ spin algebra. Since as of NLO we have higher order time derivatives of the spin appearing in the potentials (which eventually contribute in the potentials as of NNLO), it is actually  incorrect to derive the EOM for the spin using Poisson brackets.
We should resort back to the action in order to see that. For the spin EOMs, we should consider the spin dependent part of the full action obtained after the EFT computation, recalling the rotational kinetic term, e.g.~in eq.~(51) of \cite{Levi:2010zu}, which reads
\be
S_{\text{eff(spin)}}=\int dt \left[ -\sum_{I=1}^2\frac{1}{2}S_{Iab}\Omega_I^{ab}-V\left(\vec{x}_I,\dot{\vec{x}}_I, \ddot{\vec{x}}_I, \dots, S_{Iab}, \dot{S}_{Iab}, \dots \right)\right]. 
\ee
If we make an independent variation of this action with respect to the spin, and to its conjugate, the angular velocity,
we obtain the following EOM for the 4-dimensional spin tensor 
\begin{equation} 
\dot{S}^{ab}= 4 S^{c[a} \eta^{b]d} \frac{\delta\int{dt\,V}}{\delta S^{cd}}  
= 4 S^{c[a} \eta^{b]d} \left[ \frac{\partial V}{\partial S^{cd}} - \frac{d}{dt} \frac{\partial V}{\partial \dot{S}^{cd}} + \dots \right] , \label{S4EOM}
\end{equation}
see section \ref{sec:SEOM} for a similar derivation.
In the absence of time derivatives of the spin, its EOM follow from the so$(1,3)$ Poisson bracket, given by
\begin{equation}
\{ S^{ab}, S^{cd} \} = S^{ca} \eta^{bd} - S^{da} \eta^{bc}
        + S^{db} \eta^{ac} - S^{cb} \eta^{ad},
\end{equation}
and reads 
\begin{equation}
\dot{S}^{ab} = \{ S^{ab}, -V \}= 4 S^{c[a} \eta^{b]d}\frac{\partial V}{\partial S^{cd}},
\end{equation}
which is a restricted case of eq.~(\ref{S4EOM}). Therefore, like the EOM for the position, the EOM for the spin should be obtained from a variation of the action, in terms of which the EFT approach is naturally formulated.

Furthermore, according to the Routhian approach in \cite{Porto:2008tb} the $S^{0i}$ components are left as independent degrees of freedom till after the obtainment of EOM. Otherwise,  if we had no time derivatives of the spin, then if we would constrain the spin, and substitute in eq.~(\ref{covssc}) for the $S^{0i}$ components, together with another conjugate gauge constraint for the Lorentz matrix $\Lambda^{Ab}$, then the Poisson brackets would have to be replaced with the complicated Dirac brackets, which project the original symplectic structure onto the phase space hypersurface, defined by the constraints \cite{Hanson:1974qy}. Yet, according to eq.~(\ref{covssc}) in order to implement the covariant SSC, we actually need to find the 4-velocity in the local basis.
From eq.~(54) in \cite{Levi:2010zu} we get
\begin{equation} \label{locvel}
u^a_1 (\vec{x}_1) = e^a{}_{\mu}u^{\mu}_1 = \eta^{a\mu} \left( \eta_{\mu\nu} + \frac{1}{2} h_{\mu\nu}
	- \frac{1}{8} h_{\mu\rho} \eta^{\rho\sigma} h_{\sigma\nu} + \dots \right) u^{\nu}_1,
\end{equation}
where $\vec{x}_1$ is the position of object $1$, and $h_{\mu\nu}$ is the deviation of the metric from flat spacetime. Hence, we can see that additional contributions of the field in the $S^{0i}$ components are not taken into account explicitly, if the $S^{0i}$ components are left as independent degrees of freedom. It was put forward in \cite{Levi:2008nh}, that the spin gauge may be fixed prior to integrating out the fields, such that the fields from the $S^{0i}$ components would be included in the process of integrating out all field degrees of freedom, and also that these contributions are incorporated into Feynman diagrams in their proper sector. That is, according to \cite{Levi:2008nh}, if the spin gauge could be fixed at the level of the effective action, then all field contributions would be integrated out, as consistent with the philosophy of the EFT approach. Moreover, these contributions would emerge in their appropriate spin sectors, and there would not be an ambiguity of spin effects among the different spin sectors, as was noted here in the end of section \ref{EFTpot}. In addition, since the $S^{0i}$ components implicitly yield contributions of infinitely distinct PN orders, if the $S^{0i}$ components are left as independent degrees of freedom, then the power counting, which is an essential feature of an EFT approach, also breaks down. 

Therefore, we press on to find the $S^{0i}$ components, which would be required, if the EOM would be obtained from the EFT potential. Actually, since we would like to conveniently apply here a mapping to canonical variables, which was already considered in \cite{Hergt:2011ik}, and such a mapping also relies on the elimination of the $S^{0i}$ components, we will proceed to eliminate the $S^{0i}$ components at this stage.
Now, according to eq.~(\ref{covssc}), we need to obtain the local frame 4-velocity, where we note that it will also be required in the following for the transformations to canonical variables and potential. It is preferable to calculate this via the EFT approach, namely in a self-contained and consistent manner, where we actually demonstrate here the computation of the metric.

\subsection{Local frame velocity via the EFT approach}

First, we PN expand eq.~(\ref{locvel}), using $u^0_1 = 1$, $u^i_1=v^i_1$, $h_{00} = \Order(v^2) = h_{ij}$, and $h_{0i} = \Order(v^3)$. Then we obtain for the components of the local velocity $u^{(0)}$ and $u^{(i)}$ the following expressions to NNLO 
\begin{align}
u^{(0)}_1 &= 1 + \frac{1}{2} \left( h_{00} + h_{0i} v^i_1 \right) - \frac{1}{8} \left(h_{00}\right)^2 + \Order{(v^6)}, \label{u0lc}\\ 
u^{(i)}_1 &= v^i_1 - \frac{1}{2} \left( h_{ij} v^j_1 + h_{0i} \right)
	- \frac{1}{8} \left( h_{ij} h_{jk} v^k_1 + h_{ij} h_{j0} - h_{i0} h_{00} \right)
	+ \Order{(v^7)}. \label{uilc}
\end{align} 
In order to compute the local frame velocity, we should use eq.~(2) of \cite{Levi:2010zu} to rewrite eqs.~(\ref{u0lc}) and (\ref{uilc}) in terms of the NRG fields. 
We recall that the NRG fields are defined via the parametrization of the 
metric in a nonrelativistic form according to the Kaluza-Klein ansatz
\be \label{eq:kka}
d\tau^2 = g_{\mu\nu}dx^{\mu}dx^{\nu} \equiv e^{2 \phi}(dt - A_i\, dx^i)^2 -e^{-2 \phi} \gamma_{ij}dx^i dx^j~,
\ee
defining the set of nonrelativistic gravitational fields ${\phi,A_i,\gamma_{ij}\equiv\delta_{ij}+\sigma_{ij}}$.  
To the order required here for the NNLO S$_1$S$_2$ sector, we get 
\begin{align}
u^{(0)}_1 &= 1 + \phi(\vec{x}_1) - \frac{1}{2} A_i(\vec{x}_1) v^i_1 + \dots \, , \label{u0nrg}\\ 
u^{(i)}_1 &= v^i_1 - \phi(\vec{x}_1) v^i_1 + \frac{1}{2} A_i(\vec{x}_1)   + \frac{1}{2} \sigma_{ij}(\vec{x}_1) v^j_1  + \phi(\vec{x}_1) A_i(\vec{x}_1) + \dots \, . \label{uinrg}
\end{align}
Hence, we should compute the one-point functions for the fields $\phi$, $A_i$ and $\sigma_{ij}$, as well as for the composite field $\phi A_i$, in the position $\vec{x}_1$ of object $1$. For example, for $\phi$ we should compute
\begin{align}
\langle \phi(\vec{x}_1) \rangle = \int {\cal{D}}\phi {\cal{D}} A_i {\cal{D}}\sigma_{ij} \phi(\vec{x}_1) e^{iS_{eff}},
\end{align}
where $S_{eff}$ is the effective action of a binary, appearing in e.g.~eqs.~(4), (21), (51), and (52) of \cite{Levi:2010zu}. 
The EFT computation is straightforward. For example,
according to the Feynman rules in \cite{Levi:2010zu}, we have to the order required here
\begin{align} \label{phi}
\langle\phi(\vec{x}_1)\rangle &= 4\pi G \int dt_2 \delta(t_1-t_2)\left( - m_2 + v^i_2 S^{(i)(j)}_2 \partial^{\vec{x}_2}_j + S^{(0)(j)}_2 \partial^{\vec{x}_2}_j \right)
	 \int_{\vec{k}} \frac{e^{i \vec{k} (\vec{x}_1-\vec{x}_2)}}{\vec{k}^2}  \nn\\
&= - \frac{Gm_2}{r} + \frac{G}{r^2} n^j( S^{(i)(j)}_2 v^i_2 + S^{(0)(j)}_2 ) \, ,
\end{align}
where these terms correspond to contributions from the Newtonian interaction shown in figure 1a of \cite{Kol:2007bc}, and the LO SO interaction 
shown in figure 1b of \cite{Levi:2010zu}, respectively. Similarly for the leading relevant contribution of $A_i(\vec{x}_1)$, we have
\begin{align} \label{Alo}
\langle A_i(\vec{x}_1) \rangle &= -16\pi G  \int dt_2 \delta(t_1-t_2)\left( m_2 v^i_2 + \frac{1}{2} S^{(j)(i)}_2 \partial^{\vec{x}_2}_j \right)
	 \int_{\vec{k}} \frac{e^{i \vec{k} (\vec{x}_1-\vec{x}_2)}}{\vec{k}^2}  \nn \\
&= - 4 \frac{Gm_2}{r} v^i_2 + 2 \frac{G}{r^2} S^{(i)(j)}_2 n^j ,
\end{align}
where these terms correspond to contributions from the 1PN interaction shown in figure 1b of \cite{Kol:2007bc}, and the LO SO interaction shown in figure 1a of \cite{Levi:2010zu}, respectively.
For the next order of the relevant contribution of $A_i(\vec{x}_1)$, we get
\begin{align} \label{Anlo}
\begin{split}
\langle A_i(\vec{x}_1)\rangle &=\frac{G}{r^2} \left[S^{(i)(j)}_2 n^j \left( v_2^2 - 3\left(\vec{v}_2\cdot\vec{n}\right)^2 \right)
		+  2 S^{(i)(j)}_2 v^j_2 \vec{v}_2\cdot\vec{n} + 2S^{(0)(i)}_2 \vec{v}_2\cdot\vec{n} + 2S^{(0)(j)}_2 n^j v^i_2 \right]\\
		&+ \frac{G}{r}\left[ S^{(i)(j)}_2 a^k_2 \left(\delta_{jk} - n^j n^k\right)
		+ 2\dot{S}^{(0)(i)}_2 \right] -2 \frac{G^2}{r^3} S^{(i)(j)}_2 n^j \left( m_1 - m_2 \right),
\end{split}
\end{align}
where these correspond to contributions from the NLO SO interaction shown in figures 2a3, 2b1, 3b1, 4a1 and 4a2 of \cite{Levi:2010zu}.
Further, we have for $\sigma_{ij}(\vec{x}_1)$
\begin{align} \label{sigma}
\langle\sigma_{ij}(\vec{x}_1)\rangle &= 2\frac{G}{r^2}\left[S^{(k)(i)}_2 n^k v_2^j + S^{(k)(j)}_2 n^k v_2^i -2\delta_{ij}S^{(k)(l)}_2 n^k v_2^l\right],
\end{align}
where this corresponds to a contribution from the NLO SO interaction shown in figure 2a2.2 of \cite{Levi:2010zu}.
Finally, for the composite field $\phi(\vec{x}_1) A_i(\vec{x}_1)$ we have
\begin{align} \label{composite}
\langle\phi(\vec{x}_1) A_i(\vec{x}_1)\rangle &= -2\frac{G^2m_2}{r^3} S^{(i)(j)}_2 n^j,  
\end{align}
where this corresponds to a contribution from the NLO SO interaction shown in figure 3a1 of \cite{Levi:2010zu}.

Putting together eqs.~(\ref{phi}), (\ref{Alo}), (\ref{Anlo}), (\ref{sigma}) and (\ref{composite}) into eqs.~(\ref{u0nrg}) and (\ref{uinrg}), we obtain to the order required for the NNLO S$_1$S$_2$ sector, the following local frame velocity components:
\begin{align}
u^{(0)}_1 &= 1 - \frac{G m_2}{r} + \frac{G}{r^2} n^i \left( S^{(i)(j)}_2 \left(v^j_1 - v^j_2\right) + S^{(0)(i)}_2  \right), \label{u0f}\\
\begin{split}
u^{(i)}_1 &= v^i_1  + \frac{G m_2}{r} \left( v^i_1 - 2 v^i_2 \right) + \frac{G}{r^2} S_2^{(i)(j)}n^j \nl
	+ \frac{G}{r} \bigg[ \frac{n^j}{2 r} S^{(i)(j)}_2 \left( v_2^2 - 3\left(\vec{v}_2\cdot\vec{n}\right)^2 - 2 \vec{v}_1\cdot\vec{v}_2\right)
		+ \frac{n^j}{r} \left( S^{(i)(k)}_2 v^k_2 v^j_2 + S^{(j)(k)}_2 \left( v^k_1 v^i_2 - v^i_1 v^k_2 \right) \right) \nlq
		+ \frac{n^j}{r} \left( S^{(0)(i)}_2 v^j_2
		+ S^{(0)(j)}_2 \left( v^i_2 - v^i_1 \right)\right) 
		+ \frac{1}{2} S^{(i)(j)}_2 a^k_2 \left(\delta_{jk} - n^j n^k\right)
		+ \dot{S}^{(0)(i)}_2 \bigg] \\
& \quad -\frac{G^2}{r^3} S^{(i)(j)}_2 n^j \left( m_1 + m_2 \right). \label{uif} 
\end{split}
\end{align}

\paragraph{Check via the metric in harmonic coordinates.} 

Since the metric required here is that associated with the gravitational interaction up to the 
NLO SO in harmonic coordinates (recall that the EFT computation employs the  harmonic gauge), it can also be obtained from \cite{Tagoshi:2000zg}, see in particular eqs.~(3.9a-c) and appendix A there, and from the rederivation of \cite{Faye:2006gx}, which corrected several typos (note that the metric was extended to NNLO SO in \cite{Bohe:2012mr}). 
Hence, we recomputed the metric components as a check .
Evaluated for a binary system at the position of object $1$, and keeping only terms 
to the order required here, the metric is found to be
\begin{align}
h_{00}(\vec{x}_1) &= 2 \frac{G}{r} \left[ m_2 
+ \frac{n^i}{r} \left( S^{(i)(j)}_2 v^j_2 - S^{(0)(i)}_2 \right) \right] , \label{h00}\\
h_{ij}(\vec{x}_1) &= 2 \frac{G}{r} \left[ \delta_{ij} m_2 
- \delta_{ij} \frac{n^k}{r}\left( S^{(k)(l)}_2 v^l_2 + S^{(0)(k)}_2 \right)
+ \frac{n^k}{r} \left( S^{(k)(i)}_2 v^j_2 + S^{(k)(j)}_2 v^i_2 \right) \right] , \label{hij}\\
\begin{split}
h_{0i}(\vec{x}_1) &= -4 \frac{G}{r} \bigg[ m_2 v^i_2
- \frac{n^j}{4 r} \left(S^{(i)(j)}_2\left( 2 + v_2^2 - 3\left(\vec{v}_2\cdot\vec{n}\right)^2 \right)
+ 2S^{(i)(k)}_2 v^k_2 v^j_2  \right. \nlq
\left.+ 2 S^{(0)(i)}_2 v^j_2 + 2S^{(0)(j)}_2 v^i_2\right) 
+ \frac{n^j n^k - \delta_{jk}}{4} S^{(i)(j)}_2 a^k_2 
- \frac{1}{2} \dot{S}^{(0)(i)}_2 \bigg] \\
& \quad -2 \frac{G^2}{r^3} S^{(i)(j)}_2 n^j \left( m_1 + m_2 \right) .\label{h0i}
\end{split}
\end{align}
Note that these metric components should get an overall minus sign in order to conform with the signature used in the EFT convention. 
Using eqs.~(\ref{h00}), (\ref{hij}), (\ref{h0i}) in eqs.~(\ref{u0lc}) and (\ref{uilc}), we find the local frame velocity components here to be in complete agreement with those in eqs.~(\ref{u0f}), (\ref{uif}) computed via the EFT approach.

\subsection{Covariant SSC at NNLO}

For a generic PN treatment, it would be useful to write the local frame velocity as
\newcommand{\epsNLO}[1]{\epsilon_{1\text{NLO}}^{#1}}
\newcommand{\epsNLOdot}[1]{\dot{\epsilon}_{1\text{NLO}}^{#1}}
\newcommand{\epsNNLO}[1]{\epsilon_{1\text{NNLO}}^{#1}}
\begin{align}
u^{(0)}_1 &= 1 + \epsNLO{0}
        + \epsNNLO{0}
        + \Order(v^6) , \label{u0gen} \\
u^{(i)}_1 &= v^i_1 + \epsNLO{i}
        + \epsNNLO{i}
        + \Order(v^7) , \label{uigen}
\end{align}
where $\epsNLO{0}$, $\epsNLO{i}$, $\epsNNLO{0}$ and $\epsNNLO{i}$ are just placeholders. Thus, here we have 
\begin{align}
\epsNLO{0} &= - \frac{G m_2}{r}, \label{eps0nlo}\\
\epsNLO{i} &= \frac{G m_2}{r} \left(v^i_1 - 2 v^i_2\right)
        + \frac{G}{r^2} n^j S_2^{(i)(j)}, \label{epsinlo}
\end{align}   
and $\epsNNLO{0}$, $\epsNNLO{i}$ can be read off from eqs.~(\ref{u0f}), (\ref{uif}), respectively, as the remaining terms.

Using eqs.~(\ref{u0gen}), (\ref{uigen}) in eq.~(\ref{covssc}), we can then write for the NNLO SSC the following generic expression: 
\begin{align} \label{s0ipn}
S^{(0)(i)}_1 =& - S^{(i)(j)}_1 \bigg[v^j_1 +\left(\epsNLO{j}-\epsNLO{0}v_1^j\right) \nn \nlq
                +\left[\left(\epsNLO{0}\right)^2v_1^j-\epsNLO{0}\epsNLO{j}+\epsNNLO{j}-\epsNNLO{0}v_1^j\right]\bigg]+\Order(v^8),
\end{align}
or explicitly substituting eqs.~(\ref{u0f}), (\ref{uif}), the covariant SSC for the NNLO S$_1$S$_2$ sector reads
\begin{align}
\begin{split}
S^{(0)(i)}_1 =& - S^{(i)(j)}_1 \Bigg[ v^j_1  + 2\frac{G m_2}{r} \left(v^j_1 - v^j_2\right) + \frac{G}{r^2} S_2^{(j)(k)}n^k \nlq \quad
	  + \frac{G}{r} \bigg[ \frac{n^k}{2 r} S^{(j)(k)}_2 \left( v_2^2 - 3\left(\vec{v}_2\cdot\vec{n}\right)^2 - 2 \vec{v}_1\cdot\vec{v}_2\right)\nlq \quad \quad \quad
		+ \frac{n^l}{r} \left( S^{(j)(k)}_2 v^k_2 v^l_2 + S^{(l)(k)}_2 \left( v^k_1 v^j_2 - v_1^j v_1^k \right) \right) \nlq \quad \quad \quad
		+ \frac{n^k}{r} \left( S^{(0)(j)}_2 v^k_2
		+ S^{(0)(k)}_2 \left( v^j_2 - 2v^j_1 \right)\right) \nlq \quad \quad \quad
		+ \frac{1}{2} S^{(j)(k)}_2 a^l_2 \left(\delta_{kl} - n^k n^l\right)
		+ \dot{S}^{(0)(j)}_2 \bigg] \\
& \quad \quad \quad \quad -\frac{G^2 m_1}{r^3} S^{(j)(k)}_2 n^k \bigg] \,\,. 
\end{split}
\end{align}
Note that the NNLO SSC entails a recursive application of the SSC, where we would have to substitute the LO SSC in the $S^{0i}$ components on the right hand side.

To conclude this section we stress that in order to work within the EFT approach in a consistent, self-contained manner, we have to obtain the EOM for the spin from an independent variation of the action with respect to the spin and to its conjugate, the angular velocity, and not using Poisson brackets. Moreover, we note that a further EFT computation is required in order to obtain the $S^{0i}$ components, if they are left as independent degrees of freedom, to be eliminated after the derivation of the EOM. This provides indication for the incompleteness of the Routhian approach in the EFT aspect, as discussed in section \ref{EFTofspin}.

\section{Transformation to reduced canonical variables and potential} \label{VSXcan}

As explained in the previous section a possible way, which is seemingly straightforward, to compare the EFT potential in eq.~(\ref{vnnloss}) with the Hamiltonian computed in \cite{Hartung:2011ea}, would be to derive from it the EOM for the positions and spins of the objects from a variation of the action, eliminating the $S^{0i}$ components after the derivation, and to compare them with those obtained from the ADM Hamiltonian. However, in order to relate the two resulting sets of EOM, we would have to find a mapping from the EFT 
position and spin variables to the canonical variables used in the ADM Hamiltonian.

Yet, to our advantage the mapping to canonical variables has 
been already considered at the level of the effective EFT action in section 5 of \cite{Hergt:2011ik}. There, the redefinition of the spin and position 
variables emerges from the insertion of the spin constraints at the level of the action, both on the spin variable, i.e.~the insertion 
of eq.~(\ref{covssc}), and in addition, on the conjugate degrees of freedom of the spin, the Lorentz matrices $\Lambda^{Ab}$, which are absent from the EFT
spin potentials, if the effective EFT action is kept in a form of a canonical so$(1,3)$ algebra for the spin, see section \ref{Sredef} below. 
Furthermore, the EFT potential contains higher order time derivatives of position and spin, that should ultimately be eliminated, 
in order to compare with the ADM result. As we noted in section \ref{EFTpot}, a common procedure in the PN approximation, see e.g.~\cite{Barker:1980,Schafer:1984mr,Damour:1985mt,Damour:1990jh}, to eliminate these higher order time derivatives, is to substitute them 
with EOM from lower orders at the level of the Lagrangian, which also entails a redefinition of the position and spin variables, as further illustrated in section \ref{htd} here below. Therefore, we conveniently proceed to apply the transition to canonical variables at this stage, 
before any required elimination of higher order time derivatives would result further redefinitions of the variables, of which we would have to 
keep track. 

In this section, we recapitulate the transition to the canonical spin and position variables, and potential at the level of the action, 
following section 5 of \cite{Hergt:2011ik}. In particular, we improve on the redefinition of the position variable, 
following eq.~(5.42) in \cite{Hergt:2011ik} as suggested there, such that it is also formulated in closed form, 
rather than within a specific PN expansion. Thus, we employ here a fully closed form of the transformation to the canonical spin and position variables, and
potential at the level of the action, which can also be used to obtain further  
Hamiltonians, based on an EFT formulation and computation.

We would like to stress that if we just proceeded to compare the results at the level of the 
EOM, we would have not been able to relate between the variables from first principles, and 
the mapping would not have been generic and meaningful. Moreover, beyond accomplishing the comparison 
of these specific results, we here make sense of the physical connection between the EFT derivation 
and the canonical Hamiltonian. In this way we can actually generally obtain Hamiltonians, and sensible 
effective actions from EFT, which are more useful, also for the non-conservative sectors, than just the EOM.
Also from the more practical perspective the calculations at the level of the potential, 
which is just one scalar, are easier compared to calculations with the EOM, which have many 
components and are substantially less compact.

\subsection{Insertion of spin constraints and redefinition of spin} \label{Sredef}

Following the approach in section 5 of \cite{Hergt:2011ik}, working at the level of the action, we begin by considering the full EFT action
of the binary. That is, we have to restore the rotational kinetic terms, i.e.~$-\frac{1}{2}S_{ab}\Omega^{ab}$ for each spinning object, 
see eq.~(51) in \cite{Levi:2010zu}, and add them to the potentials presented in section \ref{EFTpot}. 
Here $\Omega^{ab} \equiv \Lambda_{A}^{\;\;\;a}\dot\Lambda^{Ab}$ is the angular velocity tensor, which is constructed from 
a four-dimensional Lorentz matrix $\Lambda^{Ab}$. Then, the effective EFT action can be written in the following generic form
\begin{align}\label{S4spin}
 S_{\text{eff}} = \int\dd t\left[\sum_{I=1}^2 \frac{m_I}{2} v_I^2-\sum_{I=1}^2\frac{1}{2}S_{Iab}\Omega_{I}^{ab}
	- V\left(\vct{x}_{I}, \dot{\vct{x}}_{I}, \ddot{\vct{x}}_{I},\dots, S_{Iab}, \dot{S}_{Iab}, \dots\right)\right] ,
\end{align}
where the potentials from Sec.\ \ref{EFTpot} are contained here in the potential $V$. If the higher order time derivatives of spin would be absent, then this action would correspond to a canonical so$(1,3)$ algebra Poisson bracket for the spin, and an independent variation of $\Lambda^{Ab}$ and $S_{ab}$ would then yield the same EOM for the spin as in a Routhian approach, see section \ref{EFTofspin} here, and section 5 in \cite{Hergt:2011ik}. 
Since this is an action we can directly insert spin constraints on the variables. However, here 
in addition to the spin constraint in eq.~(\ref{covssc}), we insert a corresponding conjugate gauge constraint on the Lorentz matrix, such that we substitute the temporal components of $S^{(0)(i)}$, $\Lambda^{[i](0)}$, and $\Lambda^{[0](i)}$, using the constraints
\begin{align} 
S_{ab} u^{b} &= 0   \qquad \Leftrightarrow \qquad S^{(0)(i)} = - S^{(i)(j)} \frac{u^{(j)}}{u^{(0)}},\label{supplcond}\\ 
\Lambda^{[i]a} u_a &= 0 \qquad \Leftrightarrow \qquad \Lambda_{[0]a} = \frac{u_a}{u}, \label{lorentzcond}
\end{align}
where $u \equiv \sqrt{u_a u^a}$, and square brackets on the indices denote the body-fixed frame.
We can see from the condition on the Lorentz matrix in eq.~(\ref{lorentzcond}), that this matrix describes a pure rotation 
in the rest frame of the object. Note that we have here exactly $3+3$ independent conditions to eliminate the $3+3$ redundant degrees of freedom of spin and angular velocity.
Here, we insert the constraints both on the spin as well as on its canonical conjugate in order to find a 
transformation of the variables and of the potential, such that the action takes the form 
\begin{equation}\label{canaction}
S_{\text{eff}}=\int\dd t \left[ \sum_{I=1}^2\frac{m_I}{2} \hat{v}^2_I -\sum_{I=1}^2\frac{1}{2}\hat{S}_I^{(i)(j)}\hat{\Omega}_{I}^{(i)(j)}
	- \hat{V}\left(\hat{\vct{x}}_{I}, \dot{\hat{\vct{x}}}_{I}, \ddot{\hat{\vct{x}}}_{I}, \dots, \hat{\vct{S}}_{I}, \dot{\hat{\vct{S}}}_{I}, \dots\right) \right], 
\end{equation}  
where we denote the new variables and potential by a hat. Here $\hat{\Omega}^{(i)(j)}\equiv - \hat{\Lambda}^{[k](i)} \dot{\hat{\Lambda}}^{[k](j)}$, where 
the SO$(1,3)$ Lorentz matrix is replaced by a SO$(3)$ rotation matrix, and the 4-dimensional spin tensor is reduced to a 3-dimensional one. Such a form for the action would be associated with a canonical so$(3)$ algebra for the new spin. 

Indeed, the outcome of the insertion of the constraints in eqs.~(\ref{supplcond}), (\ref{lorentzcond}) 
can be organized by introducing new variables \cite{Hergt:2011ik}.
First, the Lorentz matrix is boosted to the rest frame according to the following transformation 
\begin{align}
\Lambda^{[i](j)} &= \hat{\Lambda}^{[i](k)} \left( \eta_{(k)}{}^{(j)}
        - \frac{u_{(k)} u^{(j)}}{u(u+u^{(0)})} \right), \label{lambdatrafo} 
\end{align}
which guarantees that $\hat{\Lambda}^{[i](k)}$ is a SO$(3)$ rotation matrix.
Next, one can collect the coefficient terms of $\hat{\Omega}^{(i)(j)}$, and associate a new spin variable $\hat{S}_{(i)(j)}$ to this coefficient according to 
the transformation
\begin{align}
S_{(i)(j)} &=\hat{S}_{(i)(j)}
        -\hat{S}_{(k)(i)}\frac{u^{(j)}u^{(k)}}{u(u+u^{(0)})}
        +\hat{S}_{(k)(j)}\frac{u^{(i)}u^{(k)}}{u(u+u^{(0)})}. \label{redefinitionspin}
\end{align}
We can already see that this new spin variable coincides with the Newton-Wigner spin variable in flat spacetime, 
see eq.~(B11) in \cite{Hanson:1974qy}. Then, we are left with \cite{Hergt:2011ik} 
\begin{align}
- \frac{1}{2}S_{ab}\Omega^{ab}&=-\frac{1}{2}\hat{S}_{(i)(j)}\hat{\Omega}^{(i)(j)}+\mathcal{Z} ,
\quad \text{where }
\mathcal{Z} \equiv \hat{S}_{(i)(j)}\frac{u^{(i)}\dot{u}^{(j)}}{u(u+u^{(0)})},
\end{align}
thus the only remaining term is denoted by $\mathcal{Z}$, and contains the local acceleration $\dot{u}^{(j)}$.

\subsection{Redefinition of position and transformation of potential\label{secxredef}}

The next step is to address the contribution from $\dot{u}^{(j)}$ in $\mathcal{Z}$. 
One could keep $\mathcal{Z}$ as it stands, and take it as an addition to the potential. 
However, this addition would start to contribute already at the LO SO potential. 
This is not desirable, since $\dot{u}^{(j)}$ comprises higher order time derivatives.
Therefore, we can push these contributions to higher PN orders, based on the common 
procedure to eliminate such terms via a redefinition of the variables, see 
e.g.~\cite{Barker:1980,Schafer:1984mr,Damour:1985mt,Damour:1990jh}, and as further illustrated in section \ref{htd} here below.
Then, if we write the local-frame velocity as in eq.~(\ref{uigen}), such that $u^{(i)} = v^i + \epsilon^i$, 
we can express $\mathcal{Z}$ in the form
\begin{equation}\label{Zcal}
\mathcal{Z} =  - \left( v^j + \epsilon^j \right) \frac{\dd}{\dd t} \left[ \frac{\hat{S}_{(i)(j)} u^{(i)}}{u\left(u+u^{(0)}\right)}\right]
	        + (\text{ttd}) ,
\end{equation}
where (ttd) denotes a total time derivative. Our goal now is to eliminate the term proportional to $v^j$.
Let us then make a redefinition of the position variable, that we write as
\begin{equation}\label{xredef}
x_1^{i} = \hat{x}_1^{i} + \Delta x_1^{i}.
\end{equation}
In the kinetic part of the action, this has the effect
\begin{align}\label{dzterms}
\frac{m_1}{2} v_1^2 
	= \frac{m_1}{2} \hat{v}_1^2 
	+ m_1 \hat{v}_1^{i} \frac{\dd \Delta x_1^{i}}{\dd t}
	+ \frac{m_1}{2} \left(\frac{\dd \Delta x_1^{i}}{\dd t}\right)^2,
\end{align}
where $\hat{v}_1^{i} = \dot{\hat{x}}_1^{i}$. By setting $\Delta x_1^{i}$ as
\begin{equation}\label{redefinitionposition}
x_1^{i} = \hat{x}_1^{i} + \frac{\hat{S}_{1(j)(i)} u_1^{(j)}}{m_1 u_1 (u_1+u_1^{(0)})} ,
\end{equation}
we can indeed cancel the undesired term in $\mathcal{Z}$, using the contribution linear in $\Delta x_1^{i}$ in eq.~(\ref{dzterms}), such that we are left with 
\begin{align}\label{redefxresult}
\frac{m_1}{2} v_1^2 + \mathcal{Z}
	= \frac{m_1}{2} \hat{v}_1^2 
     -  \epsilon_1^j \frac{\dd}{\dd t} \left[ \frac{\hat{S}_{(i)(j)} u^{(i)}}{u\left(u+u^{(0)}\right)}\right]
    - \frac{m_1}{2} \left(\frac{\dd \Delta x_1^{i}}{\dd t}\right)^2.
\end{align}
Again, we can already note that the new position variable in eq.~(\ref{redefinitionposition}) coincides with the Newton-Wigner one in flat spacetime, 
see eq.~(B11) in \cite{Hanson:1974qy}.
The remaining contributions in eq.~(\ref{redefxresult}) are considered as an addition to the potential, denoted $V^{\text{extra}}$,
such that we also have a new potential
\begin{equation}\label{Vhat}
\hat{V} = V + V^{\text{extra}},
\end{equation}
where from eq.~(\ref{redefxresult}) we have that $V^{\text{extra}}$ equals
\begin{align}
\label{Vextra}
V^{\text{extra}} =& \epsilon^j_1 \frac{\dd}{\dd t} \left[ \frac{\hat{S}_{1(i)(j)} u^{(i)}_1}{u_1\left(u_1+u_1^{(0)}\right)} \right]
	+ \frac{1}{2 m_1} \left(\frac{\dd}{\dd t} \left[ \frac{\hat{S}_{1(j)(i)} u_1^{(j)}}{u_1 \left(u_1+u_1^{(0)}\right)} \right]\right)^2
+ \left[1 \longleftrightarrow 2\right].
\end{align}
The second term here is actually irrelevant for this paper, since it is at least quadratic in one of the spins. 

To conclude, after the constraints on the spin and the Lorentz matrices have been applied, and the Lorentz matrix have been transformed, 
the new potential $\hat{V}$ is obtained from the old one $V$, by a successive application of 
the spin transformation in eq.~(\ref{redefinitionspin}), 
the position transformation in eq.~(\ref{redefinitionposition}), 
and the addition of $V^{\text{extra}}$ in eq.~(\ref{Vextra}), which contains further higher order time derivatives. Note that the spin and position transformations should 
also be applied to their higher order time derivatives, and in particular to the velocities. 

Indeed, at this point the effective action 
is in the canonical form given in eq.~(\ref{canaction}). As argued in section 5 of \cite{Hergt:2011ik}, and see also in section \ref{canspin} here below, we have arrived at reduced canonical variables, which are curved spacetime generalizations of the Newton-Wigner variables.

\subsection{PN expansion of the transformation formulae}

In this section, we are going to derive the PN expansion 
of the variable transformations in eqs.~(\ref{redefinitionspin}) and (\ref{redefinitionposition}), and of the extra potential 
in eq.~(\ref{Vextra}), which as we have seen are all given in terms of the local 4-velocity. Hence, for all of these only the PN expansion of the local velocity is required.
Starting from this section, we drop the round brackets on the spins.

\paragraph{Variable transformations.}

Using eqs.~(\ref{u0gen}), (\ref{uigen}) from section \ref{spinGF}, it is straightforward to derive the PN expansion of 
eqs.~(\ref{redefinitionspin}) and (\ref{redefinitionposition}), reading
\begin{align}\label{Sijexpand}
S^{ij}_1 &= \hat{S}_1^{ij} + \hat{S}_1^{k[j} \hat{v}_1^{i]} \hat{v}_1^k
	 + \left[\left(\frac{3}{4} \hat{v}_1^2 - 2 \epsNLO{0} \right)
	        \hat{S}_1^{k[j} \hat{v}_1^{i]} \hat{v}_1^k
	 + \hat{S}_1^{k[j} \hat{v}_1^{i]} \epsNLO{k}
         + \hat{S}_1^{k[j} \epsNLO{i]} \hat{v}_1^k\right]
	 + \Order(v^7) , 
\end{align}
\begin{align}
\begin{split}\label{xexpand}
x_1^{i} &= \hat{x}_1^{i}
        - \frac{1}{2 m_1} \hat{S}_1^{ij} \hat{v}_1^j 
        - \frac{1}{2 m_1} \hat{S}_1^{ij} \left[
                \hat{v}_1^j \left( \frac{3}{4} \hat{v}_1^2 - 2 \epsNLO{0} \right) + \epsNLO{j} \right]\nl
        - \frac{1}{2 m_1} \hat{S}_1^{ij} \left[ \hat{v}_1^j \left(        
                \frac{5}{8} \hat{v}_1^4
                - 3 \hat{v}_1^2 \epsNLO{0}
                + \frac{3}{2} \hat{v}_1^k \epsNLO{k} 
                + 3 \left(\epsNLO{0}\right)^2
                \right)
        + \epsNLO{j} \left(
                \frac{3}{4} \hat{v}_1^2
                - 2 \epsNLO{0}\right)\right]\nl
        - \frac{1}{2 m_1} \hat{S}_1^{ij} \left[\epsNNLO{j}-2\hat{v}_1^j\epsNNLO{0}\right]+ \Order(v^8),
\end{split}
\end{align}
where the velocity $v$ has already been replaced by the new $\hat{v}$, neglecting spin-squared terms, which are irrelevant for this work. $\epsNLO{0}$, $\epsNLO{i}$, $\epsNNLO{0}$, and $\epsNNLO{i}$ 
are found in eqs.~(\ref{eps0nlo}), (\ref{epsinlo}), (\ref{u0f}), (\ref{uif}), respectively, where the positions, i.e.~the separation $r$, and the spin variables $S^{ij}$, 
should also be expressed in terms of the new $\hat{r}$ and $\hat{S}^{ij}$, respectively. 

\paragraph{Addition to the potential.}

For the S$_1$S$_2$ sector, we only need to take into account
the first term in eq.~(\ref{Vextra}), which is linear in the spins. 
Then the PN expansion reads
\begin{align}
\begin{split}\label{VextraExpanded}
V^{\text{extra}} &=
	 \frac{1}{2} \left( \epsNLO{j} + \epsNNLO{j} \right) \left( \hat{S}_1^{ij} \hat{a}_1^i 
	        + \dot{\hat{S}}_1^{ij} \hat{v}_1^i \right)
         + \epsNLO{j} \dot{\hat{S}}_1^{ij} \hat{v}_1^i
                \left[ \frac{3}{8} \hat{v}_1^2 - \epsNLO{0} \right] \nl
         + \epsNLO{j} \hat{S}_1^{ij} \left[
            \frac{3}{8} \hat{v}_1^2 \hat{a}_1^i
        	+ \frac{3}{4} \hat{a}_1^k \hat{v}_1^k \hat{v}_1^i
        	- \hat{a}_1^i \epsNLO{0}
        	- \hat{v}_1^i \epsNLOdot{0}
        	+ \frac{1}{2} \epsNLOdot{i}\right] \nl
	+ \left[1 \longleftrightarrow 2\right]
	+ \Order(v^{10}).
\end{split}
\end{align}
Here only the first term already contributes to NLO potentials. 
Note again that this addition contains further higher order time derivatives.

\paragraph{Cancellation of contributions from the local-frame velocity.}\label{SSCtrick}

Finally, we present here a useful and intriguing observation: It turns out, that when the $S^{0i}$ components are substituted in, and the transformations of the variables and the potential are applied, where also the extra potential is added, the contributions from $\epsNNLO{0}$ and $\epsNNLO{i}$ of the local frame velocity cancel out in the new potential.

The substitution in eq.~(\ref{s0ipn}), and the transformations of position and potential in eqs.~(\ref{xexpand}) and 
(\ref{VextraExpanded}), contain terms with $\epsNNLO{0}$ and $\epsNNLO{i}$, whereas eq.~(\ref{Sijexpand}) is independent of them. 
Eq.~(\ref{s0ipn}) contributes through the LO SO potential $(-V_{\text{SO}}^{\text{LO}})$ in eq.~(\ref{vloso}) as
\begin{equation}
-V_{\text{SO}}^{\text{LO}} \to \frac{G m_2}{\hat{r}^2} \hat{S}^{ij}_1 \hat{n}^i
        \left( \hat{v}^j_1 \epsNNLO{0} - \epsNNLO{j} \right) + [1 \longleftrightarrow 2].
\end{equation}
The transformation of the position in eq.~(\ref{xexpand}) then leads to the following contribution via the Newtonian potential
\begin{equation}
\frac{G m_1 m_2}{\hat{r}} \to \frac{G m_2}{\hat{r}^2} \hat{S}^{ij}_1 \hat{n}^i
        \left( - \hat{v}^j_1 \epsNNLO{0} + \frac{1}{2} \epsNNLO{j} \right)
        + [1 \longleftrightarrow 2].
\end{equation}
Recall that the kinetic energy $\frac{1}{2}m_1v_1^2$ was already taken into account in the extra potential in section \ref{secxredef}.
Finally, the addition to the potential in eq.~(\ref{VextraExpanded}) contributes 
\begin{align}
- V^{\text{extra}} &\to
	 - \frac{1}{2} \epsNNLO{j} \left( \hat{S}_1^{ij} \hat{a}_1^i
	        + \dot{\hat{S}}_1^{ij} \hat{v}_1^i \right) + [1 \longleftrightarrow 2] 
	        \approx \frac{G m_2}{\hat{r}^2} \hat{S}^{ij}_1 \hat{n}^i \frac{1}{2} \epsNNLO{j}
        + [1 \longleftrightarrow 2] ,
\end{align}
where we have substituted in the Newtonian EOM, see section \ref{htd} here below. Clearly, the sum of all these contributions vanishes, which is an intriguing observation. This is also very convenient, because it implies that we actually do not need to include $\epsNNLO{0}$ and $\epsNNLO{i}$ in the new potential, if we start from an EFT potential, in which the $S^{0i}$ components are left as independent degrees of freedom, as in this comparison. We should stress though that the NNLO components of the local velocity would contribute to the EOM, had we derived them in section \ref{spinGF}, through the elimination of the $S^{0i}$ components that follows. We should also make clear that this cancellation does not imply that the NNLO parts of the SSC, and of the position and potential transformations vanish, these should all still be applied to NNLO.

\section{Elimination of higher order time derivatives}\label{htd}

As we already noted in previous sections, a common procedure in the PN perturbative scheme is the elimination of higher order time derivatives, 
such as $\vct{a}_1 = \dot{\vct{v}}_1$, at the level of the Lagrangian via their substitution by lower order PN EOM 
\cite{Barker:1980,Schafer:1984mr,Damour:1985mt,Damour:1990jh}. This substitution entails a redefinition of the variables, i.e.~a change of gauge. 
This can easily be seen, considering that shifting a variable by a small variation results in a variation of the action, 
which exactly yields the EOM of the variable. This idea actually applies in any perturbative context, and we should note here the analogy 
to the concept of redundant operators in effective field theories, see e.g.~\cite{Goldberger:2007hy}. In section \ref{secxredef} 
we have already seen that higher order time derivatives can be pushed to higher PN orders via a redefinition of the position. 
Therefore at this stage we go on to eliminate the higher order time derivatives from our potential in order to simplify the comparison 
with the Hamiltonian, where higher order time derivatives are already eliminated.  

We note that in previous similar comparisons at NLO, e.g.~in \cite{Hergt:2011ik}, the higher order time derivatives 
were eliminated prior to the elimination of the $S^{0i}$ components, and the transition to the canonical potential
and variables. Yet, as we noted already in section \ref{EFTpot}, terms of the form $\dot{S}^{0i}$ appear in the potentials as of NLO, 
see e.g.~eqs.~(\ref{vnloso}) and (\ref{vnloss}), which can be avoided at NLO by flipping the time derivative. Since at NNLO we have terms which are already non linear in the $S^{0i}$ components, unlike the NLO case, the $\dot{S}^{0i}$ terms are inevitable. Therefore, since an elimination of the time derivative on the $S^{0i}$ components
entails a redefinition of the spin variable, as we show here below, performing it prior to the insertion of the spin constraint for the $S^{0i}$ component in 
eq.~(\ref{covssc}), would modify the spin constraint itself, and result in further undesirable complications. Hence, 
again we see that the unphysical degrees of freedom associated with $S^{0i}$ are a potential source of trouble, that better be eliminated as early as possible. 

In this section we review, following \cite{Damour:1990jh}, the procedure to eliminate higher order time derivatives of positions 
by a substitution of the lower order EOM, which constitutes a useful recipe in the PN approximation. Moreover, we explicitly 
extend here this procedure for a spin variable, which to the best of our knowledge was not addressed in previous literature. Finally, we provide the lower order PN EOM, which are required for the 
substitution of the higher order time derivatives, that are found in the NNLO S$_1$S$_2$ potential. Starting from this section, we work with the new variables and potential exclusively, 
so we are dropping the hat on them.

\subsection{Elimination of accelerations}

Let us consider a Lagrangian containing higher order time derivatives of the position. 
Then an infinitesimal variation of the position $\delta \vct{x}_1$ yields
\begin{equation}\label{xEOM}
\delta \left[\int \dd t \left(\sum_{I=1}^2\frac{m_I}{2} v_I^2 -V\left(\vec{x}_I, \vec{v}_I, \dot{\vec{v}}_I, \dots \right)\right)\right] =0 
\Rightarrow m_1 \vct{a}_1=
  - \frac{\partial V}{\partial \vct{x}_1}
	+ \frac{d}{dt} \frac{\partial V}{\partial \vct{v}_1}
	- \frac{d^2}{dt^2} \frac{\partial V}{\partial \dot{\vct{v}}_1}
	+ \dots ,
\end{equation} 
namely the EOM. Now, if we consider a redefinition of the position as in eq.~(\ref{xredef}), then 
similarly to linear order in $\Delta \vct{x}_1$, this adds to the Lagrangian the contribution reading 
\begin{equation}\label{xredefadd}
\Delta L \approx 
 - \left[ m_1 \vct{a}_1 + \frac{\partial V}{\partial \vct{x}_1}
	- \frac{d}{dt} \frac{\partial V}{\partial \vct{v}_1} 
	+ \frac{d^2}{dt^2} \frac{\partial V}{\partial \dot{\vct{v}}_1}
	+ \dots \right]
	\Delta \vct{x}_1 ,
\end{equation}
where total time derivatives were dropped. Here $\vct{a}_1$ is given in terms of the unshifted position, because
we neglect quadratic terms in $\Delta \vct{x}_1$. A shift of the position is therefore adding to the Lagrangian the EOM times 
the shift of position, as long as quadratic and higher orders of the position shift can be neglected. 
It is now clear from (\ref{xredefadd}), that one can set $\Delta \vct{x}_1$ so
that it cancels terms in the Lagrangian containing $\vct{a}_1$.

Let us write all contributions to the Lagrangian containing $\vct{a}_1$ as
\begin{equation}\label{La}
L_{a_1} = \vct{A}_1 \cdot \vct{a}_1 ,
\end{equation}
with some coefficient $\vct{A}_1$, which may again depend on $\vct{a}_1$.
From eq.~(\ref{xredefadd}), we see that we should set
\begin{equation}
\Delta \vct{x}_1 = \frac{\vct{A}_1}{m_1} .
\end{equation}
Notice that this implies $L_{a_1} = \Order{(\Delta \vct{x}_1)}$, so $L_{a_1}$ must be small
in some approximation scheme. If we add to $L_{a_1}$ the result of the shift,
we get
\begin{equation}\label{xredefcomplete}
L_{a_1} + \Delta L
        = \frac{\vct{A}_1}{m_1} \left[ - \frac{\partial V}{\partial \vct{x}_1}
	+ \frac{d}{dt} \frac{\partial V}{\partial \vct{v}_1} - \dots \right] .
\end{equation}
This is precisely the same result that we would get from inserting the EOM (\ref{xEOM})
into the Lagrangian (\ref{La}). Notice that one can neglect the contributions of $L_{a_1}$ itself to the substituted EOM,
because $L_{a_1} = \Order{(\Delta \vct{x}_1)}$, so its contributions to
eq.~(\ref{xredefcomplete}) via $V$ are $\Order{[(\Delta \vct{x}_1)^2]}$.
If $\Delta \vct{x}_1$ contains further higher order time derivatives, then one has to iterate
the procedure. Similarly, time derivatives of even higher order, like $\dot{\vct{a}}_1$,
can be treated by flipping of time derivatives, and see \cite{Damour:1990jh} for further details.
In conclusion, a substitution of EOM in the Lagrangian in some approximation scheme can be used to push higher order time 
derivatives of velocities to negligible orders as long as terms of order $(\Delta \vct{x}_1)^2$ can be ignored, and this corresponds to 
a shift of positions.

\subsection{Elimination of time derivatives of spin\label{sec:SEOM}}

The steps in the last section can be similarly applied to generic Lagrangians,
depending on generic variables, instead of positions. Here we will extend this construction for the case of a spin variable, which to the best of our knowledge was not addressed explicitly in previous literature, and obtain the redefinition of the spin variable, which eliminates its higher order time derivatives. 

We begin by considering a redefinition of the rotation matrix $\Lambda^{ij}_1$, which relates the corotating and the local frames of the spin.
A transformation of the local frame components, then reads
\begin{equation}
\Delta \Lambda^{ij}_1 = \Lambda^{ik}_1 \omega^{kj}_1 ,
\end{equation}
where $\omega^{kj}_1$ is a generator of a rotation matrix, i.e.~antisymmetric, and is small in some approximation.
Let us keep the redefinition of the spin $\Delta S^{ij}_1$ arbitrary for now.
This leads to an addition to the Lagrangian in eq.~(\ref{canaction}) 
of the form
\begin{equation}\label{Sredefaddpre}
\Delta L \approx  -\left[ \frac{1}{2} \dot{S}^{ij}_1 - S^{jk}_1 \Omega^{ki}_1 \right] \omega^{ij}_1
- \left[ \frac{1}{2} \Omega_1^{ij} + \frac{\partial V}{\partial S_1^{ij}} - \frac{d}{dt} \frac{\partial V}{\partial \dot{S}_1^{ij}} + \dots  \right] \Delta S_1^{ij} ,
\end{equation}
to linear order in the redefinitions, and dropping total time derivatives.
An infinitesimal independent variation of the rotation matrix $\Lambda^{ij}_1$ and the spin $S_1^{ij}$ 
yields a similar equation, where the two EOMs can be read off from eq.~(\ref{Sredefaddpre}) as
\begin{equation}
\dot{S}^{ij}_1 = - 2 S_1^{k[i} \Omega^{j]k}_1 , \qquad
\Omega_1^{ij} = - 2 \left[ \frac{\partial V}{\partial S_1^{ij}} - \frac{d}{dt} \frac{\partial V}{\partial \dot{S}_1^{ij}} + \dots  \right] .
\end{equation}
These can be combined into a single EOM for the spin, given by
\begin{equation}
\dot{S}_1^{ij} 
= - 4 S_1^{k[i} \delta^{j]l} \left[ \frac{\partial V}{\partial S_1^{kl}} - \frac{d}{dt} \frac{\partial V}{\partial \dot{S}_1^{kl}} + \dots \right] . \label{ScEOM}
\end{equation}
We see that eq.~(\ref{Sredefaddpre}) produces a $\dot{S}_1^{ij}$, but it also produces $\Omega^{ij}_1$, 
and thus $\dot{\Lambda}^{ij}_1$, which should not be introduced into the potential. However, we can cancel 
$\Omega^{ij}_1$ from eq.~(\ref{Sredefaddpre}) by specifying the spin redefinition to be
\begin{equation}
\Delta S_1^{ij} = S_1^{ik} \omega^{kj} - S_1^{jk} \omega^{ki},
\end{equation}
which means that the spin should be rotated similarly to $\Lambda^{ik}_1$.
The change in the Lagrangian in eq.~(\ref{Sredefaddpre}) then reads
\begin{equation}\label{Sredefadd}
\Delta L \approx -\left[ \frac{1}{2} \dot{S}^{ij}_1
        + 2 S_1^{ki}  \left( \frac{\partial V}{\partial S_1^{kj}} - \frac{d}{dt} \frac{\partial V}{\partial \dot{S}_1^{kj}} + \dots \right)\right] \omega_1^{ij} .
\end{equation}

Now we are ready to cancel terms containing $\dot{S}^{ij}_1$ from the Lagrangian. 
Let us write these terms as
\begin{equation}\label{RdtS}
L_{\dot{S}_1} = A^{ij}_1 \dot{S}^{ij}_1 ,
\end{equation}
where the generic coefficients $A^{ij}_1$ may again depend on time derivatives
of the spin or on accelerations.
The next step is to fix $\omega_1^{ij}$, such that the term in eq.~(\ref{RdtS}) is removed
from the Lagrangian. It is easy to see that we should fix
\begin{equation}
\omega_1^{ij} = A^{ij}-A^{ji}  ,
\end{equation}
so that we get for the Lagrangian
\begin{equation}\label{Sredefcomplete}
L_{\dot{S}_1} + \Delta L
= - 4 A^{ij}_1 S_1^{k[i} \delta^{j]l} \left[ \frac{\partial V}{\partial S_1^{kl}} - \frac{d}{dt} \frac{\partial V}{\partial \dot{S}_1^{kl}} + \dots \right] .
\end{equation}
Hence, we could have arrived at this by substituting the spin EOM in eq.~(\ref{ScEOM}) 
into eq.~(\ref{RdtS}), which is what we anticipated.

\paragraph{Spin EOM from Poisson brackets.} \label{canspin}

After the time derivatives of the spin have been eliminated, its EOM in eq.~(\ref{ScEOM})
takes on the simple form
\begin{equation}\label{canSeom}
\dot{S}_1^{ij}
= - 4 S_1^{k[i} \delta^{j]l} \frac{\partial V_s}{\partial S_1^{kl}} ,
\end{equation}
where $V_s$ denotes the potential with substituted EOM, which is actually $V_s = V - \Delta L$.
If we impose on the spin variables the equal-time Poisson brackets, namely 
\begin{equation}
\{ S_1^{ij}, S_1^{kl} \} = S_1^{ik} \delta_{jl} - S_1^{il} \delta_{jk}
        + S_1^{jl} \delta_{ik} - S_1^{jk} \delta_{il},
\end{equation}
then we can rewrite the EOM in eq.~(\ref{canSeom}) in the form
\begin{equation}
\dot{S}_1^{ij} = \{ S_1^{ij}, -V_s \}.
\end{equation}
Therefore, after the elimination of time derivatives of the spin, we arrive at canonical spin variables. Notice that the substitution of EOM implicitly redefines the spin, in addition to the
explicit redefinition in section \ref{VSXcan}. Thus, after all redefinitions
have been performed, the Poisson brackets become an alternative 
to the variation of the action for the obtainment of the spin EOM.

\subsection{Lower order equations of motion}

Finally, also here the elimination of higher order time derivatives corresponds to substituting them in the potential using lower order EOMs. However, one should keep in mind that by doing this, one is actually redefining the variables. Then, the lower order PN EOM, which are required for the NNLO S$_1$S$_2$ potential, resulting from eqs.~(\ref{xEOM}) and (\ref{ScEOM}), read
\begin{align}
a_1^i &=
- \frac{G m_2}{r^2}n^{i}
+ \frac{G m_2}{r^2}
        \left[ - n^{i} v_1^2
	 + 4 n^{i} \vec{v}_1\cdot\vec{v}_2
	 - 2 n^{i} v_2^2
	 + \frac{3}{2} n^{i} ( \vec{n}\cdot\vec{v}_2 )^2
	 + 4 \vec{n}\cdot\vec{v}_1 v_1^{i}
	 - 3 \vec{n}\cdot\vec{v}_2 v_1^{i} \right.\nnl
   - 4 \vec{n}\cdot\vec{v}_1 v_2^{i}
	 + 3 \vec{n}\cdot\vec{v}_2 v_2^{i} \bigg]
+ \frac{G^2m_2}{r^3} n^{i}
         \left(5 m_1  + 4 m_2 \right)
+ \frac{G m_2}{ r^3 m_1}
        \left[ \frac{9}{2} n^{i} \vec{S}_1\times\vec{n}\cdot\vec{v}_1 \right.\nnl
	 \left.- 6 n^{i} \vec{S}_1\times\vec{n}\cdot\vec{v}_2
	 + \frac{9}{2} n^{j} S_1^{ij} \vec{n}\cdot\vec{v}_1
	 -  \frac{9}{2} n^{j} S_1^{ij} \vec{n}\cdot\vec{v}_2
	 - 3 S_1^{ij} v_1^{j}
	 + \frac{7}{2} S_1^{ij} v_2^{j} \right] \nnl
+ \frac{G}{r^3}
        \left[ 6 n^{i} \vec{S}_2\times\vec{n}\cdot\vec{v}_1
	 - \frac{9}{2} n^{i} \vec{S}_2\times\vec{n}\cdot\vec{v}_2
	 + 6 n^{j} S_2^{ij} \vec{n}\cdot\vec{v}_1
	 - 6 n^{j} S_2^{ij} \vec{n}\cdot\vec{v}_2
	 - 4 S_2^{ij} v_1^{j}
	 + \frac{7}{2} S_2^{ij} v_2^{j} \right] \nnl
- 3\frac{G}{m_1 r^4}
        \left[ \vec{S}_1\cdot\vec{S}_2 n^i 
        - 5 \vec{S}_1\cdot\vec{n} \vec{S}_2\cdot\vec{n} n^i 
	      + \vec{S}_2\cdot\vec{n} \vec{S}_1^i 
	      + \vec{S}_1\cdot\vec{n} \vec{S}_2^i 
	  \right]
+ \Order{\left(v^6\right)} , \label{asubst}\\
\dot{S}_1^{ij} &=
\frac{G m_2}{r^2}
        \left[ 3 n^{[i} S_1^{j]k} v_1^{k}
         - 3 n^{k} v_1^{[i} S_1^{j]k}
	 + 4 n^{k} v_2^{[i} S_1^{j]k}
	 - 4 n^{[i} S_1^{j]k} v_2^{k} \right] \nnl
+ \frac{2 G}{r^3}
	 \left[ 2 S_1^{k[i} S_2^{j]k}
	 - 3 n^{k} n^{l} S_1^{k[i} S_2^{j]l}
	 - 3 n^{[i} S_1^{j]l} n^{k} S_2^{kl} \right]
+ \Order{\left(v^6\right)},
\end{align}
where the contributions required here are up to the 1PN non spinning sector, and the LO SO and S$_1$S$_2$ sectors. 
Note that the EOM from the LO SO sector are already obtained here from the canonical form of the potential, e.g.~from eq.~(70) 
in \cite{Levi:2010zu}, since we already transformed to canonical variables. A priori, the right hand side of eq.~(\ref{asubst}) contains further terms with $a_I^i$ and $\dot{S}_I^{ij}$, 
which were substituted using lower order EOM. This corresponds to an iterative redefinition of variables.
At NNLO we also encounter terms containing $\dot{a}_I^i$ and $\ddot{S}_I^{ij}$, which are produced by the transition
to the new variables. These we substitute by
\begin{equation}
\dot{a}_1^i = \frac{G m_2}{r^3} \left(3 n^{i} \vec{n}\cdot\vec{v}_1
	 - 3 n^{i} \vec{n}\cdot\vec{v}_2
	 -  {v_1}^{i}
	 + {v_2}^{i} \right)
	 + \Order{\left(v^5\right)} , \qquad
\ddot{S}_1^{ij} = 0 + \Order{\left(v^5\right)},
\end{equation}
that is with contributions only from the Newtonian level.

\section{Legendre transform and the Hamiltonians}\label{EFTH}

At this stage, after the higher order time derivatives were eliminated in the last section, 
the most direct and complete way to examine the equivalence with the ADM Hamiltonian 
is to obtain the Hamiltonian for the EFT potential, and subsequently look for 
a canonical transformation, that may relate these two results. 
Since the higher order time derivatives are now eliminated,
we can perform an ordinary Legendre transformation in the velocities $\vct{v}_I$ for this purpose.

The canonical momentum of object $1$ is defined as usual by
\begin{equation}\label{defpcan}
\vct{p}_1 = \frac{\partial L}{\partial \vct{v}_1} = m_1 \vct{v}_1 - \frac{\partial V_s} {\partial \vct{v}_1},
\end{equation}
where the part of the Lagrangian required for the obtainment of the NNLO S$_1$S$_2$ Hamiltonian 
$H_{\text{S$_1$S$_2$}}^{\text{NNLO}}$ was presented in section \ref{EFTpot}.
The Legendre transformation of the Lagrangian $L$ then leads to a Hamiltonian $H$. 
This transformation can be rewritten as
\begin{align}
H &= \vct{v}_1 \cdot \vct{p}_1 + \vct{v}_2 \cdot \vct{p}_2 - L
        = \vct{v}_1 \cdot \vct{p}_1 + \vct{v}_2 \cdot \vct{p}_2
	- \frac{1}{2} m_1 v_1^2 - \frac{1}{2} m_2 v_2^2 + V_s \nn\\
	&= \frac{p_1^2}{2 m_1} + \frac{p_2^2}{2 m_2}
	- \frac{1}{2 m_1} \left( \vct{p}_1 - m_1 \vct{v}_1 \right)^2 - \frac{1}{2 m_2} \left( \vct{p}_2 - m_2 \vct{v}_2 \right)^2 + V_s \nn\\
	&= \frac{p_1^2}{2 m_1} + \frac{p_2^2}{2 m_2}
	- \frac{1}{2 m_1} \left(\frac{\partial V_s}{\partial \vct{v}_1}\right)^2
	- \frac{1}{2 m_2} \left(\frac{\partial V_s}{\partial \vct{v}_2}\right)^2
	+ V_s . \label{legendre}
\end{align}
In this expression, we must replace the velocities by canonical momenta.
The relation $\vct{v}_I(\vct{p}_J)$ must be obtained by inverting eq.~(\ref{defpcan}) in a PN expansion. 
For completeness, we provide the explicit result here, reading
\begin{align}\label{vtop}
v_1^i &= 
\frac{p_1^{i}}{m_1}
- \frac{p_1^{i} p_1^2 }{2 m_1^3}
+ \frac{G}{r} \left[- \frac{3 m_2 p_1^{i}}{m_1}
	 + \frac{7}{2} p_2^{i}
	 + \frac{1}{2} n^{i} \vec{n}\cdot\vec{p}_2 \right]
- \frac{3 G m_2 n^{j} S_1^{ij}}{2 m_1 r^2}
- \frac{2 G n^{j} S_2^{ij}}{r^2} \nnl
+ \frac{G}{r^2} \left[\frac{p_2^{j} S_1^{ij} \vec{n}\cdot\vec{p}_1 }{4 m_1^2}
	 + \frac{5 m_2 n^{j} S_1^{ij} p_1^2 }{8 m_1^3}
	 - \frac{p_2^{j} S_1^{ij} \vec{n}\cdot\vec{p}_2 }{m_1 m_2}
	 + \frac{9 n^{j} S_1^{ij} \vec{n}\cdot\vec{p}_1 \vec{n}\cdot\vec{p}_2 }{4 m_1^2}
	 + \frac{n^{j} S_1^{ij} \vec{p}_1\cdot\vec{p}_2 }{4 m_1^2} \right.\nnl
	 - \frac{n^{j} S_1^{ij} p_2^2 }{4 m_1 m_2}
	 + \vec{S}_1\times\vec{n}\cdot\vec{p}_1
	 \left(- \frac{5 m_2 p_1^{i}}{4 m_1^3}
	 - \frac{p_2^{i}}{4 m_1^2}
	 - \frac{9 n^{i} \vec{n}\cdot\vec{p}_2 }{4 m_1^2} \right) 
	 + \frac{n^{i} \vec{S}_1\times\vec{p}_1\cdot\vec{p}_2 }{4 m_1^2} \nnl
         + \vec{S}_1\times\vec{n}\cdot\vec{p}_2 \left(\frac{p_2^{i}}{m_1 m_2}
	 + \frac{3 n^{i} \vec{n}\cdot\vec{p}_2 }{m_1 m_2} \right)  \Bigg]
+ \frac{G}{r^2} \left[- \frac{p_2^{j} S_2^{ij} \vec{n}\cdot\vec{p}_1 }{m_1 m_2}
	 + \frac{p_2^{j} S_2^{ij} \vec{n}\cdot\vec{p}_2 }{4 m_2^2} \right.\nnl
	 + \frac{3 n^{j} S_2^{ij} \vec{n}\cdot\vec{p}_1 \vec{n}\cdot\vec{p}_2 }{m_1 m_2}
	 + \frac{n^{j} S_2^{ij} \vec{p}_1\cdot\vec{p}_2 }{m_1 m_2}
	 - \frac{n^{i} \vec{S}_2\times\vec{p}_1\cdot\vec{p}_2 }{m_1 m_2}
	 - \vec{S}_2\times\vec{n}\cdot\vec{p}_1 
	 \left(\frac{p_2^{i}}{m_1 m_2}
        	 +\frac{3 n^{i} \vec{n}\cdot\vec{p}_2 }{m_1 m_2} \right)  \nnl
	 + \vec{S}_2\times\vec{n}\cdot\vec{p}_2
	 \left(- \frac{p_1^{i}}{2 m_1 m_2}
	 + \frac{p_2^{i}}{4 m_2^2}
	 + \frac{9 n^{i} \vec{n}\cdot\vec{p}_2 }{4 m_2^2} \right)  \Bigg]
+ \frac{G^2}{r^3} n^{j} S_1^{ij} \left[\frac{11}{2} m_2 + \frac{5 m_2^2}{m_1}\right] \nnl
+ \frac{G^2}{r^3} n^{j} S_2^{ij} \left[7 m_1 + 6 m_2 \right]
+ \frac{G}{r^3} \left[- \frac{p_2^{j} S_1^{jk} S_2^{ik}}{2 m_1 m_2}
	 + \frac{p_2^{j} S_1^{ik} S_2^{jk}}{4 m_1 m_2}
	 + \frac{3 n^{j} S_1^{ik} S_2^{jk} \vec{n}\cdot\vec{p}_1 }{2 m_1^2}\right.\nnl
	 - \frac{3 n^{j} S_1^{jk} S_2^{ik} \vec{n}\cdot\vec{p}_2 }{2 m_1 m_2}
	 - \frac{3 n^{j} S_1^{ik} S_2^{jk} \vec{n}\cdot\vec{p}_2 }{2 m_1 m_2}
	 + \frac{3 n^{j} S_2^{ij} \vec{S}_1\times\vec{n}\cdot\vec{p}_1 }{2 m_1^2}
	 - \frac{3 n^{j} S_2^{ij} \vec{S}_1\times\vec{n}\cdot\vec{p}_2 }{m_1 m_2} \nnl
	 + \frac{3 n^{j} S_1^{ij} \vec{S}_2\times\vec{n}\cdot\vec{p}_1 }{2 m_1^2}
	 - \frac{3 n^{j} S_1^{ij} \vec{S}_2\times\vec{n}\cdot\vec{p}_2 }{4 m_1 m_2}
	 + \left(- \frac{p_2^{i}}{2 m_1 m_2}
	 - \frac{3 n^{i} \vec{n}\cdot\vec{p}_2 }{2 m_1 m_2} \right) S_1^{kl} S_2^{kl} \nnl
	 + \frac{3 n^{i} n^k p_1^l S_1^{lm} S_2^{km}}{2 m_1^2}
	 - \frac{3 n^{i} n^k p_2^l S_1^{lm} S_2^{km}}{2 m_1 m_2}
	 + \left(\frac{3 p_2^{i}}{2 m_1 m_2}
	 + \frac{15 n^{i} \vec{n}\cdot\vec{p}_2 }{2 m_1 m_2} \right) n^k n^l S_1^{km} S_2^{lm} \nnl
	 - \frac{3 n^{i} n^k p_2^l S_1^{km} S_2^{lm}}{2 m_1 m_2}\Bigg] ,
\end{align}
where we have also dropped the terms coming from 2PN, because they are not
relevant for the NNLO S$_1$S$_2$ sector.

To conclude, the potential in the Hamiltonian arises from adding the terms containing 
partial derivatives in eq.~(\ref{legendre}) to the potential
$V_s$, and by replacing velocities with the conjugate momenta using eq.~(\ref{vtop}).
The resulting Hamiltonians are collected below.

\subsection{EFT Hamiltonians}

Here we present the EFT Hamiltonians, which follow from applying all of the previous steps, that we described, on the EFT Lagrangians.
The non spinning Hamiltonians up to 2PN are given by
\begin{align}
H_{\text{N}} &=
	 \frac{p_1^2}{2 m_1}
	 + \frac{p_2^2 }{2 m_2}
         - \frac{G m_1 m_2}{r} , \\
H_{\text{1PN}} &=
	 - \frac{p_1^4}{8 m_1^3} - \frac{p_2^4}{8 m_2^3}
	 + \frac{G}{2 m_1 m_2 r} \left(
        	 - 3 m_2^2 p_1^2
        	 + m_1 m_2 \vec{n}\cdot\vec{p}_1  \vec{n}\cdot\vec{p}_2
        	 + 7 m_1 m_2 \vec{p}_1\cdot\vec{p}_2
        	 - 3 m_1^2 p_2^2 \right) \nnl
        + \frac{G^2}{2 r^2} \left(m_1^2 m_2 + m_1 m_2^2 \right) , \\
H_{\text{2PN}} &=
\frac{p_1^6}{16 m_1^5} + \frac{p_2^6}{16 m_2^5}
+ \frac{G}{8 m_1^3 m_2^3 r}
         \left( 5 m_2^4 p_1^4
         + 4 m_1 m_2^3 \vec{n}\cdot\vec{p}_1  p_1^2  \vec{n}\cdot\vec{p}_2
         - 3 m_1^2 m_2^2 ( \vec{n}\cdot\vec{p}_1 )^2 ( \vec{n}\cdot\vec{p}_2 )^2 \right.\nnl
         + m_1^2 m_2^2 p_1^2 ( \vec{n}\cdot\vec{p}_2 )^2
         - 4 m_1 m_2^3 p_1^2  \vec{p}_1\cdot\vec{p}_2
         - 12 m_1^2 m_2^2 \vec{n}\cdot\vec{p}_1  \vec{n}\cdot\vec{p}_2  \vec{p}_1\cdot\vec{p}_2
         - 2 m_1^2 m_2^2 ( \vec{p}_1\cdot\vec{p}_2 )^2 \nnl
         \left. + m_1^2 m_2^2 ( \vec{n}\cdot\vec{p}_1 )^2 p_2^2
         - 3 m_1^2 m_2^2 p_1^2 p_2^2
         + 4 m_1^3 m_2 \vec{n}\cdot\vec{p}_1  \vec{n}\cdot\vec{p}_2  p_2^2
         - 4 m_1^3 m_2 \vec{p}_1\cdot\vec{p}_2  p_2^2
         + 5 m_1^4 p_2^4 \right) \nnl
+ \frac{G^2}{4 m_1 m_2 r^2}
         \left(-2 m_2^3 ( \vec{n}\cdot\vec{p}_1 )^2
	 + 17 m_1 m_2^2 p_1^2
	 + 11 m_2^3 p_1^2
	 - 2 m_1^3 ( \vec{n}\cdot\vec{p}_2 )^2
	 - 28 m_1^2 m_2 \vec{p}_1\cdot\vec{p}_2 \right.\nnl
	 \left.- 28 m_1 m_2^2 \vec{p}_1\cdot\vec{p}_2
	 + 11 m_1^3 p_2^2
	 + 17 m_1^2 m_2 p_2^2 \right)
- \frac{G^3}{4 r^3}
        \left( 2 m_1^3 m_2
	 + 5 m_1^2 m_2^2
	 + 2 m_1 m_2^3 \right) .
\end{align}
The LO linear in spin Hamiltonians are given by
\begin{align}
H_{\text{SO}}^{\text{LO}} &=
	 \frac{G}{r^2} \vec{S}_1\times\vec{n}\cdot
	        \left( \frac{3 m_2 \vec{p}_1 }{2 m_1} - 2 \vec{p}_2 \right)
	 + \left(1 \longleftrightarrow 2\right), \\
H_{\text{S$_1$S$_2$}}^{\text{LO}} &=
	 \frac{G}{r^3} \left( 3 \vec{n}\cdot\vec{S}_1  \vec{n}\cdot\vec{S}_2
	 -  \vec{S}_1\cdot\vec{S}_2 \right).
\end{align}
The NLO linear in spin Hamiltonians are given by 
\begin{align}
H_{\text{SO}}^{\text{NLO}} &=
\frac{G}{r^2}
	  \left[ \frac{\vec{S}_1\times\vec{n}\cdot\vec{p}_1 }{8 m_1^3 m_2}
        	  \left(-5 m_2^2 p_1^2
	         - 18 m_1 m_2 \vec{n}\cdot\vec{p}_1  \vec{n}\cdot\vec{p}_2
        	 - 2 m_1 m_2 \vec{p}_1\cdot\vec{p}_2
        	 + 2 m_1^2 p_2^2 \right) \right.\nnl	  
	 \left. + \frac{\vec{S}_1\times\vec{n}\cdot\vec{p}_2 }{m_1 m_2}
	        \left(3 \vec{n}\cdot\vec{p}_1  \vec{n}\cdot\vec{p}_2
	         + \vec{p}_1\cdot\vec{p}_2 \right)
	 + \frac{\vec{S}_1\times\vec{p}_1\cdot\vec{p}_2 }{4 m_1^2 m_2}
        	 \left(m_2 \vec{n}\cdot\vec{p}_1
        	 - 4 m_1 \vec{n}\cdot\vec{p}_2 \right) \right] \nnl
+ \frac{G^2}{r^3} \vec{S}_1\times\vec{n}\cdot
         \left[ \frac{\vec{p}_1}{2 m_1}
                \left(-11 m_1 m_2 - 10 m_2^2\right)
	 + \vec{p}_2 \left(6 m_1 + 7 m_2\right) \right]
	 + \left[1 \longleftrightarrow 2\right] , \\
H_{\text{S$_1$S$_2$}}^{\text{NLO}} &=
\frac{G}{r^3}
        \left[ \frac{3 \vec{n}\cdot\vec{S}_1  \vec{n}\cdot\vec{S}_2 }{4 m_1^2 m_2^2}
                \left( 2 m_2^2 p_1^2
        	 - 10 m_1 m_2 \vec{n}\cdot\vec{p}_1  \vec{n}\cdot\vec{p}_2
	         - 7 m_1 m_2 \vec{p}_1\cdot\vec{p}_2
        	 + 2 m_1^2 p_2^2 \right) \right.\nnl
	 + \frac{3 \vec{p}_1\cdot\vec{S}_1  \vec{n}\cdot\vec{S}_2 }{2 m_1^2 m_2}
	        \left( 3 m_1 \vec{n}\cdot\vec{p}_2 - m_2 \vec{n}\cdot\vec{p}_1 \right)
	 -  \frac{3 \vec{p}_2\cdot\vec{S}_1  \vec{n}\cdot\vec{S}_2 }{4 m_1 m_2^2}
	        \left( 4 m_1 \vec{n}\cdot\vec{p}_2 - 3 m_2 \vec{n}\cdot\vec{p}_1 \right) \nnl
	 + \frac{3 \vec{n}\cdot\vec{S}_1  \vec{p}_1\cdot\vec{S}_2 }{4 m_1^2 m_2}
	        \left( 3 m_1 \vec{n}\cdot\vec{p}_2 - 4 m_2 \vec{n}\cdot\vec{p}_1 \right)
	 + \frac{3 \vec{p}_1\cdot\vec{S}_1  \vec{p}_1\cdot\vec{S}_2 }{2 m_1^2}
	 - \frac{ \vec{p}_2\cdot\vec{S}_1  \vec{p}_1\cdot\vec{S}_2 }{m_1 m_2} \nnl
	 - \frac{3 \vec{n}\cdot\vec{S}_1  \vec{p}_2\cdot\vec{S}_2 }{2 m_1 m_2^2}
	        \left( m_1 \vec{n}\cdot\vec{p}_2 - 3 m_2 \vec{n}\cdot\vec{p}_1 \right)
	 -  \frac{5 \vec{p}_1\cdot\vec{S}_1  \vec{p}_2\cdot\vec{S}_2 }{2 m_1 m_2}
	 + \frac{3 \vec{p}_2\cdot\vec{S}_1  \vec{p}_2\cdot\vec{S}_2 }{2 m_2^2} \nnl
	 + \frac{\vec{S}_1\cdot\vec{S}_2 }{4 m_1^2 m_2^2}
	        \left( 12 m_2^2 ( \vec{n}\cdot\vec{p}_1 )^2
        	 - 6 m_2^2 p_1^2
        	 - 21 m_1 m_2 \vec{n}\cdot\vec{p}_1  \vec{n}\cdot\vec{p}_2
	         + 12 m_1^2 ( \vec{n}\cdot\vec{p}_2 )^2
	         - 6 m_1^2 p_2^2 \right.\nnl
   \left.+ 16 m_1 m_2 \vec{p}_1\cdot\vec{p}_2 \right) \Bigg]
+ \frac{G^2}{r^4} \left[
        6 \left(m_1 + m_2\right) \vec{S}_1\cdot\vec{S}_2
        - 12 \left(m_1 + m_2\right) \vec{n}\cdot\vec{S}_1  \vec{n}\cdot\vec{S}_2 \right] .
\end{align}
Here we have avoided double scalar triple products by using the identity
\begin{equation}\label{epsepsformula}
\epsilon_{ijk} \epsilon_{lmn} =
        \delta_{il} \delta_{jm} \delta_{kn} - \delta_{il} \delta_{jn} \delta_{km}
        + \delta_{im} \delta_{jn} \delta_{kl} - \delta_{im} \delta_{jl} \delta_{kn}
        + \delta_{in} \delta_{jl} \delta_{km} - \delta_{in} \delta_{jm} \delta_{kl} .
\end{equation}
Finally, the NNLO S$_1$S$_2$ Hamiltonian is given by
\begin{align} 
H_{\text{S$_1$S$_2$}}^{\text{NNLO}} &=
\frac{G}{r^3} \left[
        - \frac{3 \vec{n}\cdot\vec{S}_1  \vec{n}\cdot\vec{S}_2 }{16 m_1^4 m_2^4}
                \left[ 6 m_2^4 p_1^4
        	 - 70 m_1^2 m_2^2 ( \vec{n}\cdot\vec{p}_1 )^2 ( \vec{n}\cdot\vec{p}_2 )^2
        	 + 10 m_1^2 m_2^2 p_1^2 ( \vec{n}\cdot\vec{p}_2 )^2 \right.\right.\nnl
        	 - 15 m_1 m_2^3 p_1^2  \vec{p}_1\cdot\vec{p}_2
        	 - 50 m_1^2 m_2^2 \vec{n}\cdot\vec{p}_1  \vec{n}\cdot\vec{p}_2  \vec{p}_1\cdot\vec{p}_2
        	 + 10 m_1^2 m_2^2 ( \vec{p}_1\cdot\vec{p}_2 )^2 \nnl
        	 \left.+ 10 m_1^2 m_2^2 ( \vec{n}\cdot\vec{p}_1 )^2 p_2^2
        	 - 2 m_1^2 m_2^2 p_1^2  p_2^2
	         - 15 m_1^3 m_2 \vec{p}_1\cdot\vec{p}_2  p_2^2
        	 + 6 m_1^4 p_2^4 \right] \nnl
	      - \frac{3 \vec{p}_1\cdot\vec{S}_1  \vec{n}\cdot\vec{S}_2 }{8 m_1^4 m_2^2}
	        \left[ -3 m_2^2 \vec{n}\cdot\vec{p}_1  p_1^2
	         - 10 m_1 m_2 ( \vec{n}\cdot\vec{p}_1 )^2 \vec{n}\cdot\vec{p}_2
	         + 6 m_1 m_2 p_1^2  \vec{n}\cdot\vec{p}_2 \right.\nnl
	         \left.+ 30 m_1^2 \vec{n}\cdot\vec{p}_1 ( \vec{n}\cdot\vec{p}_2 )^2
	         + 4 m_1 m_2 \vec{n}\cdot\vec{p}_1  \vec{p}_1\cdot\vec{p}_2
	         - 5 m_1^2 \vec{n}\cdot\vec{p}_2  \vec{p}_1\cdot\vec{p}_2
        	 + 2 m_1^2 \vec{n}\cdot\vec{p}_1  p_2^2 \right] \nnl
	 + \frac{3 \vec{p}_2\cdot\vec{S}_1  \vec{n}\cdot\vec{S}_2 }{16 m_1^3 m_2^4}
	        \left[ m_2^3 \vec{n}\cdot\vec{p}_1  p_1^2
	         + 10 m_1 m_2^2 ( \vec{n}\cdot\vec{p}_1 )^2 \vec{n}\cdot\vec{p}_2
	         - 6 m_1 m_2^2 p_1^2  \vec{n}\cdot\vec{p}_2 \right.\nnl
	         \left.+ 6 m_1 m_2^2 \vec{n}\cdot\vec{p}_1  \vec{p}_1\cdot\vec{p}_2
	         - 16 m_1^2 m_2 \vec{n}\cdot\vec{p}_2  \vec{p}_1\cdot\vec{p}_2
	         - 3 m_1^2 m_2 \vec{n}\cdot\vec{p}_1  p_2^2
        	 + 16 m_1^3 \vec{n}\cdot\vec{p}_2  p_2^2 \right] \nnl
	 + \frac{3 \vec{n}\cdot\vec{S}_1  \vec{p}_1\cdot\vec{S}_2 }{16 m_1^4 m_2^3}
	        \left[ 16 m_2^3 \vec{n}\cdot\vec{p}_1  p_1^2
	         - 3 m_1 m_2^2 p_1^2  \vec{n}\cdot\vec{p}_2
	         + 10 m_1^2 m_2 \vec{n}\cdot\vec{p}_1 ( \vec{n}\cdot\vec{p}_2 )^2 \right.\nnl
	         \left.- 16 m_1 m_2^2 \vec{n}\cdot\vec{p}_1  \vec{p}_1\cdot\vec{p}_2
	         + 6 m_1^2 m_2 \vec{n}\cdot\vec{p}_2  \vec{p}_1\cdot\vec{p}_2
        	 - 6 m_1^2 m_2 \vec{n}\cdot\vec{p}_1  p_2^2
	         + m_1^3 \vec{n}\cdot\vec{p}_2  p_2^2 \right] \nnl
	 + \frac{\vec{p}_1\cdot\vec{S}_1  \vec{p}_1\cdot\vec{S}_2 }{8 m_1^4 m_2^2}
	        \left[ -11 m_2^2 p_1^2
	         + 12 m_1 m_2 \vec{n}\cdot\vec{p}_1  \vec{n}\cdot\vec{p}_2
	         - 21 m_1^2 ( \vec{n}\cdot\vec{p}_2 )^2
	         + 4 m_1 m_2 \vec{p}_1\cdot\vec{p}_2 \right.\nnl
        	 \left.+ 17 m_1^2 p_2^2 \right]
	 + \frac{\vec{p}_2\cdot\vec{S}_1  \vec{p}_1\cdot\vec{S}_2 }{8 m_1^3 m_2^3}
	        \left[ m_2^2 p_1^2
	         - 12 m_1 m_2 \vec{n}\cdot\vec{p}_1  \vec{n}\cdot\vec{p}_2
	         - 15 m_1 m_2 \vec{p}_1\cdot\vec{p}_2
        	 + m_1^2 p_2^2 \right] \nnl
	 + \frac{3 \vec{n}\cdot\vec{S}_1  \vec{p}_2\cdot\vec{S}_2 }{8 m_1^2 m_2^4}
	        \left[ -30 m_2^2 ( \vec{n}\cdot\vec{p}_1 )^2 \vec{n}\cdot\vec{p}_2
	         - 2 m_2^2 p_1^2  \vec{n}\cdot\vec{p}_2
	         + 10 m_1 m_2 \vec{n}\cdot\vec{p}_1 ( \vec{n}\cdot\vec{p}_2 )^2 \right.\nnl
	         \left.+ 5 m_2^2 \vec{n}\cdot\vec{p}_1  \vec{p}_1\cdot\vec{p}_2
	         - 4 m_1 m_2 \vec{n}\cdot\vec{p}_2  \vec{p}_1\cdot\vec{p}_2
	         - 6 m_1 m_2 \vec{n}\cdot\vec{p}_1  p_2^2
        	 + 3 m_1^2 \vec{n}\cdot\vec{p}_2  p_2^2 \right] \nnl
	 + \frac{\vec{p}_1\cdot\vec{S}_1  \vec{p}_2\cdot\vec{S}_2 }{8 m_1^3 m_2^3}
	        \left[ -18 m_2^2 ( \vec{n}\cdot\vec{p}_1 )^2
	         + 10 m_2^2 p_1^2
	         + 84 m_1 m_2 \vec{n}\cdot\vec{p}_1  \vec{n}\cdot\vec{p}_2
	         - 18 m_1^2 ( \vec{n}\cdot\vec{p}_2 )^2 \right.\nnl
	         \left.- 23 m_1 m_2 \vec{p}_1\cdot\vec{p}_2
        	 + 10 m_1^2 p_2^2 \right]
	 + \frac{\vec{p}_2\cdot\vec{S}_1  \vec{p}_2\cdot\vec{S}_2 }{8 m_1^2 m_2^4}
	        \left[ -21 m_2^2 ( \vec{n}\cdot\vec{p}_1 )^2
	         + 17 m_2^2 p_1^2 \right.\nnl
	         \left.+ 12 m_1 m_2 \vec{n}\cdot\vec{p}_1  \vec{n}\cdot\vec{p}_2
	         + 4 m_1 m_2 \vec{p}_1\cdot\vec{p}_2
        	 - 11 m_1^2 p_2^2 \right]
	 + \frac{\vec{S}_1\cdot\vec{S}_2 }{16 m_1^4 m_2^4}
	        \left[ -48 m_2^4 ( \vec{n}\cdot\vec{p}_1 )^2 p_1^2 \right.\nnl
	         + 22 m_2^4 p_1^4
	         + 9 m_1 m_2^3 \vec{n}\cdot\vec{p}_1  p_1^2  \vec{n}\cdot\vec{p}_2
	         + 60 m_1^2 m_2^2 p_1^2 ( \vec{n}\cdot\vec{p}_2 )^2
	         + 48 m_1 m_2^3 ( \vec{n}\cdot\vec{p}_1 )^2 \vec{p}_1\cdot\vec{p}_2 \nnl
	         - 34 m_1 m_2^3 p_1^2  \vec{p}_1\cdot\vec{p}_2
	         - 162 m_1^2 m_2^2 \vec{n}\cdot\vec{p}_1  \vec{n}\cdot\vec{p}_2  \vec{p}_1\cdot\vec{p}_2
	         + 48 m_1^3 m_2 ( \vec{n}\cdot\vec{p}_2 )^2 \vec{p}_1\cdot\vec{p}_2 \nnl
	         + 50 m_1^2 m_2^2 ( \vec{p}_1\cdot\vec{p}_2 )^2
	         + 60 m_1^2 m_2^2 ( \vec{n}\cdot\vec{p}_1 )^2 p_2^2
	         - 40 m_1^2 m_2^2 p_1^2  p_2^2
	         + 9 m_1^3 m_2 \vec{n}\cdot\vec{p}_1  \vec{n}\cdot\vec{p}_2  p_2^2 \nnl
	         \left.- 48 m_1^4 ( \vec{n}\cdot\vec{p}_2 )^2 p_2^2
	         - 34 m_1^3 m_2 \vec{p}_1\cdot\vec{p}_2  p_2^2
        	 + 22 m_1^4 p_2^4 \right] \Bigg] \nnl
+ \frac{G^2}{r^4} \left[
        \frac{\vec{n}\cdot\vec{S}_1  \vec{n}\cdot\vec{S}_2 }{4 m_1^2 m_2^2}
                \left[ 66 m_1 m_2^2 ( \vec{n}\cdot\vec{p}_1 )^2
	         - 85 m_1 m_2^2 p_1^2
	         - 35 m_2^3 p_1^2
	         + 72 m_1^2 m_2 \vec{n}\cdot\vec{p}_1  \vec{n}\cdot\vec{p}_2 \right.\right.\nnl
	         + 72 m_1 m_2^2 \vec{n}\cdot\vec{p}_1  \vec{n}\cdot\vec{p}_2
	         + 66 m_1^2 m_2 ( \vec{n}\cdot\vec{p}_2 )^2
	         + 143 m_1^2 m_2 \vec{p}_1\cdot\vec{p}_2
	         + 143 m_1 m_2^2 \vec{p}_1\cdot\vec{p}_2 \nnl
	         \left.- 35 m_1^3 p_2^2
        	 - 85 m_1^2 m_2 p_2^2 \right]
	 -  \frac{8 \vec{p}_1\cdot\vec{S}_1  \vec{p}_1\cdot\vec{S}_2 }{m_1^2} \left(m_1 + m_2\right)
	 + \frac{3 \vec{p}_2\cdot\vec{S}_1  \vec{p}_1\cdot\vec{S}_2 }{2 m_1 m_2} \left(m_1 + m_2)\right. \nnl
	 + \frac{\vec{p}_1\cdot\vec{S}_1  \vec{n}\cdot\vec{S}_2 }{4 m_1^2 m_2}
	        \left[ 37 m_1 m_2 \vec{n}\cdot\vec{p}_1
	         + 36 m_2^2 \vec{n}\cdot\vec{p}_1
	         - 106 m_1^2 \vec{n}\cdot\vec{p}_2
        	 - 83 m_1 m_2 \vec{n}\cdot\vec{p}_2 \right] \nnl
	 + \frac{\vec{p}_2\cdot\vec{S}_1  \vec{n}\cdot\vec{S}_2 }{4 m_1 m_2^2}
	        \left[ -27 m_1 m_2 \vec{n}\cdot\vec{p}_1
	         + 20 m_2^2 \vec{n}\cdot\vec{p}_1
	         + 58 m_1^2 \vec{n}\cdot\vec{p}_2
        	 + 7 m_1 m_2 \vec{n}\cdot\vec{p}_2 \right] \nnl
	 + \frac{\vec{n}\cdot\vec{S}_1  \vec{p}_1\cdot\vec{S}_2 }{4 m_1^2 m_2}
	        \left[ 7 m_1 m_2 \vec{n}\cdot\vec{p}_1
	         + 58 m_2^2 \vec{n}\cdot\vec{p}_1
	         + 20 m_1^2 \vec{n}\cdot\vec{p}_2
        	 - 27 m_1 m_2 \vec{n}\cdot\vec{p}_2 \right] \nnl
	 + \frac{\vec{n}\cdot\vec{S}_1  \vec{p}_2\cdot\vec{S}_2 }{4 m_1 m_2^2}
	        \left[ -83 m_1 m_2 \vec{n}\cdot\vec{p}_1
	         - 106 m_2^2 \vec{n}\cdot\vec{p}_1
	         + 36 m_1^2 \vec{n}\cdot\vec{p}_2
        	 + 37 m_1 m_2 \vec{n}\cdot\vec{p}_2 \right] \nnl
	 + \frac{61 \vec{p}_1\cdot\vec{S}_1  \vec{p}_2\cdot\vec{S}_2 }{4 m_1 m_2} \left(m_1 + m_2\right)
	 -  \frac{8 \vec{p}_2\cdot\vec{S}_1  \vec{p}_2\cdot\vec{S}_2 }{m_2^2} \left(m_1 + m_2\right)\nnl
	 + \frac{\vec{S}_1\cdot\vec{S}_2 }{4 m_1^2 m_2^2}
	        \left[ - 58 m_2^3 ( \vec{n}\cdot\vec{p}_1 )^2 
	         - 50 m_1 m_2^2 ( \vec{n}\cdot\vec{p}_1 )^2
	         + 51 m_1 m_2^2 p_1^2
	         + 31 m_2^3 p_1^2 \right.\nnl
	         + 68 m_1^2 m_2 \vec{n}\cdot\vec{p}_1  \vec{n}\cdot\vec{p}_2 
	         + 68 m_1 m_2^2 \vec{n}\cdot\vec{p}_1  \vec{n}\cdot\vec{p}_2
	         - 58 m_1^3 ( \vec{n}\cdot\vec{p}_2 )^2
	         - 50 m_1^2 m_2 ( \vec{n}\cdot\vec{p}_2 )^2 \nnl
	         \left.- 94 m_1^2 m_2 \vec{p}_1\cdot\vec{p}_2 
	         - 94 m_1 m_2^2 \vec{p}_1\cdot\vec{p}_2
	         + 31 m_1^3 p_2^2
        	 + 51 m_1^2 m_2 p_2^2 \right] \Bigg]\nnl
+ \frac{G^3}{r^5} \left[
        \frac{5 \vec{n}\cdot\vec{S}_1 \vec{n}\cdot\vec{S}_2}{2} \left(11 m_1^2 + 28 m_1 m_2+ 11 m_2^2\right)\right.\nnl
        \left.- \frac{\vec{S}_1\cdot\vec{S}_2}{2} \left(31 m_1^2 + 60 m_1 m_2 + 31 m_2^2\right)\right] , \label{HNNLOSS}
\end{align}
where again we have made use of eq.~(\ref{epsepsformula}).

\subsection{ADM Hamiltonians}

Here, we also collect all of the ADM Hamiltonians required for the comparison.
The Hamiltonians $H_{\text{N}}$, $H_{\text{1PN}}$, $H_{\text{SO}}^{\text{LO}}$,
$H_{\text{S$_1$S$_2$}}^{\text{LO}}$, and $H_{\text{S$_1$S$_2$}}^{\text{NLO}}$, 
which was obtained in \cite{Steinhoff:2007mb,Steinhoff:2008zr},
are the same as in the EFT formalism, thus are not repeated here.

The 2PN Hamiltonian obtained in \cite{Ohta:1973je,Ohta:1974kp,Ohta:1974pq,Damour:1985mt,Damour:1988mr},
see e.g.~eq.~(2.5) in \cite{Damour:1988mr}, is given by 
\begin{align}
H_{\text{2PN}} &=
\frac{p_1^6}{16 m_1^5} + \frac{p_2^3}{16 m_2^5}
+ \frac{G}{8 m_1^3 m_2^3 r}
         \left[ 5 m_2^4 p_1^4
	 - 3 m_1^2 m_2^2 ( \vec{n}\cdot\vec{p}_1 )^2 ( \vec{n}\cdot\vec{p}_2 )^2
	 + 5 m_1^2 m_2^2 p_1^2 ( \vec{n}\cdot\vec{p}_2 )^2 \right.\nnl
	 - 12 m_1^2 m_2^2 \vec{n}\cdot\vec{p}_1  \vec{n}\cdot\vec{p}_2  \vec{p}_1\cdot\vec{p}_2
	 - 2 m_1^2 m_2^2 ( \vec{p}_1\cdot\vec{p}_2 )^2
	 + 5 m_1^2 m_2^2 ( \vec{n}\cdot\vec{p}_1 )^2 p_2^2
	 - 11 m_1^2 m_2^2 p_1^2  p_2^2 \nnl
	 \left.+ 5 m_1^4 ( p_2^2 )^2 \right]
+ \frac{G^2}{4 m_1 m_2 r^2}
        \left[ 19 m_1 m_2^2 p_1^2
	 + 10 m_2^3 p_1^2
	 - 6 m_1^2 m_2 \vec{n}\cdot\vec{p}_1  \vec{n}\cdot\vec{p}_2 \right.\nnl
	 \left.- 6 m_1 m_2^2 \vec{n}\cdot\vec{p}_1  \vec{n}\cdot\vec{p}_2
	 - 27 m_1^2 m_2 \vec{p}_1\cdot\vec{p}_2
	 - 27 m_1 m_2^2 \vec{p}_1\cdot\vec{p}_2
	 + 10 m_1^3 p_2^2
	 + 19 m_1^2 m_2 p_2^2 \right] \nnl
- \frac{G^3}{4 r^3} \left(m_1^3 m_2 + 5 m_1^2 m_2^2 + m_1 m_2^3\right).
\end{align} 
The NLO SO Hamiltonian obtained in \cite{Damour:2007nc,Steinhoff:2008zr}, is given by
\begin{align}
H_{\text{SO}}^{\text{NLO}} &=
\frac{G}{r^2}
         \left[ \frac{\vec{S}_1\times\vec{n}\cdot\vec{p}_1 }{8 m_1^3 m_2}
                \left[ -5 m_2^2 p_1^2
	         - 6 m_1 m_2 \vec{n}\cdot\vec{p}_1  \vec{n}\cdot\vec{p}_2
	         - 12 m_1^2 ( \vec{n}\cdot\vec{p}_2 )^2
	          \right.\right.\nnl
        	 \left. - 6 m_1 m_2 \vec{p}_1\cdot\vec{p}_2+ 6 m_1^2 p_2^2 \right]
	 + \frac{\vec{S}_1\times\vec{n}\cdot\vec{p}_2 }{m_1 m_2}
	        \left[ 3 \vec{n}\cdot\vec{p}_1  \vec{n}\cdot\vec{p}_2
        	 + \vec{p}_1\cdot\vec{p}_2 \right] \nnl
	 \left.+ \frac{\vec{S}_1\times\vec{p}_1\cdot\vec{p}_2 }{4 m_1^2 m_2}
	        \left[ 3 m_2 \vec{n}\cdot\vec{p}_1
	- 8 m_1 \vec{n}\cdot\vec{p}_2 \right] \right]\nnl
	+ \frac{G^2}{r^3}
	 \vec{S}_1\times\vec{n}\cdot \left[ \frac{3 \vec{p}_2}{2} \left(4 m_1 + 5 m_2\right)
	 - \frac{\vec{p}_1}{2 m_1} \left(11 m_1 m_2 + 10 m_2^2\right) \right]
	 + \left[1 \longleftrightarrow 2\right].
\end{align}
Finally, the NNLO S$_1$S$_2$ Hamiltonian obtained in \cite{Hartung:2011ea,Hartung:2013dza} is given by
\begin{align} 
H_{\text{S$_1$S$_2$}}^{\text{NNLO}} &=
\frac{G}{r^3}
        \left[ - \frac{3 \vec{n}\cdot\vec{S}_1  \vec{n}\cdot\vec{S}_2 }{16 m_1^4 m_2^4}
                \left[ 6 m_2^4 p_1^4
	         - 20 m_1 m_2^3 \vec{n}\cdot\vec{p}_1  p_1^2  \vec{n}\cdot\vec{p}_2
	         - 70 m_1^2 m_2^2 ( \vec{n}\cdot\vec{p}_1 )^2 ( \vec{n}\cdot\vec{p}_2 )^2 \right.\right.\nnl
	         + 30 m_1^2 m_2^2 p_1^2 ( \vec{n}\cdot\vec{p}_2 )^2
	         - 11 m_1 m_2^3 p_1^2  \vec{p}_1\cdot\vec{p}_2
	         - 50 m_1^2 m_2^2 \vec{n}\cdot\vec{p}_1  \vec{n}\cdot\vec{p}_2  \vec{p}_1\cdot\vec{p}_2 \nnl
	         + 10 m_1^2 m_2^2 ( \vec{p}_1\cdot\vec{p}_2 )^2
	         + 30 m_1^2 m_2^2 ( \vec{n}\cdot\vec{p}_1 )^2 p_2^2
	         - 10 m_1^2 m_2^2 p_1^2  p_2^2
	         - 20 m_1^3 m_2 \vec{n}\cdot\vec{p}_1  \vec{n}\cdot\vec{p}_2  p_2^2 \nnl
	         \left.- 11 m_1^3 m_2 \vec{p}_1\cdot\vec{p}_2  p_2^2
        	 + 6 m_1^4 p_2^4 \right]
	 - \frac{3 \vec{p}_1\cdot\vec{S}_1  \vec{n}\cdot\vec{S}_2 }{8 m_1^4 m_2^3}
	        \left[ -3 m_2^3 \vec{n}\cdot\vec{p}_1  p_1^2
	         + 6 m_1 m_2^2 p_1^2  \vec{n}\cdot\vec{p}_2 \right.\nnl
	        \left. + 20 m_1^2 m_2 \vec{n}\cdot\vec{p}_1 ( \vec{n}\cdot\vec{p}_2 )^2
	         -  m_1^2 m_2 \vec{n}\cdot\vec{p}_2  \vec{p}_1\cdot\vec{p}_2
        	 - 4 m_1^2 m_2 \vec{n}\cdot\vec{p}_1  p_2^2
	         + 6 m_1^3 \vec{n}\cdot\vec{p}_2  p_2^2 \right] \nnl
	 + \frac{3 \vec{p}_2\cdot\vec{S}_1  \vec{n}\cdot\vec{S}_2 }{16 m_1^3 m_2^4}
	        \left[ -3 m_2^3 \vec{n}\cdot\vec{p}_1  p_1^2
	         + 10 m_1 m_2^2 ( \vec{n}\cdot\vec{p}_1 )^2 \vec{n}\cdot\vec{p}_2
	         + 2 m_1 m_2^2 p_1^2  \vec{n}\cdot\vec{p}_2 \right.\nnl
	         \left.- 10 m_1 m_2^2 \vec{n}\cdot\vec{p}_1  \vec{p}_1\cdot\vec{p}_2
	         - 3 m_1^2 m_2 \vec{n}\cdot\vec{p}_1  p_2^2
        	 + 12 m_1^3 \vec{n}\cdot\vec{p}_2  p_2^2 \right] \nnl
	 - \frac{3 \vec{n}\cdot\vec{S}_1  \vec{p}_1\cdot\vec{S}_2 }{16 m_1^4 m_2^3}
	        \left[ -12 m_2^3 \vec{n}\cdot\vec{p}_1  p_1^2
	         + 3 m_1 m_2^2 p_1^2  \vec{n}\cdot\vec{p}_2
	         - 10 m_1^2 m_2 \vec{n}\cdot\vec{p}_1 ( \vec{n}\cdot\vec{p}_2 )^2 \right.\nnl
	         \left.+ 10 m_1^2 m_2 \vec{n}\cdot\vec{p}_2  \vec{p}_1\cdot\vec{p}_2
	         - 2 m_1^2 m_2 \vec{n}\cdot\vec{p}_1  p_2^2
        	 + 3 m_1^3 \vec{n}\cdot\vec{p}_2  p_2^2 \right]
	 - \frac{3 \vec{p}_1\cdot\vec{S}_1  \vec{p}_1\cdot\vec{S}_2 }{8 m_1^4 m_2^2}
	        \left[ 3 m_2^2 p_1^2 \right.\nnl
	         \left.+ m_1^2 ( \vec{n}\cdot\vec{p}_2 )^2
        	 -  m_1^2 p_2^2 \right]
	 + \frac{\vec{p}_2\cdot\vec{S}_1  \vec{p}_1\cdot\vec{S}_2 }{8 m_1^3 m_2^3}
	        \left[ m_2^2 p_1^2
	         - 12 m_1 m_2 \vec{n}\cdot\vec{p}_1  \vec{n}\cdot\vec{p}_2
	         + m_1 m_2 \vec{p}_1\cdot\vec{p}_2 \right.\nnl
        	 \left.+ m_1^2 p_2^2 \right]
	 + \frac{3 \vec{n}\cdot\vec{S}_1  \vec{p}_2\cdot\vec{S}_2 }{8 m_1^3 m_2^4}
	        \left[ - 6 m_2^3 \vec{n}\cdot\vec{p}_1  p_1^2
	         - 20 m_1 m_2^2 ( \vec{n}\cdot\vec{p}_1 )^2 \vec{n}\cdot\vec{p}_2
	         + 4 m_1 m_2^2 p_1^2  \vec{n}\cdot\vec{p}_2 \right.\nnl
	        \left. + m_1 m_2^2 \vec{n}\cdot\vec{p}_1  \vec{p}_1\cdot\vec{p}_2
	         - 6 m_1^2 m_2 \vec{n}\cdot\vec{p}_1  p_2^2
        	 + 3 m_1^3 \vec{n}\cdot\vec{p}_2  p_2^2 \right]
	 + \frac{\vec{p}_1\cdot\vec{S}_1  \vec{p}_2\cdot\vec{S}_2 }{8 m_1^3 m_2^3}
	        \left[ 10 m_2^2 p_1^2 \right.\nnl
	         \left.+ 36 m_1 m_2 \vec{n}\cdot\vec{p}_1  \vec{n}\cdot\vec{p}_2
	         - 7 m_1 m_2 \vec{p}_1\cdot\vec{p}_2
        	 + 10 m_1^2 p_2^2 \right]
	 - \frac{3 \vec{p}_2\cdot\vec{S}_1  \vec{p}_2\cdot\vec{S}_2 }{8 m_1^2 m_2^4}
	        \left[ m_2^2 ( \vec{n}\cdot\vec{p}_1 )^2 \right.\nnl
        	 \left.-  m_2^2 p_1^2
	         + 3 m_1^2 p_2^2 \right]
	 + \frac{\vec{S}_1\cdot\vec{S}_2 }{16 m_1^4 m_2^4}
	        \left[ -36 m_2^4 ( \vec{n}\cdot\vec{p}_1 )^2 p_1^2
        	 + 18 m_2^4 p_1^4
        	 + 33 m_1 m_2^3 \vec{n}\cdot\vec{p}_1  p_1^2  \vec{n}\cdot\vec{p}_2 \right.\nnl
        	 + 24 m_1^2 m_2^2 p_1^2 ( \vec{n}\cdot\vec{p}_2 )^2
        	 - 26 m_1 m_2^3 p_1^2  \vec{p}_1\cdot\vec{p}_2
        	 - 66 m_1^2 m_2^2 \vec{n}\cdot\vec{p}_1  \vec{n}\cdot\vec{p}_2  \vec{p}_1\cdot\vec{p}_2 \nnl
        	 + 18 m_1^2 m_2^2 ( \vec{p}_1\cdot\vec{p}_2 )^2
        	 + 24 m_1^2 m_2^2 ( \vec{n}\cdot\vec{p}_1 )^2 p_2^2
        	 - 16 m_1^2 m_2^2 p_1^2  p_2^2
        	 + 33 m_1^3 m_2 \vec{n}\cdot\vec{p}_1  \vec{n}\cdot\vec{p}_2  p_2^2 \nnl
        	 \left.- 36 m_1^4 ( \vec{n}\cdot\vec{p}_2 )^2 p_2^2
        	 - 26 m_1^3 m_2 \vec{p}_1\cdot\vec{p}_2  p_2^2
        	 + 18 m_1^4 p_2^4 \right] \Bigg] \nnl
+ \frac{G^2}{r^4}
         \left[ \frac{\vec{n}\cdot\vec{S}_1  \vec{n}\cdot\vec{S}_2 }{4 m_1^2 m_2^2}
                \left[ - 88 m_1 m_2^2 p_1^2 - 48 m_1 m_2^2 ( \vec{n}\cdot\vec{p}_1 )^2
	         - 36 m_2^3 p_1^2\right.\right.\nnl
	         + 255 m_1^2 m_2 \vec{n}\cdot\vec{p}_1  \vec{n}\cdot\vec{p}_2
           + 255 m_1 m_2^2 \vec{n}\cdot\vec{p}_1  \vec{n}\cdot\vec{p}_2 
	         - 48 m_1^2 m_2 ( \vec{n}\cdot\vec{p}_2 )^2\nnl
	         \left.+ 145 m_1^2 m_2 \vec{p}_1\cdot\vec{p}_2
        	 + 145 m_1 m_2^2 \vec{p}_1\cdot\vec{p}_2
        	 - 36 m_1^3 p_2^2
        	 - 88 m_1^2 m_2 p_2^2 \right] \nnl
	 + \frac{\vec{p}_1\cdot\vec{S}_1  \vec{n}\cdot\vec{S}_2 }{4 m_1^2 m_2}
	        \left[ 64 m_1 m_2 \vec{n}\cdot\vec{p}_1
	         + 36 m_2^2 \vec{n}\cdot\vec{p}_1
	         - 125 m_1^2 \vec{n}\cdot\vec{p}_2
        	 - 125 m_1 m_2 \vec{n}\cdot\vec{p}_2 \right] \nnl
	 + \frac{\vec{p}_2\cdot\vec{S}_1  \vec{n}\cdot\vec{S}_2 }{4 m_1 m_2^2}
	        \left[ -83 m_1 m_2 \vec{n}\cdot\vec{p}_1
	         - 85 m_2^2 \vec{n}\cdot\vec{p}_1
	         + 72 m_1^2 \vec{n}\cdot\vec{p}_2
        	 + 114 m_1 m_2 \vec{n}\cdot\vec{p}_2 \right] \nnl
	 + \frac{\vec{n}\cdot\vec{S}_1  \vec{p}_1\cdot\vec{S}_2 }{4 m_1^2 m_2}
	        \left[ 114 m_1 m_2 \vec{n}\cdot\vec{p}_1
	         + 72 m_2^2 \vec{n}\cdot\vec{p}_1
	         - 85 m_1^2 \vec{n}\cdot\vec{p}_2
        	 - 83 m_1 m_2 \vec{n}\cdot\vec{p}_2 \right] \nnl
	 - \frac{9 \vec{p}_1\cdot\vec{S}_1 \vec{p}_1\cdot\vec{S}_2 }{8 m_1^2} \left(13 m_1 + 8 m_2\right)
	 + \frac{83 \vec{p}_2\cdot\vec{S}_1 \vec{p}_1\cdot\vec{S}_2 }{8 m_1 m_2} \left(m_1 + m_2\right) \nnl
	 + \frac{\vec{n}\cdot\vec{S}_1  \vec{p}_2\cdot\vec{S}_2 }{4 m_1 m_2^2}
	        \left[ -125 m_1 m_2 \vec{n}\cdot\vec{p}_1
        	 - 125 m_2^2 \vec{n}\cdot\vec{p}_1
        	 + 36 m_1^2 \vec{n}\cdot\vec{p}_2
        	 + 64 m_1 m_2 \vec{n}\cdot\vec{p}_2 \right] \nnl
	 + \frac{71 \vec{p}_1\cdot\vec{S}_1 \vec{p}_2\cdot\vec{S}_2 }{4 m_1 m_2} \left(m_1 + m_2\right)
	 - \frac{9 \vec{p}_2\cdot\vec{S}_1 \vec{p}_2\cdot\vec{S}_2 }{8 m_2^2} \left(8 m_1 + 13 m_2\right) \nnl
	 + \frac{\vec{S}_1\cdot\vec{S}_2 }{8 m_1^2 m_2^2}
	        \left[ -196 m_1 m_2^2 ( \vec{n}\cdot\vec{p}_1 )^2
	         - 144 m_2^3 ( \vec{n}\cdot\vec{p}_1 )^2
	         + 145 m_1 m_2^2 p_1^2
	         + 72 m_2^3 p_1^2 \right.\nnl
	        + 286 m_1^2 m_2 \vec{n}\cdot\vec{p}_1  \vec{n}\cdot\vec{p}_2
	         + 286 m_1 m_2^2 \vec{n}\cdot\vec{p}_1  \vec{n}\cdot\vec{p}_2
	         - 144 m_1^3 ( \vec{n}\cdot\vec{p}_2 )^2 \nnl
	         - 196 m_1^2 m_2 ( \vec{n}\cdot\vec{p}_2 )^2
	          - 259 m_1^2 m_2 \vec{p}_1\cdot\vec{p}_2
	         - 259 m_1 m_2^2 \vec{p}_1\cdot\vec{p}_2
	         + 72 m_1^3 p_2^2 \nnl
        	 \left.+ 145 m_1^2 m_2 p_2^2 \right] \Bigg]
+ \frac{G^3}{r^5}
        \left[ \frac{\vec{n}\cdot\vec{S}_1 \vec{n}\cdot\vec{S}_2}{4}
        \left(105 m_1^2 + 289 m_1 m_2 + 105 m_2^2\right) \right.\nnl 
	      \left. -\frac{\vec{S}_1\cdot\vec{S}_2}{4} 
	      \left(63 m_1^2 + 145 m_1 m_2 + 63 m_2^2\right)\right] , \label{HADMs1s2}
\end{align}
where in contrast to previous presentations of this Hamiltonian in \cite{Hartung:2011ea,Hartung:2013dza},
we have again eliminated double scalar triple products using eq.~(\ref{epsepsformula}).
This is preferable for the comparison.

\section{Resolution via canonical transformations} \label {CT} 

If the EFT Hamiltonian obtained here in eq.~(\ref{HNNLOSS}) is physically equivalent to that of \cite{Hartung:2011ea} in eq.~(\ref{HADMs1s2}), that is if it yields the same EOM, 
there should exist an infinitesimal generator $g$ of a canonical transformation such that 
\bea\label{eq:gH} 
\Delta H&=&\{H,g\}=\{H_{\text{N}}+H_{\text{1PN}}+H^{\text{SO}}_{\text{LO}}+H^{\text{S$_1$S$_2$}}_{\text{LO}},\,\,\,g_{\text{NNLO}}^{\text{S$_1$S$_2$}}+g_{\text{NLO}}^{\text{S$_1$S$_2$}}+g_{\text{NLO}}^{\text{SO}}+g_{\text{2PN}}^{}\}
\nn\\
&=&
\Delta H_{\text{2PN}}+\Delta H_{\text{3PN}}+\Delta H_{\text{SO}}^{\text{NLO}}+\Delta H_{\text{SO}}^{\text{NNLO}}+\Delta H_{\text{S$_1$S$_2$}}^{\text{NLO}}+\Delta H_{\text{S$_1$S$_2$}}^{\text{NNLO}}, 
\eea
where here we have dropped contributions to sectors beyond linear in spin and beyond NNLO, and where
\be \label{hdiff}
\Delta H=H_{\text{EFT}}-H_{\text{ADM}}.
\ee
Thus, the contribution to the NNLO S$_1$S$_2$ sector comprises   
\be \label{ctHs1s2}
\Delta H_{\text{S$_1$S$_2$}}^{\text{NNLO}}=\{H_N,g_{\text{NNLO}}^{\text{S$_1$S$_2$}}\}+\{H_{\text{1PN}},g_{\text{NLO}}^{\text{S$_1$S$_2$}}\}+\{H^{\text{SO}}_{\text{LO}},g_{\text{NLO}}^{\text{SO}}\}+\{H^{\text{S$_1$S$_2$}}_{\text{LO}},g_{\text{2PN}}^{}\},
\ee
and we also have here contributions to orders below NNLO (recall that the 2PN non spinning sector is a NLO correction to Newtonian gravity), given by 
\begin{align} \label{nloctHs1s2}
\Delta H_{\text{S$_1$S$_2$}}^{\text{NLO}}&=\{H_{\text{N}},g_{\text{NLO}}^{\text{S$_1$S$_2$}}\},\nn\\
\Delta H_{\text{SO}}^{\text{NLO}}&=\{H_{\text{N}},g_{\text{NLO}}^{\text{SO}}\},\nn\\
\Delta H_{\text{2PN}}&=\{H_{\text{N}},g_{\text{2PN}}^{}\},
\end{align}
so actually we should require that the canonical transformation is also consistent with the equivalence at NLO of the S$_1$S$_2$, 
SO, and 2PN non spinning Hamiltonians.

Then, let us construct a suitable infinitesimal generator of PN canonical transformations for the NNLO S$_1$S$_2$ sector, $g_{\text{NNLO}}^{\text{S$_1$S$_2$}}$. 
Such a generator must be a scalar, linear in the spin vectors, and constructed out of the spins, $\vec{p}_1$, $\vec{p}_2$, and $\vec{n}$ vectors. 
There may be only two orders of $G$ in the generator, $O(G^1p^3)$ and $O(G^2p^1)$, which generate the $O(G)$, $O(G^2)$, and $O(G^3)$ terms 
of the form that appears in the NNLO S$_1$S$_2$ sector. Hence, we consider the following ansatz for the generator 
\begin{align} \label{gnnlos1s2}
g_{\text{NNLO}}^{\text{S$_1$S$_2$}}=\frac{G}{r^2}&\left[g_1\left(\vec{S}_1 \cdot \frac{\vec{p}_1}{m_1} \vec{S}_2 \cdot \vec{n}\frac{p_1^2}{m_1^2}-\vec{S}_2 \cdot \frac{\vec{p}_2}{m_2} \vec{S}_1 \cdot \vec{n}\frac{p_2^2}{m_2^2}\right)\right.
\nn\\
&
+g_2\left(\vec{S}_1 \cdot \frac{\vec{p}_1}{m_1} \vec{S}_2 \cdot \vec{n}\frac{\vec{p}_1}{m_1} \cdot \frac{\vec{p}_2}{m_2}-\vec{S}_2 \cdot \frac{\vec{p}_2}{m_2} \vec{S}_1 \cdot \vec{n}\frac{\vec{p}_1}{m_1} \cdot \frac{\vec{p}_2}{m_2}\right)
\nn\\
&
+g_3\left(\vec{S}_1 \cdot \frac{\vec{p}_1}{m_1} \vec{S}_2 \cdot \vec{n}\frac{p_2^2}{m_2^2}-\vec{S}_2 \cdot \frac{\vec{p}_2}{m_2} \vec{S}_1 \cdot \vec{n}\frac{p_1^2}{m_1^2}\right)
\nn\\
&
+g_4\left(\vec{S}_1 \cdot \frac{\vec{p}_1}{m_1} \vec{S}_2 \cdot \vec{n}\frac{\vec{p}_1}{m_1} \cdot \vec{n}\frac{\vec{p}_1}{m_1} \cdot \vec{n}-\vec{S}_2 \cdot \frac{\vec{p}_2}{m_2} \vec{S}_1 \cdot \vec{n}\frac{\vec{p}_2}{m_2} \cdot \vec{n}\frac{\vec{p}_2}{m_2} \cdot \vec{n}\right) 
\nn\\
&
+g_5\left(\vec{S}_1 \cdot \frac{\vec{p}_1}{m_1} \vec{S}_2 \cdot \vec{n}\frac{\vec{p}_1}{m_1} \cdot \vec{n}\frac{\vec{p}_2}{m_2} \cdot \vec{n}-\vec{S}_2 \cdot \frac{\vec{p}_2}{m_2} \vec{S}_1 \cdot \vec{n}\frac{\vec{p}_1}{m_1} \cdot \vec{n}\frac{\vec{p}_2}{m_2} \cdot \vec{n}\right) 
\nn\\ 
&
+g_6\left(\vec{S}_1 \cdot \frac{\vec{p}_1}{m_1} \vec{S}_2 \cdot \vec{n}\frac{\vec{p}_2}{m_2} \cdot \vec{n}\frac{\vec{p}_2}{m_2} \cdot \vec{n}-\vec{S}_2 \cdot \frac{\vec{p}_2}{m_2} \vec{S}_1 \cdot \vec{n}\frac{\vec{p}_1}{m_1} \cdot \vec{n}\frac{\vec{p}_1}{m_1} \cdot \vec{n}\right)
\nn\\
&
+g_7\left(\vec{S}_1 \cdot \frac{\vec{p}_2}{m_2} \vec{S}_2 \cdot \vec{n}\frac{p_1^2}{m_1^2}-\vec{S}_2 \cdot \frac{\vec{p}_1}{m_1} \vec{S}_1 \cdot \vec{n}\frac{p_2^2}{m_2^2}\right)
\nn\\
&
+g_8\left(\vec{S}_1 \cdot \frac{\vec{p}_2}{m_2} \vec{S}_2 \cdot \vec{n}\frac{\vec{p}_1}{m_1} \cdot \frac{\vec{p}_2}{m_2}-\vec{S}_2 \cdot \frac{\vec{p}_1}{m_1} \vec{S}_1 \cdot \vec{n}\frac{\vec{p}_1}{m_1} \cdot \frac{\vec{p}_2}{m_2}\right)
\nn\\
&
+ g_9\left(\vec{S}_1 \cdot \frac{\vec{p}_2}{m_2} \vec{S}_2 \cdot \vec{n}\frac{p_2^2}{m_2^2}-\vec{S}_2 \cdot \frac{\vec{p}_1}{m_1} \vec{S}_1 \cdot \vec{n}\frac{p_1^2}{m_1^2}\right)
\nn\\
&
+ g_{10}\left(\vec{S}_1 \cdot \frac{\vec{p}_2}{m_2} \vec{S}_2 \cdot \vec{n}\frac{\vec{p}_1}{m_1} \cdot \vec{n}\frac{\vec{p}_1}{m_1} \cdot \vec{n}-\vec{S}_2 \cdot \frac{\vec{p}_1}{m_1} \vec{S}_1 \cdot \vec{n}\frac{\vec{p}_2}{m_2} \cdot \vec{n}\frac{\vec{p}_2}{m_2} \cdot \vec{n}\right) 
\nn\\
&
+g_{11}\left(\vec{S}_1 \cdot \frac{\vec{p}_2}{m_2} \vec{S}_2 \cdot \vec{n}\frac{\vec{p}_1}{m_1} \cdot \vec{n}\frac{\vec{p}_2}{m_2} \cdot \vec{n}-\vec{S}_2 \cdot \frac{\vec{p}_1}{m_1} \vec{S}_1 \cdot \vec{n}\frac{\vec{p}_1}{m_1} \cdot \vec{n}\frac{\vec{p}_2}{m_2} \cdot \vec{n}\right) 
\nn\\ 
&
+g_{12}\left(\vec{S}_1 \cdot \frac{\vec{p}_2}{m_2} \vec{S}_2 \cdot \vec{n}\frac{\vec{p}_2}{m_2} \cdot \vec{n}\frac{\vec{p}_2}{m_2} \cdot \vec{n}-\vec{S}_2 \cdot \frac{\vec{p}_1}{m_1} \vec{S}_1 \cdot \vec{n}\frac{\vec{p}_1}{m_1} \cdot \vec{n}\frac{\vec{p}_1}{m_1} \cdot \vec{n}\right)
\nn\\
&+
g_{13}\left(\vec{S}_1 \cdot \vec{S}_2 \frac{\vec{p}_1}{m_1} \cdot \vec{n}\frac{p_1^2}{m_1^2}-\vec{S}_1 \cdot \vec{S}_2 \frac{\vec{p}_2}{m_2} \cdot \vec{n}\frac{p_2^2}{m_2^2}\right)
\nn\\
&+ 
g_{14}\left(\vec{S}_1 \cdot \vec{S}_2 \frac{\vec{p}_1}{m_1} \cdot \vec{n}\frac{\vec{p}_1}{m_1} \cdot \frac{\vec{p}_2}{m_2}-\vec{S}_1 \cdot \vec{S}_2 \frac{\vec{p}_2}{m_2} \cdot \vec{n}\frac{\vec{p}_1}{m_1} \cdot \frac{\vec{p}_2}{m_2}\right)
\nn\\
&+
g_{15}\left(\vec{S}_1 \cdot \vec{S}_2 \frac{\vec{p}_1}{m_1} \cdot \vec{n}\frac{p_2^2}{m_2^2}-\vec{S}_1 \cdot \vec{S}_2 \frac{\vec{p}_2}{m_2} \cdot \vec{n}\frac{p_1^2}{m_1^2}\right)
\nn\\
&+ 
g_{16}\left(\vec{S}_1 \cdot \vec{S}_2 \frac{\vec{p}_1}{m_1} \cdot \vec{n}\frac{\vec{p}_1}{m_1} \cdot \vec{n}\frac{\vec{p}_1}{m_1} \cdot \vec{n}-\vec{S}_1 \cdot \vec{S}_2 \frac{\vec{p}_2}{m_2} \cdot \vec{n}\frac{\vec{p}_2}{m_2} \cdot \vec{n}\frac{\vec{p}_2}{m_2} \cdot \vec{n}\right)
\nn\\ 
&+
g_{17}\left(\vec{S}_1 \cdot \vec{S}_2 \frac{\vec{p}_1}{m_1} \cdot \vec{n}\frac{\vec{p}_1}{m_1} \cdot \vec{n}\frac{\vec{p}_2}{m_2} \cdot \vec{n}-\vec{S}_1 \cdot \vec{S}_2 \frac{\vec{p}_2}{m_2} \cdot \vec{n}\frac{\vec{p}_1}{m_1} \cdot \vec{n}\frac{\vec{p}_2}{m_2} \cdot \vec{n}\right) 
\nn\\ 
&+
g_{18}\left(\vec{S}_1 \cdot \vec{n}\vec{S}_2 \cdot \vec{n} \frac{\vec{p}_1}{m_1} \cdot \vec{n}\frac{p_1^2}{m_1^2}-\vec{S}_1 \cdot \vec{n}\vec{S}_2 \cdot \vec{n} \frac{\vec{p}_2}{m_2} \cdot \vec{n}\frac{p_2^2}{m_2^2}\right)
\nn\\
&+g_{19}\left(\vec{S}_1 \cdot \vec{n} \vec{S}_2 \cdot \vec{n} \frac{\vec{p}_1}{m_1} \cdot \vec{n}\frac{\vec{p}_1}{m_1} \cdot \frac{\vec{p}_2}{m_2}-\vec{S}_1 \cdot \vec{n}\vec{S}_2 \cdot \vec{n} \frac{\vec{p}_2}{m_2} \cdot \vec{n}\frac{\vec{p}_1}{m_1} \cdot \frac{\vec{p}_2}{m_2}\right)
\nn\\
&+
g_{20}\left(\vec{S}_1 \cdot \vec{n}\vec{S}_2 \cdot \vec{n} \frac{\vec{p}_1}{m_1} \cdot \vec{n}\frac{p_2^2}{m_2^2}-\vec{S}_1 \cdot \vec{n}\vec{S}_2  \cdot \vec{n} \frac{\vec{p}_2}{m_2} \cdot \vec{n}\frac{p_1^2}{m_1^2}\right)
\nn\\
&+ 
g_{21}\left(\vec{S}_1 \cdot \vec{n}\vec{S}_2 \cdot \vec{n} \frac{\vec{p}_1}{m_1} \cdot \vec{n}\frac{\vec{p}_1}{m_1} \cdot \vec{n}\frac{\vec{p}_1}{m_1} \cdot \vec{n}-\vec{S}_1 \cdot \vec{n}\vec{S}_2 \cdot \vec{n}\frac{\vec{p}_2}{m_2} \cdot \vec{n}\frac{\vec{p}_2}{m_2} \cdot \vec{n}\frac{\vec{p}_2}{m_2} \cdot \vec{n}\right) 
\nn\\
&+
g_{22}\left(\vec{S}_1 \cdot \vec{n}\vec{S}_2 \cdot \vec{n}\frac{\vec{p}_1}{m_1} \cdot \vec{n}\frac{\vec{p}_1}{m_1} \cdot \vec{n}\frac{\vec{p}_2}{m_2} \cdot \vec{n}-\vec{S}_1 \cdot \vec{n}\vec{S}_2 \cdot \vec{n} \frac{\vec{p}_2}{m_2} \cdot \vec{n}\frac{\vec{p}_1}{m_1} \cdot \vec{n}\frac{\vec{p}_2}{m_2} \cdot \vec{n}\right) 
\nn\\ 
&+
g_{23}\left(\vec{S}_1 \cdot \frac{\vec{p}_1}{m_1} \vec{S}_2 \cdot \frac{\vec{p}_1}{m_1}\frac{\vec{p}_1}{m_1}\cdot\vec{n}-\vec{S}_2 \cdot \frac{\vec{p}_2}{m_2} \vec{S}_1 \cdot \frac{\vec{p}_2}{m_2}\frac{\vec{p}_2}{m_2} \cdot \vec{n}\right)
\nn\\
&+
g_{24}\left(\vec{S}_1 \cdot \frac{\vec{p}_1}{m_1} \vec{S}_2 \cdot \frac{\vec{p}_1}{m_1}\frac{\vec{p}_2}{m_2}\cdot\vec{n}-\vec{S}_2 \cdot \frac{\vec{p}_2}{m_2} \vec{S}_1 \cdot \frac{\vec{p}_2}{m_2}\frac{\vec{p}_1}{m_1} \cdot \vec{n}\right)
\nn\\
&+
g_{25}\left(\vec{S}_1 \cdot \frac{\vec{p}_1}{m_1} \vec{S}_2 \cdot \frac{\vec{p}_2}{m_2}\frac{\vec{p}_1}{m_1}\cdot\vec{n}-\vec{S}_2 \cdot \frac{\vec{p}_2}{m_2} \vec{S}_1 \cdot \frac{\vec{p}_1}{m_1}\frac{\vec{p}_2}{m_2} \cdot \vec{n}\right)
\nn\\
&\left. +
g_{26}\left(\vec{S}_1 \cdot \frac{\vec{p}_2}{m_2} \vec{S}_2 \cdot \frac{\vec{p}_1}{m_1}\frac{\vec{p}_2}{m_2}\cdot\vec{n}-\vec{S}_1 \cdot \frac{\vec{p}_2}{m_2} \vec{S}_2 \cdot \frac{\vec{p}_1}{m_1}\frac{\vec{p}_1}{m_1} \cdot \vec{n}\right)
\right]
\nn\\
+\frac{G^2}{r^3}&\left[g_{27}\left(\vec{S}_1 \cdot \frac{m_2}{m_1}\vec{p}_1 \vec{S}_2 \cdot \vec{n}-\vec{S}_2 \cdot \frac{m_1}{m_2}\vec{p}_2 \vec{S}_1 \cdot \vec{n}\right)
\right.
\nn\\
&+g_{28}\left(\vec{S}_1 \cdot \vec{p}_1 \vec{S}_2 \cdot \vec{n}-\vec{S}_2 \cdot \vec{p}_2 \vec{S}_1 \cdot \vec{n}\right)
\nn\\
&+g_{29}\left(\vec{S}_1 \cdot \frac{m_1}{m_2}\vec{p}_2 \vec{S}_2 \cdot \vec{n}-\vec{S}_2 \cdot \frac{m_2}{m_1}\vec{p}_1 \vec{S}_1 \cdot \vec{n}\right)
\nn\\
&+g_{30}\left(\vec{S}_1 \cdot \vec{p}_2 \vec{S}_2 \cdot \vec{n}-\vec{S}_2 \cdot \vec{p}_1 \vec{S}_1 \cdot \vec{n}\right)
\nn\\
& +g_{31}\left(\vec{S}_1 \cdot \vec{S}_2 \frac{m_2}{m_1}\vec{p}_1 \cdot \vec{n}-\vec{S}_1 \cdot \vec{S}_2 \frac{m_1}{m_2}\vec{p}_2\cdot \vec{n}\right)
\nn\\
& +g_{32}\left(\vec{S}_1 \cdot \vec{S}_2 \vec{p}_1 \cdot \vec{n}-\vec{S}_1 \cdot \vec{S}_2 \vec{p}_2\cdot \vec{n}\right)
\nn\\
&+g_{33}\left(\vec{S}_1 \cdot \vec{n} \vec{S}_2\cdot \vec{n} \frac{m_2}{m_1}\vec{p}_1\cdot \vec{n}-\vec{S}_1 \cdot \vec{n} \vec{S}_2 \cdot \vec{n} \frac{m_1}{m_2}\vec{p}_2\cdot \vec{n}\right)
\nn\\
&+\left. g_{34}\left(\vec{S}_1 \cdot \vec{n} \vec{S}_2\cdot \vec{n} \vec{p}_1\cdot \vec{n}-\vec{S}_1 \cdot \vec{n} \vec{S}_2 \cdot \vec{n} \vec{p}_2\cdot \vec{n}\right)
\right].
\end{align}

We should also have the generators contributing first to lower orders, as noted in eq.~(\ref{ctHs1s2}), 
so that their coefficients are already set from eq.~(\ref{nloctHs1s2}).
For the NLO S$_1$S$_2$ sector generator $g_{\text{NLO}}^{\text{S$_1$S$_2$}}$, we have the following general form \cite{Steinhoff:2007mb}
\begin{align}\label{gnlos1s2}
\begin{split}
g_{\text{NLO}}^{\text{S$_1$S$_2$}}=&\frac{G}{r^2}\left[g_{1}\left(\vec{S}_1 \cdot \frac{1}{m_1}\vec{p}_1 \vec{S}_2 \cdot \vec{n}-\vec{S}_2 \cdot \frac{1}{m_2}\vec{p}_2 \vec{S}_1 \cdot \vec{n}\right)
\right.\\
&+g_{2}\left(\vec{S}_1 \cdot \frac{1}{m_2}\vec{p}_2 \vec{S}_2 \cdot \vec{n}-\vec{S}_2 \cdot \frac{1}{m_1}\vec{p}_1 \vec{S}_1 \cdot \vec{n}\right)
+g_{3}\left(\vec{S}_1 \cdot \vec{S}_2 \frac{1}{m_1}\vec{p}_1 \cdot \vec{n}-\vec{S}_1 \cdot \vec{S}_2 \frac{1}{m_2}\vec{p}_2\cdot \vec{n}\right)
\\
&\left.+g_{4}\left(\vec{S}_1 \cdot \vec{n} \vec{S}_2\cdot \vec{n} \frac{1}{m_1}\vec{p}_1\cdot \vec{n}-\vec{S}_1 \cdot \vec{n} \vec{S}_2 \cdot \vec{n} \frac{1}{m_2}\vec{p}_2\cdot \vec{n}\right)\right],
\end{split}
\end{align}
with its coefficients set to the values
\begin{align}\label{gnlos1s2fix}
g_1&=0,\,\,g_2=0,\,\,g_3=0,\,\,g_4=0.
\end{align} 
For the NLO SO sector we have the following general form for the generator $g_{\text{NLO}}^{\text{SO}}$ \cite{Levi:2010zu}
\be \label{gnloso}
g_{\text{NLO}}^{\text{SO}}= \frac{Gm_2}{r}\vec{S}_1 \cdot \left[g_1\frac{\vec{p}_1\times\vec{p}_2}{m_1m_2}+\frac{\vec{p}_1\times\vec{n}}{m_1}\left(g_2\frac{\vec{p}_1\cdot\vec{n}}{m_1}+g_3\frac{\vec{p}_2\cdot\vec{n}}{m_2}\right)
+\frac{\vec{p}_2\times\vec{n}}{m_2}\left(g_4\frac{\vec{p}_1\cdot\vec{n}}{m_1}+g_5\frac{\vec{p}_2\cdot\vec{n}}{m_2}\right)\right],
\ee
where this generator should be taken with $1\leftrightarrow2$, and with its coefficients set to the values
\begin{align}\label{gnlosofix}
g_1&=-\frac{1}{2},\,\,g_2=0,\,\,g_3=\frac{1}{2},\,\,g_4=0,\,\,g_5=0.
\end{align} 
Finally, we should also take into account the generator, which contributes first at the 2PN non spinning sector, with the following general form
\begin{align} \label{g2pn}
\begin{split}
g_{\text{2PN}}^{}=Gm_1m_2&\left[g_1\frac{p_1^2}{m_1^2}\frac{\vec{p}_1}{m_1}\cdot\vec{n}+g_2\frac{p_2^2}{m_2^2}\frac{\vec{p}_1}{m_1}\cdot\vec{n}+g_3\frac{\vec{p}_1\cdot\vec{p}_2}{m_1m_2}\frac{\vec{p}_1}{m_1}\cdot\vec{n}+g_4\frac{\left(\vec{p}_1\cdot\vec{n}\right)^3}{m_1^3}\right.\\
&\left.+g_5\frac{\left(\vec{p}_2\cdot\vec{n}\right)^2}{m_2^2}\frac{\vec{p}_1}{m_1}\cdot\vec{n}\right]+\frac{G^2m_1m_2}{r}\left[\frac{\vec{p}_1}{m_1}\cdot\vec{n}\left(g_6m_1+g_7m_2\right)\right],
\end{split}
\end{align}
where this generator should also be taken with $1\leftrightarrow2$, and with its coefficients set to 
\begin{align} \label{g2pnfix}
g_1&=0,\,\,g_2=-\frac{1}{2},\,\,g_3=0,\,\,g_4=0,\,\,g_5=0,\,\,g_6=0,\,\,g_7=-\frac{1}{4}.
\end{align} 

Thus, we plug in eq.~(\ref{ctHs1s2}) our ansatz for $g_{\text{NNLO}}^{\text{S$_1$S$_2$}}$ from eq.~(\ref{gnnlos1s2}), together with the fixed generators in 
eqs.~(\ref{gnlos1s2}), (\ref{gnlos1s2fix}), (\ref{gnloso}), (\ref{gnlosofix}), (\ref{g2pn}), (\ref{g2pnfix}), and we compare that to eq.~(\ref{hdiff}). 
Comparing $O(G)$ terms fixes the $O(G)$ coefficients of $g_{\text{NNLO}}^{\text{S$_1$S$_2$}}$ to the values
\begin{align}
g_1&=0, &g_2&=-\frac{1}{4}, &g_3&=-\frac{3}{4}, &g_4&=0, 
&g_5&=\frac{3}{4}, &g_6&=0, &g_7&=\frac{1}{4}, &g_8&=1, \nn\\
g_9&=-\frac{1}{4}, &g_{10}&=0, &g_{11}&=0, &g_{12}&=0,   
&g_{13}&=-\frac{1}{4}, &g_{14}&=1, &g_{15}&=\frac{3}{4}, &g_{16}&=0, \nn\\ 
g_{17}&=0, &g_{18}&=0, &g_{19}&= 0, &g_{20}&=\frac{3}{4},  
&g_{21}&=0, &g_{22}&=0, &g_{23}&=0, &g_{24}&=\frac{3}{4}, \nn\\ 
g_{25}&=-\frac{3}{4}, &g_{26}&=0.
\end{align}
This eliminates all of the $O(G)$ terms in the difference.
Comparing the remaining $O(G^2)$ terms in the difference fixes the $O(G^2)$ coefficients of $g_{\text{NNLO}}^{\text{S$_1$S$_2$}}$ to the values
\begin{align}
g_{27}=0,\,\,g_{28}=-\frac{1}{4},\,\,g_{29}=1,\,\,g_{30}=\frac{45}{8},\,\,g_{31}=1,\,\,g_{32}=\frac{25}{8},\,\,g_{33}=0,\,\,g_{34}=\frac{19}{4}.
\end{align}
This eliminates all of the $O(G^2)$ terms, as well as all terms at $O(G^3)$ in the difference.
Hence, we have shown that the ADM Hamiltonian and the EFT Potential at NNLO S$_1$S$_2$ are completely equivalent.

\section{Gauge invariant relations to 4PN order with spins} \label{GI}

Neither the EFT potentials nor the Hamiltonians presented before are gauge invariant.
However, the Hamiltonians are related to the total energy of the binary system, 
which is a global gauge invariant quantity. Similarly, the total angular
momentum of the binary is gauge invariant. 
For the case of non spinning binaries in quasi-circular orbit, 
the energy depends only on the magnitude of the angular momentum 
(and on the mass ratio). This relation between energy and angular momentum
is gauge invariant, and is therefore a useful tool for evaluating different analytic and numerical
descriptions of the binary dynamics, such as the analytic post-Newtonian \cite{Damour:2014jta},
effective-one-body \cite{Damour:2011fu,Taracchini:2013rva},
self-force \cite{LeTiec:2011dp}, and test-particle \cite{Steinhoff:2012rw} approximations,
and also numerical simulations of black hole \cite{Damour:2011fu,Taracchini:2013rva}
or neutron star binaries \cite{Bernuzzi:2012ci,Bernuzzi:2013rza}. 
Another gauge invariant relation that can be derived is the energy as a function of 
the orbital frequency. This relation, together with the energy flux of the gravitational waves, 
can be used to derive the phasing of the gravitational waves.

For spinning binaries, the energy will not only depend on the total angular momentum,
but also on the spins, and in particular on their orientation. 
Here we will consider the so-called ``aligned spins'' case, 
where both spins are aligned with the orbital angular momentum, 
in addition to the orbits being circular. 
This specific configuration is actually the orbital configuration, to which the binary is driven 
throughout the inspiral phase, and to which the GW detectors have the highest signal-to-noise ratio  \cite{Taracchini:2013rva,Bernuzzi:2013rza}.  
In this section we are going to derive the binding energy as a function of the angular momentum, 
and of the orbital angular frequency for this orbital configuration.

In the following we rescale all variables such that they become dimensionless.
First, we introduce the total mass $m \equiv m_1 + m_2$, 
and the reduced mass $\mu \equiv m_1 m_2 / m$ of the binary.
From these, we can define the mass ratio $q$, 
and the symmetric mass ratio $\nu$, as
\begin{equation}
q \equiv \frac{m_1}{m_2} , \qquad
\nu \equiv \frac{m_1 m_2}{m^2} = \frac{\mu}{m} = \frac{q}{(1+q)^2},
\end{equation}
in terms of which all of the results can be expressed.
The center-of-mass frame is defined by a vanishing total linear momentum.
In the present case, radiation does not contribute to the linear momentum, so we require the
sum of the particle momenta to vanish, $\vec{p}_1 + \vec{p}_2 = 0$, or
\begin{equation}
\vec{p} \equiv \vec{p}_1 = - \vec{p}_2 .
\end{equation}
The orbital angular momentum is defined by $\vec{L} \equiv r \vec{n} \times \vec{p}$.
Then we rescale the variables as follows (recall that $c\equiv1$):
\begin{equation} 
\tilde{H} \equiv \frac{H}{\mu} , \qquad
\tilde{r} \equiv \frac{r}{G m} , \qquad
\tilde{L} \equiv \frac{L}{G m \mu}, \qquad 
\tilde{S}_I \equiv \frac{S_I}{G m \mu},
\end{equation}
where the rescaled variables are denoted by a tilde, and in general every unit of mass is 
divided by $\mu$ and every unit of length is divided by $Gm$.
Note that we rescaled our spins just as we rescaled the orbital angular momentum.
Since we are considering here the total angular momentum of the binary, $J\equiv L+S_1+S_2$, 
it makes sense to treat the spins and orbital angular momentum on the same footing. 
However, it should be noted that, e.g.~in numerical simulations \cite{Taracchini:2013rva,Bernuzzi:2013rza}, 
it is customary to rescale the spin variables differently such that 
\be
\bar{S}_I \equiv \frac{S_I}{G m_I^2},
\ee
where this common dimensionless spin variable $\bar{S}_I$ 
is simply related to our rescaled spin variable $\tilde{S}_I$ by
\be 
\tilde{S}_1= q \bar{S}_1 , \qquad
\tilde{S}_2= \bar{S}_2/q.
\ee
\def\LADM{\tilde{L}}
\def\rADM{\tilde{r}}
\def\Sf{\tilde{S}_1}
\def\Ss{\tilde{S}_2}
\def\HADM{\tilde{H}}

Circular orbits are defined by $r=\text{const}$. This holds if
the radial momentum vanishes, such that 
\be \label{prnull}
p_r \equiv \vec{p} \cdot \vec{n}=0 \Rightarrow p^2=L^2/r^2.
\ee
Eq.~(\ref{prnull}) is preserved in time, if, using Hamilton's equations, we require that
\begin{equation}\label{circularcondition}
\dot{\tilde{p}}_r = - \frac{\partial \HADM(\rADM, \LADM)}{\partial \rADM}=0.
\end{equation}
This condition can be solved to yield the constant radius $\rADM$ in terms of $\LADM$.
Notice that the spins are omitted as variables of the Hamiltonian, because only
the spin length, which is constant, neglecting dissipative effects from the absorption 
of gravitational waves by the compact objects as considered in e.g.~\cite{Poisson:2004cw}, 
enters for the considered orbital configuration.

We have another equation of motion for the orbital phase $\phi$. 
The orbital angular frequency follows then from Hamilton's equation 
\begin{equation}\label{orbitalfrequency}
\frac{d\phi}{d\tilde{t}}\equiv\tilde{\omega} = \frac{\partial \HADM(\rADM, \LADM)}{\partial \LADM} ,
\end{equation}
since $\LADM$ is its canonical conjugate. 
From the orbital frequency it is customary to define the gauge-invariant PN parameter
\begin{equation}\label{xgi}
x \equiv \tilde{\omega}^{2/3} ,
\end{equation}
inferred from Kepler's third law for circular orbits in the Newtonian limit, $\omega^2=Gm/r^3$. 
Thus, $1/x$ can be thought of 
as a measure for the orbital separation of the binary, but since it is defined 
from the orbital frequency of the binary, which is observable asymptotically 
through the emitted gravitational waves, it is gauge-invariant, as opposed to 
the separation $\rADM$, which depends on the coordinate system. 
Moreover, in the expansion of, e.g.~the binding energy, in terms of $x$, 
the exponent of $x$ corresponds to the $(n-1)$ PN order.

Finally, the spins are aligned with the orbital angular momentum, if in addition
\begin{equation}
\vec{S} \cdot \vec{n} = \vec{S} \cdot \vec{p}=0, 
\end{equation}
holds for both spins.

\paragraph{Simplified Hamiltonians.}

Here we provide all relevant Hamiltonians in the center-of-mass frame 
for the restricted case of circular orbits and aligned spins.
We start here from the ADM Hamiltonians, which also include at present the NNLO SO sector.

Under the aforementioned assumptions, the Newtonian, 1PN, and 2PN ADM Hamiltonians are given by
\begin{align}
\HADM_{\text{N}} &= \frac{1}{\rADM} \left[ - 1 + \frac{\LADM^2}{2 \rADM} \right] , \\
\HADM_{\text{1PN}} &= \frac{1}{2 \rADM^2} \left[
        1
        - \frac{\LADM^2}{\rADM} \left(\nu + 3\right)
        + \frac{\LADM^4}{4 \rADM^2} \left(3 \nu -1\right) \right] , \\
\HADM_{\text{2PN}} &= \frac{1}{2 \rADM^3} \left[
        - \frac{1}{2}\left(3\nu+1\right)
        + \frac{\LADM^2}{\rADM} \left(8 \nu +5\right)
        - \frac{\LADM^4}{4\rADM^2} \left( 3 \nu^2 + 20\nu - 5 \right)
        + \frac{\LADM^6}{8 \rADM^3} \left(5 \nu^2 - 5 \nu + 1\right)
        \right] .
\end{align}
The LO SO, S$_1$S$_2$, and spin-squared Hamiltonians are given by
\begin{align}
\HADM_{\text{LO}}^{\text{SO}} &= \frac{\nu \LADM}{\rADM^3}
        \left[ 2 \left(\Sf+\Ss \right) + \frac{3}{2} \left( \Sf / q + \Ss q \right) \right] , \\
\HADM_{\text{LO}}^{\text{S$_1$S$_2$}} &=
        -\frac{\nu  \Sf \Ss}{\rADM^3} , \\        
\HADM_{\text{LO}}^{\text{S$^2$}} &= -\frac{\nu}{2 \rADM^3}
        \left[ C_{Q1} \Sf^2 / q + C_{Q2} \Ss^2 q \right].
\end{align}
Here $C_{Q1}$ and $C_{Q2}$ are the constants describing the quadrupole
deformation due to spin, similar to the notation in \cite{Hergt:2010pa}. 
They are identical to the parameter $a$ in \cite{Poisson:1997ha}, 
or to $C_{ES^2}$ introduced in \cite{Porto:2008jj}.
For black holes $C_Q = 1$.

The NLO spin-dependent Hamiltonians are
\begin{align}
\begin{split}
\HADM_{\text{NLO}}^{\text{SO}} &= \frac{\nu \LADM}{\rADM^4} \left[
        \left(- 2 \nu - 6 + \frac{19 \nu \LADM^2}{8 \rADM} \right) \left(\Sf+\Ss\right) \right.\nlq
        \left.+\left(- 2 \nu - 5 + \frac{\LADM^2}{8 \rADM } \left(16 \nu -5\right) \right) \left( \Sf / q + \Ss q \right) \right] ,
\end{split}\\
\HADM_{\text{NLO}}^{\text{S$_1$S$_2$}} &= \frac{\nu  \Sf \Ss}{\rADM^4}
        \left[ 6 - \frac{\LADM^2}{2\rADM} \left( 2\nu + 3 \right) \right], \\ 
\begin{split}
\HADM_{\text{NLO}}^{\text{S$^2$}} &= \frac{\nu \Sf^2}{2 \rADM^4} \left[
        \frac{1}{q} \left(
                2\nu + 2 + C_{Q1} \left(-3\nu + 4\right)
                + \frac{\LADM^2}{2\rADM} \left(
                        - 4 \nu + 5
                        + C_{Q1} \left(\nu - 5\right) \right)
        \right) \right.\nlq
        \left. + \nu \left(
                2 - 3 C_{Q1}
                + \frac{\LADM^2}{2\rADM} \left(- 5 + 2 C_{Q1}\right)
        \right) \right] + \left[1 \longleftrightarrow 2 \right]. 
\end{split} 
\end{align}
Note that for the exchange of particles $\left[1 \longleftrightarrow 2\right]$, the mass
ratio $q$ is substituted by $1/q$, but the symmetric mass ratio $\nu$ is invariant.
The NLO spin-squared Hamiltonian can be found in \cite{Steinhoff:2008ji,Hergt:2008jn,Hergt:2010pa}. 

Finally, the NNLO SO \cite{Hartung:2011te,Hartung:2013dza} and S$_1$S$_2$ Hamiltonians are given by
\begin{align}
\begin{split}
\HADM_{\text{NNLO}}^{\text{SO}} &= \frac{\nu  \LADM}{2 \rADM^5} \left[
        \left(
                21 \left(\nu + 1\right)
                - \frac{\nu \LADM^2}{8 \rADM} \left(39 \nu + 314\right)
                + \frac{\nu \LADM^4}{4 \rADM^2} \left(22 \nu - 9\right)
        \right) \left(\Sf+\Ss \right) \right.\nl
        \left.+ \left(
                \frac{82 \nu + 75}{4}
                - \frac{3 \LADM^2}{8 \rADM} \left(13\nu^2 +86 \nu - 18\right)
                + \frac{\LADM^4}{8 \rADM^2} \left(39\nu^2 - 37\nu +7\right)
        \right) \left(\Sf / q + \Ss q\right) \right] ,
\end{split}\\
\HADM_{\text{NNLO}}^{\text{S$_1$S$_2$}} &= \frac{\nu \Sf \Ss}{2 \rADM^5} \left[
        - \frac{19\nu + 63}{2}
        + \frac{\LADM^2}{\rADM} \left(47 \nu +18\right)
        - \frac{\LADM^4}{4 \rADM^2} \left(7\nu^2 + 23\nu - 9\right) \right].
\end{align}
In the following the Hamiltonian $\HADM$ stands for the sum of all the Hamiltonians provided here.
Notice that we are not including the relativistic rest-mass contribution to the energy,
so that $\HADM$ is identical to the dimensionless binding energy $e \equiv \HADM$.
For completeness to 4PN order, we should include the NNLO spin-squared Hamiltonian, but this one was not computed yet.
We are also not including here the cubic and quartic in spin Hamiltonians from \cite{Hergt:2007ha,Hergt:2008jn}, since they are only valid for black holes. 
Moreover, the quartic in spin Hamiltonian is incomplete, see \cite{Steinhoff:2012rw}.

\subsection{Binding energy and angular momentum}

The simplified expressions we provided for the Hamiltonian or binding energy are
still gauge dependent, since they contain the radial coordinate $\rADM$, which
depends on the ADM gauge. In order to eliminate the radial coordinate, we should 
use the condition in eq.~(\ref{circularcondition}) for circular orbits to obtain $\rADM(\LADM)$.
We may then substitute this back into the Hamiltonian to obtain the binding energy
as a function of the orbital angular momentum $e(\LADM) = \HADM(\rADM(\LADM), \LADM)$.
From this result, the gauge invariant relation $e(\tilde{J})$ between the binding
energy and the total angular momentum $\tilde{J}$ follows simply by inserting
$\LADM = \tilde{J} - \Sf - \Ss$. Since the lengths of the spins are conserved
in our analytical approximation, which is neglecting dissipative effects, 
$e(\LADM)$ can also be considered gauge invariant.

Thus, from eq.~(\ref{circularcondition}) we can write the solution for $\rADM(\LADM)$ as 
\begin{align}\label{rcirc}
\frac{1}{\rADM} &= \frac{1}{\LADM^2} + \frac{4}{\LADM^4}
+ \frac{1}{\LADM^6} \left[ - \frac{43 \nu }{8} + \frac{101}{4} \right] \nnl
+ \left(\Sf+\Ss\right) \frac{\nu}{\LADM^5} \left[ - 6
        + \frac{1}{\LADM^2} \left( \frac{41 \nu}{8}-81 \right)
        + \frac{1}{\LADM^4} \left( -\frac{17 \nu^2}{16}+\frac{4201 \nu}{16}-\frac{3855}{4} \right) \right] \nnl
+ \left(\Sf / q + \Ss q\right) \frac{\nu}{\LADM^5} \left[ - \frac{9}{2}
        + \frac{1}{\LADM^2} \left(\frac{19 \nu}{4}-\frac{445}{8}\right)
        + \frac{1}{\LADM^4} \left(-\frac{9 \nu^2}{8}+\frac{3379 \nu}{16}-\frac{10095}{16}\right) \right] \nnl
+ \frac{\nu \Sf \Ss}{\LADM^6} \left[ 3   
				+\frac{1}{\LADM^2} \left( \frac{19 \nu}{2} +144 \right)
        +\frac{1}{\LADM^4} \left( -7 \nu^2-\frac{369 \nu}{8} +\frac{11301}{4} \right)\right]\nnl
+ \frac{\nu \Sf^2}{\LADM^6} \left[
        \frac{1}{q} \left( \frac{3 C_{Q1}}{2}
                + \frac{1}{4\LADM^2} \left(112 \nu+121\right)
                + \frac{C_{Q1}}{2\LADM^2} \left(5 \nu+49\right) \right)
        + \frac{\nu}{\LADM^2} \left( \frac{135}{4} + \frac{7 C_{Q1}}{2} \right) \right] \nnl
+ \frac{\nu \Ss^2}{\LADM^6} \left[
        q \left( \frac{3 C_{Q2}}{2}
                + \frac{1}{4\LADM^2} \left(112 \nu+121\right)
                + \frac{C_{Q2}}{2\LADM^2} \left(5 \nu+49\right) \right)
        + \frac{\nu}{\LADM^2} \left( \frac{135}{4} + \frac{7 C_{Q2}}{2} \right) \right].
\end{align} 
After inserting this back into the Hamiltonian, we obtain the spin-dependent part of the
gauge-invariant relation $e(\LADM)$ as
\begin{align} \label{EJ}
e_{\text{spin}}(\LADM) &=
\left(\Sf+\Ss\right) \frac{\nu}{\LADM^5} \left[ 2
        + \frac{1}{\LADM^2} \left( \frac{3 \nu}{8}+18 \right)
        + \frac{1}{\LADM^4} \left( \frac{5 \nu^2}{16}-27 \nu+162 \right) \right] \nnl
+ \left(\Sf / q + \Ss q\right) \frac{\nu}{\LADM^5} \left[ \frac{3}{2}
        + \frac{99}{8 \LADM^2}
        - \frac{1}{\LADM^4} \left(\frac{195 \nu}{8}-\frac{1701}{16}\right) \right] \nnl
-\frac{\nu  \Sf \Ss}{\LADM^6} \left[ 1
        + \frac{1}{\LADM^2} \left(\frac{13 \nu}{4}+\frac{69}{2}\right)
        + \frac{1}{\LADM^4} \left(\frac{25 \nu^2}{8}+\frac{507 \nu}{16}+\frac{4041}{8}\right) \right]\nnl        
- \frac{\nu \Sf^2}{\LADM^6} \left[
        \frac{1}{q} \left( \frac{C_{Q1}}{2}
        +\frac{1}{8\LADM^2} \left(54\nu+63\right)
        +\frac{C_{Q1}}{4\LADM^2} \left(5\nu+21\right) \right)
        + \frac{\nu}{\LADM^2} \left( \frac{65}{8} + C_{Q1} \right) \right] \nnl
- \frac{\nu \Ss^2}{\LADM^6} \left[
        q \left( \frac{C_{Q2}}{2}
        +\frac{1}{8\LADM^2} \left(54\nu+63\right)
        +\frac{C_{Q2}}{4\LADM^2} \left(5\nu+21\right) \right)
        + \frac{\nu}{\LADM^2} \left( \frac{65}{8} + C_{Q2} \right) \right].
\end{align}
The point-mass (non spinning) contributions up to 4PN can be found in eq.~(5.3) of \cite{Damour:2014jta},
and are not repeated here.

\subsection{Binding energy and orbital frequency}

In this section we are going to derive the binding energy as a function of the orbital frequency, or equivalently of the PN parameter $x$ from eq.~(\ref{xgi}), in the form $e(x)$. If the energy flux of the emitted
gravitational waves is also known as a function of the frequency, then $e(x)$ can be used
to translate the energy loss to the change in $x$ or the frequency $\tilde{\omega}$.
This means one can obtain the change of the frequency of the gravitational wave
over time, that is the phasing of the wave. This is the most important application of $e(x)$.
We will also obtain here the angular momentum as a function of the orbital frequency.

Using the definition in eq.~(\ref{xgi}) for the parameter $x$, we start by computing the orbital angular frequency 
from eq.~(\ref{orbitalfrequency}). Then, upon the substitution of eq.~(\ref{rcirc}), we obtain the relation $x(\LADM)$,
and we invert it to obtain $\LADM(x)$ with the result
\begin{align}
\frac{1}{\LADM^2} &= x - x^2 \left[\frac{\nu }{3}+3\right] + \frac{25 \nu  x^3}{4} \nnl
+ \nu x^{5/2} \left(\Sf+\Ss\right) \left[ \frac{20}{3}
        - x \left(\frac{337 \nu}{36}+16\right)
        + \nu x^2 \left(\frac{121 \nu}{24}+\frac{17}{2}\right) \right] \nnl
+ \nu x^{5/2} \left(\Sf / q + \Ss q\right) \left[ 5
        - x \left(\frac{25 \nu}{3}+\frac{69}{4}\right)
        + x^2 \left(5 \nu^2+\frac{63 \nu}{4}+\frac{27}{8}\right) \right] \nnl
+ \nu x^3 \Sf \Ss \left[ - 4
        + x \left(\frac{95 \nu}{18}+\frac{196}{3}\right)
        + x^2 \left(\frac{155 \nu^2}{108}-\frac{8177 \nu}{72}-\frac{313}{2}\right)
        \right] \nnl        
+ \nu x^3 \Sf^2 \left[ \frac{1}{q} \left( - 2 C_{Q1}
        + x \left(\frac{59 \nu}{6}+\frac{107}{4}\right)
        - x C_{Q1} \left(3 \nu - 5\right) \right)
        + \nu x \left(\frac{365}{36}-\frac{16 C_{Q1}}{3}\right) \right] \nnl
+ \nu x^3 \Ss^2 \left[ q \left( - 2 C_{Q2}
        + x \left(\frac{59 \nu}{6}+\frac{107}{4}\right)
        - x C_{Q2} \left(3 \nu - 5\right) \right)
        + \nu x \left(\frac{365}{36}-\frac{16 C_{Q2}}{3}\right) \right].
\end{align}
We insert this into the result for $e(\LADM)$ in eq.~(\ref{EJ}), where the point-mass contributions up to 2PN should also be taken into account. 
This finally leads to the spin-dependent part of the gauge-invariant relation $e(x)$ given by 
\begin{align}
e_{\text{spin}}(x) &=
\nu x^{5/2} \left(\Sf+\Ss\right) \left[ - \frac{4}{3}
        + x \left( \frac{31 \nu}{18} - 4 \right)
        - x^2 \left( \frac{7 \nu^2}{12}-\frac{211 \nu}{8}+\frac{27}{2} \right) \right] \nnl
+ \nu x^{5/2} \left(\Sf / q + \Ss q\right) \left[ - 1
        + x \left( \frac{5 \nu}{3}-\frac{3}{2} \right)
        - x^2 \left( \frac{5 \nu^2}{8}-\frac{39 \nu}{2}+\frac{27}{8} \right) \right] \nnl
+ \nu x^3 \Sf \Ss \left[ 1
        + x \left(\frac{5 \nu}{18}+\frac{5}{6}\right)
        - x^2 \left(\frac{371 \nu^2}{216}+\frac{1001 \nu}{72}-\frac{35}{8}\right) \right] \nnl        
+ \nu x^3 \Sf^2 \left[ \frac{1}{q} \left( \frac{C_{Q1}}{2}
        + \frac{5 x}{6} \left(\nu-3\right)
        + \frac{5 x C_{Q1}}{4} \left(\nu+1\right) \right)
        + \nu x \left(\frac{25}{18}+\frac{5 C_{Q1}}{3}\right) \right] \nnl
+ \nu x^3 \Ss^2 \left[ q \left( \frac{C_{Q2}}{2}
        + \frac{5 x}{6} \left(\nu-3\right)
        + \frac{5 x C_{Q2}}{4} \left(\nu+1\right) \right)
        + \nu x \left(\frac{25}{18}+\frac{5 C_{Q2}}{3}\right) \right].
\end{align}
Again, the point-mass contributions up to 4PN can be found in eq.~(5.5) of \cite{Damour:2014jta},
and are not repeated here.
We note that in \cite{Tessmer:2012xr} there is an interpolation between the post-Newtonian 
results and the test-spin case for a modified binding energy $e(x)$ in eq.~(9), where 
the binding energy is denoted there by $E(x)$, and their $e(x)$ is related to $E(x)$ by eq.~(5) there.
Using eqs.~(5) and (9) in \cite{Tessmer:2012xr} then, we compared our expression for $e(x)$ with 
their result, and found agreement up to the $S_2^2$ terms, which were dropped in \cite{Tessmer:2012xr}
for the test spin case. 

\section{Conclusions} \label{theendmyfriend}

In this paper we have shown the complete equivalence of the NNLO S$_1$S$_2$ Hamiltonian computed in \cite{Hartung:2011ea,Hartung:2013dza} 
with the potential computed via the EFT approach in \cite{Levi:2011eq} at 4PN. 
Such high PN orders are required for the successful detection of gravitational radiation. 

Due to the level of complexity that the NNLO spin-dependent sector displays, the comparison 
was carried out at the level of the potentials as we explain throughout its different stages. First, it is clear that a mapping between the canonical and the EFT variables is essential. To our advantage, such a mapping to canonical variables was already considered in \cite{Hergt:2011ik} at the level of the effective EFT action. This mapping is obtained through the insertion of gauge constraints for the spin, and also for its conjugate degrees of freedom, at the level of the EFT action, as the spin gauge is not fixed prior to the EFT computation. Instead, the unphysical redundant degrees of freedom, associated with the temporal components of the spin tensor $S^{i0}$, are treated as independent degrees of freedom, and are carried along to the EFT potential, to be eliminated after the obtainment of the EOM. Yet, the NNLO potential also comprises  higher order time derivatives of velocities and spins, that should be eliminated. This actually set the manner in which the comparison was executed, since the elimination of higher order time derivatives entails a redefinition of the variables, that would have modified the mapping to canonical variables, and possibly even the spin constraint for the $S^{0i}$ components.
Therefore we started by eliminating the temporal spin components $S^{i0}$, after which we could conveniently apply the transformations to the canonical variables and potential, and only then we eliminated the higher order time derivatives. After all of these imperative steps were carried out the most direct and complete way to compare the two results was just to obtain the corresponding EFT Hamiltonian via a straightforward Legendre transform.

The main ingredients in this work included first the obtainment of the temporal spin components $S^{0i}$ in the covariant SSC, where we have computed them at NNLO, via an additional EFT computation of the metric. Such additional EFT computation would also be required for the EOM in the EFT for spin, which indicates the incompleteness of the Routhian approach \cite{Porto:2008tb} in the EFT sense. We have also clarified that the EOM for spin should be obtained via an independent variation of the action with respect to the angular velocity and the spin, and not via the Poisson bracket, as advertised in the Routhian approach \cite{Porto:2008tb}, due to the higher order time derivatives of spin, 
which formally appear in the potentials as of NLO. After the elimination of the temporal spin components $S^{i0}$ was made, we could apply the transformation to the reduced canonical spin and position variables, together with a transformation of the potential, at the level of the EFT action, based on the work in \cite{Hergt:2011ik}. We improved on the redefinition of the position variable, such that we actually arrived to curved spacetime generalizations of the Newton-Wigner variables in closed form, together with the transition to the canonical potential, all available at the level of the action. At the next stage, we have made use of variable redefinitions to eliminate higher order time derivatives of velocities and spins, via their substitution with lower order EOM, where we extended explicitly the treatment for the case of a spin variable. After Legendre transforming, we resolved the difference between the resulting EFT Hamiltonian and the ADM one, via canonical transformations, where we had to take into account contributions coming from several sectors, and also maintain consistency with NLO sectors, which were both spinning and non spinning ones. Finally, we made use of our validated result, and derived gauge invariant relations among the binding energy, angular momentum, and orbital frequency of an inspiralling binary with generic compact spinning components to 4PN order, including all known results up to date.

In conclusion, our proof of equivalence here has actually also served to elucidate and highlight the obstacles and subtleties that are found in the current EFT formulation for spin, and to shed more light on possibilities to overcome them. Recalling \cite{Levi:2008nh}, it may be a good idea to fix the spin gauge prior to the EFT computation, such that all of the field degrees of freedom, would be integrated out of the resulting effective action.
Here the ADM approach to spin \cite{Steinhoff:2009ei,Steinhoff:2010zz} may serve as a guide,             
because there the spin gauge is completely fixed prior to eliminating the field.                         
It should be stressed, that in general, the EFT formulation and computation framework is a very robust and efficient one, all the more so with the employment of the NRG field decomposition. Therefore, considering also the great usefulness of the canonical Hamiltonian form for the GW detection efforts, the closed form transformation to canonical variables and potential, at the level of the effective EFT action from section \ref{VSXcan} here, can be used to obtain further Hamiltonians, based on an EFT formulation and computation of the potential, and hence may allow to have the best of both worlds.

\acknowledgments

We would like to thank Johannes Hartung for collaboration in early stages of this work, 
in particular for his help in initiating the automatizing of the comparison.
We would also like to thank Thibault Damour for helpful and pleasant discussion. 
This work has been done within the Labex ILP (reference ANR-10-LABX-63) part of the Idex SUPER, 
and received financial French state aid managed by the Agence Nationale de la Recherche, 
as part of the programme Investissements d'Avenir under the reference ANR-11-IDEX-0004-02.
It was also supported by FCT (Portugal) through projects SFRH/BI/52132/2013 and PCOFUND-GA-2009-246542
(co-funded by Marie Curie Actions).
JS is grateful for the kind hospitality at IAP where part of this work was done.

\bibliographystyle{jhep}
\bibliography{gwbibtex}

\end{document}